\documentclass[12pt]{article}

\usepackage{a4,graphicx,amssymb,slashed}
\usepackage[footnotesize,bf]{caption}

\usepackage{MnSymbol}

\usepackage{hyperref}
\hypersetup{
    colorlinks,
    linkcolor={red!50!black},
    citecolor={blue!50!black},
    urlcolor={blue!80!black}
}

\usepackage{xcolor}

\newcommand{\ind}[1]{\rm\scriptscriptstyle #1}

\def\lsim{\mathrel{\rlap{\lower4pt\hbox{\hskip1pt$\sim$}}
    \raise1pt\hbox{$<$}}}                
\def\gsim{\mathrel{\rlap{\lower4pt\hbox{\hskip1pt$\sim$}}
    \raise1pt\hbox{$>$}}}                

\begin{document}

\title{
\vspace{-3.25cm}
\flushright{\small ADP-23-12/T1221} \\
\vspace{-0.35cm}
{\small DESY-23-059} \\
\vspace{-0.35cm}
{\small Liverpool LTH 1340} \\
\vspace{-0.35cm}
{\small MIT-CTP/5582} \\
\vspace{-0.35cm}
{\small August 21, 2023} \\
\vspace{0.5cm}
\centering{\Large \bf Feynman--Hellmann approach to transition matrix 
                      elements and quasi-degenerate energy states}}

\author{\large
        M. Batelaan$^a$, K.~U. Can$^a$, R. Horsley$^b$, Y. Nakamura$^c$, \\
        P.~E.~L. Rakow$^d$, G. Schierholz$^e$, H. St\"uben$^f$, \\ 
        R.~D. Young$^{a,g}$ and J.~M. Zanotti$^a$ \\[1em]
        \small -- QCDSF-UKQCD-CSSM Collaboration -- \\[1em]
        \footnotesize $^a$ CSSM, Department of Physics,
               University of Adelaide, \\[-0.5em]
        \footnotesize Adelaide SA 5005, Australia \\[0.25em]
        \footnotesize $^b$ School of Physics and Astronomy,
               University of Edinburgh, \\[-0.5em]
        \footnotesize Edinburgh EH9 3FD, UK \\[0.25em]
        \footnotesize $^c$ RIKEN Center for Computational Science, \\[-0.5em]
        \footnotesize Kobe, Hyogo 650-0047, Japan \\[0.25em]
        \footnotesize $^d$ Theoretical Physics Division,
               Department of Mathematical Sciences, \\[-0.5em]
        \footnotesize University of Liverpool,
               Liverpool L69 3BX, UK \\[0.25em]
        \footnotesize $^e$ Deutsches Elektronen-Synchrotron DESY, \\[-0.5em]
        \footnotesize Notkestr. 85, 22607 Hamburg, Germany  \\[0.25em]
        \footnotesize $^f$ Universit\"at Hamburg,
               Regionales Rechenzentrum, \\[-0.5em]
        \footnotesize 20146 Hamburg, Germany \\[0.25em]
        \footnotesize $^g$ Center for Theoretical Physics, 
                           Massachusetts Institute of Technology, \\[-0.5em]
        \footnotesize Cambridge, MA 02139, USA}

\date{}

\maketitle


\begin{abstract}
The Feynman--Hellmann approach to computing matrix elements in lattice
QCD by first adding a perturbing operator to the action is described 
using the transition matrix and the Dyson expansion formalism. 
This perturbs the energies in the two-point baryon correlation
function, from which the matrix element can be obtained.
In particular at leading order in the perturbation
we need to diagonalise a matrix of near-degenerate energies. 
While the method is general for all hadrons, we apply it here to a 
study of a Sigma to Nucleon baryon transition vector matrix 
element.
\end{abstract}


\clearpage




\section{Introduction} 


Quantum Chromodynamics (QCD) -- the theory of quarks and gluons has
been spectacularly successful in describing inelastic scattering of 
particles at very high energies, as witnessed in particle accelerators. 
In this region the coupling constant decreases and allows the application
of perturbation theory for quarks and gluons. However at
lower energies these bind into hadrons. This is a non-perturbative
effect and presently the most successful method to try to describe
this is via numerical Monte Carlo simulations of a discretised
version of QCD -- lattice QCD. While this approach has been
pursued from the early days of QCD, it is only recently that
computer speeds have improved to such an extent that reasonably
accurate numerical results are possible. The general situation of the field
is given in \cite{Aoki:2021kgd}. While many early computations
were for the mass spectrum, more recently the focus is now on
matrix elements, particularly for the nucleon or more generally
for the baryon octet%
\footnote{While we shall concentrate on the baryon
octet in this article, the results presented here are more general 
and applicable to all hadrons.}.

Most of these baryon matrix elements are needed at non-zero momentum 
transfer. Typical examples are those relevant to lepton--hadron 
scattering processes leading to form factors, e.g.\ see 
\cite{Meyer:2022mix,Djukanovic:2021qxp,Can:2021ehb} for recent reviews,
or to inelastic processes such as DIS (deep inelastic scattering) with the
associated parton distribution functions, PDFs, e.g.\ 
\cite{Constantinou:2020hdm,Cichy:2018mum} or alternatively via the related 
hadron tensor or Compton amplitude to give the structure function 
\cite{Gambino:2022dvu,Chambers:2017dov,Liu:2016djw}.
(Often the Operator Product Expansion, OPE, is used as it is simpler 
to determine moments of structure functions, which are also related 
to matrix elements.) Alternatively matrix elements at low energies 
for baryon (or meson) semi-leptonic decays are of interest.
Indeed these matrix elements are becoming increasingly important,
as they provide crucial input into the precision determination
of elements of the Cabibbo-Kobayashi-Maskawa (CKM)
\cite{Seng:2021nar,Gottlieb:2020zsa}, nuclear physics \cite{Davoudi:2020ngi}
and the search for beyond-the-standard-model-effects in neutron $\beta$-decay, 
\cite{Severijns:2006dr,Gonzalez-Alonso:2018omy,Cirigliano:2020yhp,
Smail:2023eyk}, and as such are to be regarded as complementary to 
searches at the Large Hadron Collider, LHC.

Traditionally matrix elements have been computed from three-point
correlation functions.
On the lattice these require a (baryon) source and sink together 
with an operator between them. To avoid excited state contamination
and to achieve ground state dominance the distances between the source,
operator and sink must be large enough for this to be numerically achieved.
However given the lattice sizes at present available and coupled with 
the fact that higher-point correlation functions are numerically noisier,
this can be difficult to achieve. More recently an alternative approach based
on the Feynman--Hellmann theorem has emerged
\cite{QCDSF:2012mkm,CSSM:2014uyt,Chambers:2015bka,Bouchard:2016heu}. 
This perturbs the QCD action with a given operator leading to the 
required matrix element residing in the resulting energy shift. 
This can be determined from a two-point correlation function with 
just a source and sink, rather than a three-point correlation function.

While this approach has been successfully applied to elastic form
factors \cite{Chambers:2017tuf}, as described in more detail later
this needed an application of (degenerate) perturbation theory for matrix 
elements with baryons having the same energy. While possible, in practice
this restricts the approach. In this article we shall generalise 
previous results from needing degenerate energies to `near-degenerate'
or `quasi-degenerate' energies. As discussed here this then allows
us to consider processes such as the decays of baryons, where it is
otherwise difficult to achieve degeneracy of their respective energies.

Derivations of the the Feynman--Hellmann approach can be given
based on the $2$-point Green's function defined from the partition function,
or from the transfer matrix viewpoint. In this article we shall adopt 
the latter approach. In principle this can allow for a better discussion 
of the source and sink wavefunctions to be used. In section~\ref{transfer}
we briefly describe the transfer matrix technique, mainly to introduce
our notation and modification of the QCD Hamiltonian to include 
a perturbing operator. We shall take the spectrum of the QCD Hamiltonian
to have a set of isolated quasi-degenerate energy states.
In section~\ref{dyson_series} we shall consider a two-point baryon correlation 
function which upon using the Dyson expansion for the transfer matrix
leads, for large source--sink separation times $t$, to the result given 
in eq.~(\ref{C_fo}), namely a sum of exponential decays in $t$ with 
coefficients given by various perturbed energies. The energies are related
to eigenvalues from the diagonalisation of a matrix in the space of 
quasi-degenerate states and leads to the phenomenon of `avoided'
energy levels.The simplest case is of two quasi-degenerate states,
leading to the solution of a quadratic equation for the eigenvalues. 
To resolve these energy states we regard the associated $2$-point 
correlation function matrix as a Generalised Eigen-Value Problem, GEVP
(equivalent to a variational approach) 
\cite{Luscher:1990ck,Blossier:2009kd,Owen:2012ts}.
(This is further discussed in section~\ref{lattice} when we consider
the numerical implementation.) Furthermore, incorporating the spin index,
as also discussed in this section, leads to a doubling of the 
eigenvalue matrix. However, due to the spin structure of the baryon 
matrix elements under consideration this does not complicate 
the determination of the eigenvalues significantly.

The results are rather general, and in this article in 
section~\ref{quasi_degeb_N_energy_ela} and the Appendices we consider several
examples. They are all variations where the kinematic geometry is chosen 
so that initial baryon, $B$, moves with $3$-momentum $\vec{p}$ and the 
final baryon, $B^\prime$ with momentum $\vec{p}^{\,\prime} = \vec{p}+\vec{q}$ 
(or alternatively $\vec{p}-\vec{q}$) where $\vec{q}$ is the momentum transfer
chosen such that $E_B(\vec{p}) \approx E_{B^\prime}(\vec{p}^{\,\prime})$. 
Taking $B^\prime = B$ for flavour diagonal baryons describes the lepton
scattering case, while $B^\prime \not= B$ gives the flavour changing
decay case appropriate to investigating weak decays.
In section~\ref{lattice} we discuss specific lattice arrangements.
We first discuss our proposal for including the (quark) operator in the 
action and the subsequent matrix inversion. This effectively inserts 
the operator in quark lines between the source and sink baryons and so we
consider here valence insertions only. (To include sea quarks for
flavour diagonal matrix elements would require special purpose generation
of configurations or re-weighting with trace estimates.) The explicit 
example of the vector current decay $\Sigma \to N$, 
\cite{Guadagnoli:2006gj,Shanahan:2015dka,Sasaki:2017jue,Bickerton:2021yzn},
is then considered, whose transition matrix elements are flavour off-diagonal. 
Some numerical results follow, which we also compare with the conventional 
$3$-point correlation function determination of the matrix element. 
Section~\ref{conclusions} gives our conclusions. 

The Appendices give some further details of
the methods employed in this article. Appendix~\ref{euclid_FF} briefly 
discusses the Euclideanised matrix elements, for completeness, of all 
local bilinear currents. To evaluate these we need in turn the spinor 
bilinear terms, which are given in Appendix~\ref{spinor_results}. 
In Appendix~\ref{alt_energy_states} we give an alternative derivation 
of the energy results including spin for the examples 
considered here. In Appendix~\ref{corr_fun} we describe all the 
correlation functions needed for this article, while the last appendix
(\ref{fermion_inversion}) gives some more details
of the fermion matrix inversion employed here. Preliminary results have
appeared in \cite{Horsley:2022ouc}.


\section{The transfer matrix}
\label{transfer}


\subsection{Background}
\label{background}


In this article we shall consider the Euclidean $2$-point correlation 
function with a Hamiltonian which includes a perturbing operator with a 
possible $3$-momentum transfer $\vec{q}$
\begin{eqnarray}
   C_{\lambda\,B^\prime B}(t)
     &=& \langle \hat{\tilde{B}}^{\,\prime}(t;\vec{p}^{\,\prime}) 
                           \hat{\bar{B}}(0,\vec{0}) \rangle_\lambda
                                                             \nonumber  \\
     &\equiv& 
         {\rm tr}\, [\hat{\tilde{B}}^\prime(t;\vec{p}^{\,\prime}) 
                    \hat{\bar{B}}(0,\vec{0}) \hat{S}_\lambda(\vec{q})^T ]
           \, / \,\, {\rm tr} \, \hat{S}_\lambda(\vec{q})^T  \,,
\label{baryon_2pt}
\end{eqnarray}
where $T$ is the temporal box size and with $\hat{\bar{B}}(0,\vec{0})$ 
the initial baryon state at time $0$ and spatial origin $\vec{x}_0 = \vec{0}$
together with $\hat{\tilde{B}}^\prime(t;\vec{p}^{\,\prime})$ the final 
baryon state at time $t$ and momentum $\vec{p}^{\,\prime}$. 
Presently we shall ignore any complications arising from the baryon 
spin structure, and include this later by generalising appropriately 
the formulae obtained. (Other hadrons, for example mesons, 
could thus be considered.) The final baryon state%
\footnote{For simplicity we use a mixture of continuum notation and
discrete notation. We shall not consider any possible lattice artifacts 
effects here. So for example we shall use
\begin{eqnarray}
   {(2\pi)^3 \over V}\delta^3(\vec{p}-\vec{q}) 
      \equiv \delta_{\vec{p},\vec{q}} = \left\{ \begin{array}{ll}
                                               1 & \vec{p} = \vec{q} \\
                                               0 & \vec{p} \not= \vec{q}
                                            \end{array}
                                    \right. \,.
                                                             \nonumber
\end{eqnarray}}
\begin{eqnarray}
   \hat{\tilde{B}}^\prime(t;\vec{p}^{\,\prime}) 
      = \int_{\vec{x}} e^{-i\vec{p}^{\,\prime}\cdot\vec{x}} \hat{B}^\prime(t,\vec{x})\,,
\label{B_f}
\end{eqnarray}
is a function of momentum $\vec{p}^{\,\prime}$. We shall not consider 
any possible lattice discretisation effects in this article, so we shall use a 
continuum notation in all dimensions. As the initial baryon state is taken 
at the source position $\vec{x}_0 = \vec{0}$ it contains all momenta and thus
\begin{eqnarray}
   \hat{\bar{B}}(0,\vec{0}) 
    = \int_{\vec{p}} \, \hat{\tilde{\bar{B}}}(0;\vec{p}) \,.
\end{eqnarray}
This arrangement is adopted because when numerically finding the 
correlation function we invert the Dirac operator for the Green's function 
on a spatial source point. 

The transfer matrix $\hat{S}_\lambda$ is defined by
\begin{eqnarray}
   \hat{S}_\lambda(\vec{q}) = e^{-\hat{H}_\lambda(\vec{q})} \,,
\end{eqnarray}
where we assume the Hamiltonian, $\hat{H}_\lambda$, exists together with 
the associated complete set of energy eigenstates%
\footnote{Strictly speaking, even for the unperturbed action considered 
later here, see section~\ref{lattice_details}, positivity is lost, 
but a transfer matrix can still be defined, \cite{Luscher:1984is}. 
Practically this is not a problem and we ignore this point here.},
in particular a unique vacuum state.

We shall consider a perturbed Hamiltonian, here given by
\begin{eqnarray}
   \hat{H}_\lambda(\vec{q}) = \hat{H}_0 
         + \sum_\alpha \lambda_\alpha \hat{\tilde{{\cal O}}}_\alpha(\vec{q}) \,,
\label{pert_ham}
\end{eqnarray}
with momentum $\vec{q}$, as an expansion in $\lambda_\alpha$ where $\alpha$ 
is to be regarded as just a label (so for example can be a single
Lorentz index or a collection of indices). The perturbing operator
$\hat{\tilde{{\cal O}}}_\alpha(\vec{q})$ is defined by
\begin{eqnarray}
   \hat{\tilde{{\cal O}}}_\alpha(\vec{q})
      = \int_{\vec{x}} \left( \hat{O}_\alpha(\vec{x})e^{i\vec{q}\cdot\vec{x}}
                            + \hat{O}_\alpha^\dagger(\vec{x}) e^{-i\vec{q}\cdot\vec{x}}
                     \right) \,,
\label{op_def}
\end{eqnarray}
where $\hat{O}_\alpha(\vec{x})$ may be taken to be a bilinear in the
quark fields, i.e.\ a generalised current, see Appendix~\ref{euclid_FF} 
for some more details. $\hat{H}_0$ conserves momentum, but $\hat{H}_\lambda$
only conserves momentum modulo $\vec{q}$. In this form 
$\hat{\tilde{{\cal O}}}_\alpha(\vec{q})$ is Hermitian. 
Note that in \cite{Can:2020sxc} we considered
just the case where $\hat{O}_\alpha(\vec{x})$ is also Hermitian. 
(It is possible to generalise to non-Hermitian operators, see 
\cite{Chambers:2015bka}, however we shall not consider this further here.)
As the previous equations indicate, we are considering
operators defined in Euclidean space. The Hermiticity relation for 
bilinear operators between the Euclidean and Minkowski spaces is also briefly
discussed in Appendix~\ref{euclid_FF}. It is also easy to include 
covariant derivatives, for example in \cite{Best:1997qp} eq.~(23)
where the general relation between the Minkowski and Euclidean operators
for the vector and axial currents was given. We do not discuss this case
further here.

We can also incorporate the generalisation to complex $\lambda$ by writing 
$\lambda$ in polar form, $\lambda_\alpha = |\lambda_\alpha| e^{i\phi_{\alpha}}$
and absorb the phase into the definition of the operator. Thus we have
\begin{eqnarray}
   \lambda_\alpha\hat{\tilde{{\cal O}}}_\alpha(\vec{q})
      &\to& \lambda_\alpha\int_{\vec{x}} \hat{O}_\alpha(\vec{x})e^{i\vec{q}\cdot\vec{x}}
            + \lambda_\alpha^*\int_{\vec{x}}
                    \hat{O}_\alpha^\dagger(\vec{x})e^{-i\vec{q}\cdot\vec{x}}
                                                       \nonumber    \\
      &=& |\lambda_\alpha|\int_{\vec{x}} 
           (e^{i\phi_{\alpha}}\hat{O}_\alpha(\vec{x}))e^{i\vec{q}\cdot\vec{x}}
            + |\lambda_\alpha|\int_{\vec{x}}
              (e^{i\phi_{\alpha}}\hat{O}_\alpha(\vec{x}))^\dagger
                                            e^{-i\vec{q}\cdot\vec{x}} \,.
\label{complex_lam}
\end{eqnarray}
This can be useful if we are considering the $O(\lambda^2)$ terms 
which gives the Compton Amplitude
\cite{Chambers:2017dov,Can:2020sxc,Can:2022rgi},
as indicated here in section \ref{pert_en} (real $\lambda$ gives the
symmetric part of the amplitude while complex $\lambda$
enables the antisymmetric part of the Compton Amplitude to be 
determined). However as we are only interested in the $O(\lambda)$ 
result here, for simplicity of notation in future we just take 
$\lambda_\alpha$ as real. In addition for this case then the index 
$\alpha$ is redundant, as we are practically just considering 
one operator. So we shall usually suppress it, but it can easily be reinstated
if necessary. (Again, if we are interested in the $O(\lambda^2)$
or higher order terms then the index is relevant, as cross terms 
of operators appear.)

First using 
$\hat{\tilde{B}}^\prime(t;\vec{p}) 
 = \hat{S}_\lambda^{\dagger}(\vec{q})^t\hat{\tilde{B}}^\prime(0;\vec{p}) 
                                                \hat{S}_\lambda(\vec{q})^t$
and then inserting a complete set of states (in the presence of the 
perturbation) and taking the temporal box size large picks out the 
vacuum state and gives the usual result
\begin{eqnarray}
   C_{\lambda\,B^\prime B}(t)
      = {}_\lambda\langle 0| \hat{\tilde{B}}^\prime(0;\vec{p}^\prime) 
                           \hat{S}_\lambda(\vec{q})^t 
                           \hat{\bar{B}}(0,\vec{0}) |0\rangle_\lambda \,,
\label{C_basic}
\end{eqnarray}
where $|0\rangle_\lambda$ is the vacuum in the presence of the perturbation
and the spectrum of $\hat{H}$ is now normalised with respect to this
vacuum. As all the operators are at time $t=0$, in future we drop this 
argument. Eq.~(\ref{C_basic}) is the basic equation we shall consider in 
this article.


\subsection{Quasi-degenerate energy states}
\label{quasi_degen}


We shall first derive a general expression, and then consider particular
cases. In particular we shall consider discrete degenerate energy states, 
i.e.\ $E_{B_r}(\vec{p}_r) = E_{B_s}(\vec{p}_s)$ or near-degenerate energy 
states $E_{B_r}(\vec{p}_r) \approx E_{B_s}(\vec{p}_s)$, both possibilities
labelled by $r = 1, 2, \ldots$ (similarly for $s$) each with a given 
fixed momentum. Collectively we call this set $S$ of `quasi-degenerate energy' 
states, the total number being $d_S$.

\begin{figure}[!htbp]
   \begin{center}
     \includegraphics[width=3.50cm]{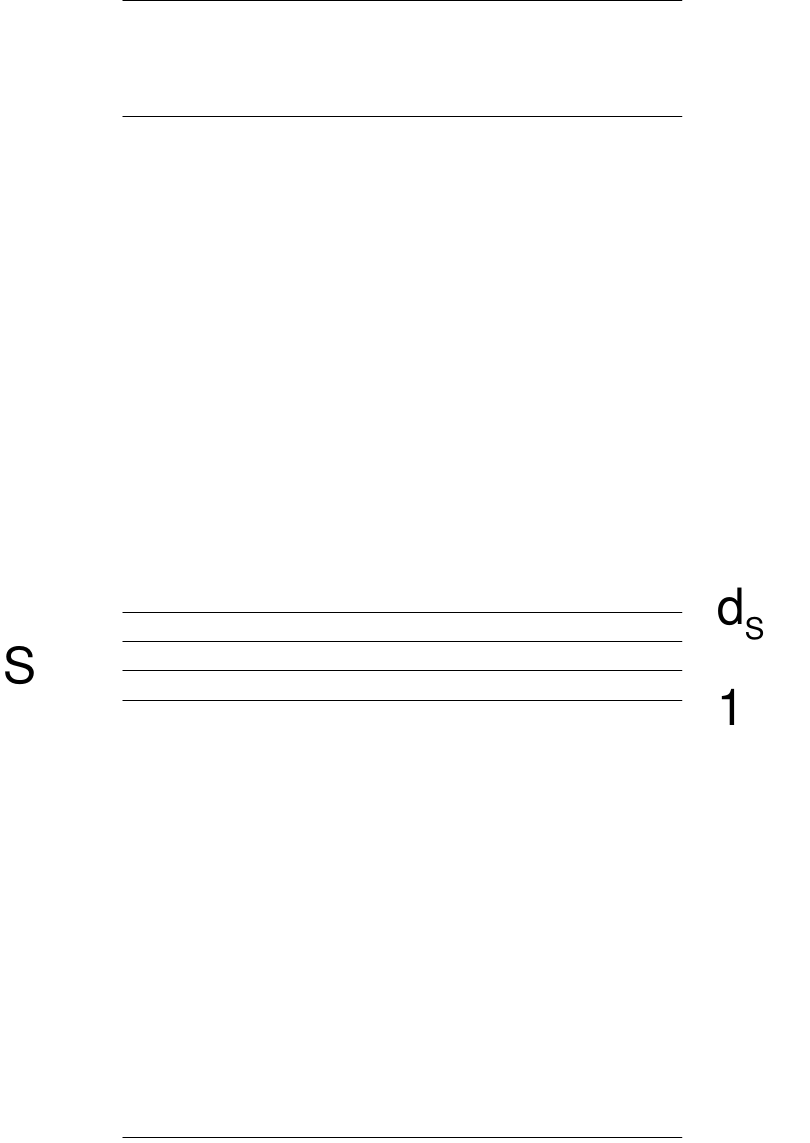}
   \end{center}
   \caption{A sketch of the energy levels. The set of 
            quasi-degenerate energy states are denoted by $S$, labelled
            from $1$ to $d_S$. These states are well separated from other
            states.}
\label{sketch_energy_levels}
\end{figure}

In this scenario, as we shall see, simple perturbation theory as it 
stands breaks down and we have to consider degenerate perturbation theory. 
This also ensures smooth behaviour in $\lambda$. In the following 
we shall assume that these energy states are the only possible 
quasi-degenerate states and well separated from other states, 
as sketched in Fig.~\ref{sketch_energy_levels}.
We shall later argue that other states are either more damped (those 
with higher energies in the figure), or for any lower state(s)
a GEVP must be applied. However in this article we will only consider the
quasi-degenerate energy states as the ground states.

The spectrum of the unperturbed Hamiltonian, $\hat{H}_0$, is given by 
\begin{eqnarray}
   \hat{H}_0 |X(\vec{p}_X) \rangle = E_X(\vec{p}_X) |X(\vec{p}_X) \rangle \,.
\label{energy_spectrum}
\end{eqnarray}
Let $S$ be the discrete set of quasi-degenerate energy states and have 
$d_S$ elements labelled by $r$. More concretely we write for these states
\begin{eqnarray}
   E_{B_r}(\vec{p}_r) 
      = \bar{E} + \epsilon_r \,, \qquad r = 1\,, \ldots \,, d_S \,,
\label{almost_degen_energy}
\end{eqnarray}
where $\bar{E}$ is some suitable energy close to all the quasi-degenerate
energies. It could be taken as the average over the quasi-degenerate 
energy states $\bar{E}  = ( E_{B_1} + \ldots + E_{B_{d_S}})/d_S$ where we would 
have $\epsilon_1 + \ldots + \epsilon_{d_S} = 0$ but this is not necessary 
in the following. (Alternatively we could choose one of the quasi-degenerate 
energy states, such as the one with lowest energy.)
Writing $\epsilon_r = \epsilon c_r$ where $c_r \sim O(1)$ then 
$\epsilon \sim |E_{B_r}(\vec{p}_r) - E_{B_s}(\vec{p}_s)|$
effectively represents the difference in energies between the 
quasi-degenerate states  where $\epsilon$ is small and is taken in the 
following to be another expansion parameter in addition to $\lambda$.
The corresponding states are denoted by $|B_r(\vec{p}_r) \rangle$.
For these quasi-degenerate states we have the energies $E_{B_r}(\vec{p}_r)$
defined by
\begin{eqnarray}
   \hat{H}_0 |B_r(\vec{p}_r) \rangle 
         = E_{B_r}(\vec{p}_r) |B_r(\vec{p}_r) \rangle \,.
\end{eqnarray}
The set of unperturbed states obeys the completeness condition
where we sum over all states and momenta. We often explicitly isolate
the quasi-degenerate states so
\begin{eqnarray}
   \lefteqn{\sumint_{X(\vec{p}_X)}
      |X(\vec{p}_X) \rangle \, \langle X(\vec{p}_X)| \equiv}
      & &                                                 \nonumber    \\
      & & \hspace*{0.75in}
   \sum_r \, 
      |B_r(\vec{p}_r) \rangle \, \langle B_r(\vec{p}_r)| +
   \sumint_{X(\vec{p}_X) \,\, \not\in \,\, S}  
      |X(\vec{p}_X) \rangle \, \langle X(\vec{p}_X)|
        = \hat{1} \,.
\label{complete_unit}
\end{eqnarray}
We use the lattice normalisation, namely
\begin{eqnarray}
   \langle X(\vec{p}_X)|Y(\vec{p}_Y)\rangle 
       = \delta_{X,Y}\,\delta_{\vec{p}_X,\vec{p}_Y} \,.
\end{eqnarray}
However all the formulae and results are such that they can be easily 
converted to another normalisation by the substitution for all states
\begin{eqnarray}
   |X(\vec{p}_X)\rangle 
      \to { |X(\vec{p}_X)\rangle \over
             \sqrt{\langle X(\vec{p}_X)|X(\vec{p}_X) \rangle} }\,,\qquad
   |0\rangle \to |0\rangle \,.
\label{normalisation}
\end{eqnarray}
The usual case, of course, is the relativistic normalisation
\begin{eqnarray}
   \langle X(\vec{p}_X)|Y(\vec{p}_Y)\rangle_{\rm rel} 
      = 2E_X(\vec{p}_X)\delta_{X,Y}\delta_{\vec{p}_X,\vec{p}_Y} \,,
\label{foot_norm}
\end{eqnarray}
which we shall later use when discussing the numerical results.

Now inserting two complete sets of unperturbed states before 
and after $\hat{S}_\lambda^t$ in eq.~(\ref{C_basic}) gives
\begin{eqnarray}
   \lefteqn{C_{\lambda\,B^\prime B}(t) =}
     & &                                     \label{intermediate_C}  \\
     & & \hspace*{-0.15in}
         \sumint_{X(\vec{p}_X)} \sumint_{Y(\vec{p}_Y)} 
            {}_\lambda\langle 0 |
                 \hat{\tilde{B}}^\prime(\vec{p}^\prime) 
                                              | X(\vec{p}_X) \rangle \,
        \langle X(\vec{p}_X) | \hat{S}_\lambda(\vec{q})^t 
                               | Y(\vec{p}_Y) \rangle \, 
        \langle Y(\vec{p}_Y) | \hat{\bar{B}}(\vec{0}) 
                                |0\rangle_\lambda  \,.
                                                             \nonumber
\end{eqnarray}
From eq.~(\ref{B_f}), as $\hat{B}^\prime$ has a definite momentum, 
$\vec{p}^\prime$, we can take the geometry to be such that we
have a good overlap with just one of the $d_S$ quasi-degenerate states 
$|B_r(\vec{p}_r)\rangle$ as depicted in Fig.~\ref{sketch_energy_levels}.
For $\hat{B}(\vec{0})$ we also choose an operator with
a good overlap with one of the quasi-degenerate states noting that
it contains all momenta. We shall further discuss this in the next section,
but initially we shall keep the operators general.


\section{Dyson series and the correlation function}
\label{dyson_series}


\subsection{Perturbed energies}
\label{pert_en}


We wish to determine $C_{\lambda\,B^\prime B}(t)$ to $O(\lambda)$.
To this end, first for any two operators $\hat{A}$, $\hat{B}$ consider 
the function defined by $f(t) = e^{-t\hat{A}} e^{t(\hat{A}+\hat{B})}$.
By the usual technique of differentiating and then integrating $f(t)$ 
with respect to $t$ we soon find the operator identity
\begin{eqnarray}
   e^{t(\hat{A}+\hat{B})} 
      = e^{t\hat{A}} 
           + \int_0^t dt^\prime \, e^{(t-t^\prime)\hat{A}} \hat{B}
                                  \, e^{t^\prime(\hat{A}+\hat{B})} \,.
\end{eqnarray}
Regarding $\hat{B}$ as `small', this can be iterated.
From eq.~(\ref{pert_ham}) we thus set $\hat{A} \to - \hat{H}_0$ and 
$\hat{B} \to -\lambda_\alpha \hat{\tilde{{\cal O}}}_\alpha$.
This gives to $O(\lambda^2)$,
\begin{eqnarray}
   e^{-(\hat{H}_0 + \lambda_\alpha\hat{\tilde{{\cal O}}}_\alpha)t}
      &=& e^{-\hat{H}_0t} 
           - \lambda_\alpha\, \int_0^t dt^\prime\, e^{-\hat{H}_0(t - t^\prime)} \, 
                    \hat{\tilde{{\cal O}}}_\alpha \, e^{-\hat{H}_0 t^\prime}
                                                          \nonumber    \\
      & & \phantom{e^{-\hat{H}_0t}}
           + \lambda_\alpha\lambda_\beta \, 
               \int_0^t dt^\prime\,\int_0^{t^\prime} dt^{\prime\prime} \,
                 e^{-\hat{H}_0(t-t^\prime)} \, \hat{\tilde{{\cal O}}}_\alpha \, 
                 e^{-\hat{H}_0(t^\prime-t^{\prime\prime})}
                    \, \hat{\tilde{{\cal O}}}_\beta e^{-\hat{H}_0 t^{\prime\prime}}
                                                          \nonumber    \\
      & & \phantom{e^{-\hat{H}_0t}}
            + O(\lambda^3) \,,
\label{dyson_expansion}
\end{eqnarray}
which is equivalent to the Dyson expansion. 
We note that the term quadratic in $\lambda$ can be manipulated 
into a form appropriate for the Compton amplitude. An alternative
derivation using the path integral is discussed in 
\cite{Chambers:2017dov,Can:2020sxc}. A recent review is given in
\cite{Can:2022chd}. For the specific approach using the transfer matrix 
given here see also \cite{Can:2022rgi}.

To evaluate $C_{\lambda\,B^\prime B}(t)$ we apply the Dyson 
expansion of eq.~(\ref{dyson_expansion}) to eq.~(\ref{intermediate_C})
after splitting the completeness relation as given in eq.~(\ref{complete_unit}).
As mentioned before we shall consider the case where $E_X$ (and $E_Y$) 
are much greater than all the isolated quasi-degenerate states as depicted in 
Fig.~\ref{sketch_energy_levels}, i.e.\ $E_X, E_Y \gg \bar{E}$ in 
eq.~(\ref{complete_unit}). There are four terms and dropping temporarily
the momentum arguments gives
\begin{eqnarray}
   \langle B_r| e^{-(\hat{H}_0 + \lambda\hat{\tilde{{\cal O}}})t} | B_s\rangle
      &=& \phantom{\lambda}e^{-\bar{E}t} 
             \left(\delta_{rs} - tD_{rs} +O(2)\right) \,,
                                                          \nonumber    \\
   \langle B_r| e^{-(\hat{H}_0 + \lambda\hat{\tilde{{\cal O}}})t} | Y\rangle
      &=& -e^{-\bar{E}t} \left( 
              \lambda { \langle B_r|\hat{\tilde{{\cal O}}}|Y\rangle
                  \over
                       E_Y-E_{B_r} } + O(2)
                      \right)
                   +  \begin{array}{c}
                         \mbox{more}    \\
                         \mbox{damped}  
                      \end{array} \,,
                                                          \nonumber    \\
   \langle X| e^{-(\hat{H}_0 + \lambda\hat{\tilde{{\cal O}}})t} | B_s\rangle
      &=& -e^{-\bar{E}t} \left( 
              \lambda { \langle X|\hat{\tilde{{\cal O}}}|B_s\rangle
                  \over
                       E_X-E_{B_s} } + O(2)
                      \right)
                   +  \begin{array}{c}
                         \mbox{more}    \\
                         \mbox{damped}  
                      \end{array} \,,
                                                          \nonumber    \\
   \langle X| e^{-(\hat{H}_0 + \lambda\hat{\tilde{{\cal O}}})t} | Y\rangle
      &=& \begin{array}{c}
             \mbox{more}    \\
             \mbox{damped}  
          \end{array} \,,
\label{MErs_trans}
\end{eqnarray}
where we have defined the $d_S\times d_S$ matrix%
\footnote{$D$ is a function of the momenta, but as with
$C_{\lambda\,B^\prime B}(t)$ we shall suppress this dependence.}
\begin{eqnarray}
   D_{rs} =  \epsilon_r \delta_{rs}
           + \lambda a_{rs} \,, \quad \mbox{with} \quad
               a_{rs} = \langle B_r(\vec{p}_r) | \hat{\tilde{{\cal O}}}(\vec{q}) 
                                              | B_s(\vec{p}_s) \rangle \,.
\label{matrix_cr_rs}
\end{eqnarray}
Note that from eq.~(\ref{almost_degen_energy}) we have 
$\epsilon_r = E_{B_r}-\bar{E}$. In eq.~(\ref{MErs_trans}) ``more damped'' 
means that these terms drop off as $\propto e^{-E_Xt}$, i.e.\ faster 
then $e^{-\bar{E}t}$. The kept terms (i.e.\ $D$) means terms of the form 
$O(1)$ or $O(\epsilon t)$, $O(\lambda t)$ while $O(2)$ means terms 
of the form $O(\epsilon^2 t^2)$, $O(\lambda^2 t^2)$, $O(\epsilon t\lambda t)$. 
Thus for this expansion to be valid we need 
$\lambda t \ll 1$, $|\epsilon_r| t \ll 1$ and $t \gg 0$ (for the damped 
terms to be negligible) thus
\begin{eqnarray}
   0 \ll t \ll {1 \over \lambda}\,, \qquad \mbox{and} \qquad
   0 \ll t \ll {1 \over \max |E_{B_r}(\vec{p}_r)-E_{B_s}(\vec{p}_s)|} \,.
\label{t_condition}
\end{eqnarray}
Furthermore defining $|B_s(\vec{p}_s)\rangle_\lambda$ as
\begin{eqnarray}
   |B_s(\vec{p}_s)\rangle_\lambda
      = |B_s(\vec{p}_s)\rangle
         - \lambda \sumint_{E_Y \gg \bar{E}}
                    |Y(\vec{p}_Y)\rangle  \,
                    { \langle Y(\vec{p}_Y)|\hat{\tilde{{\cal O}}}(\vec{q})
                      |B_s(\vec{p}_s)\rangle  
                    \over
                    E_Y - E_{B_s} } \,,
\label{state_lambda}
\end{eqnarray}
then we can re-write eq.~(\ref{intermediate_C}) as
\begin{eqnarray}
   \lefteqn{C_{\lambda\,B^\prime B}(t)} \hspace*{0.25in}
     & &                                                  \nonumber    \\
     &=& \sum_{rs} \, {}_\lambda\langle 0 |
                 \hat{\tilde{B}}^\prime(\vec{p}^{\,\prime}) 
                                       | B_r(\vec{p}_r) \rangle_\lambda \,\,
          \langle B_r|e^{-(\hat{H}_0 + \lambda\hat{\tilde{{\cal O}}})t} 
                     | B_s\rangle \,\,
         {}_\lambda\langle B_s(\vec{p}_s) | \hat{\bar{B}}(\vec{0}) 
                                |0\rangle_\lambda \,,
\label{2pt_correl_degen_final}
\end{eqnarray}
where
\begin{eqnarray}
   \langle B_r(\vec{p}_r)|e^{-(\hat{H}_0 + \lambda\hat{\tilde{{\cal O}}})t} 
                         | B_s(\vec{p}_s)\rangle
      = (\delta_{rs} - tD_{rs}) \times e^{-\bar{E}t} \,.
\label{ME_1+D}
\end{eqnarray}
Note that we have achieved a factorisation where any unwanted 
$|Y\rangle$ states, with $E_Y \gg E_{B_s}$, have been absorbed into the time 
independent renormalisation of the wavefunction and do not need 
to be further considered.

The matrix $D$ given in eq.~(\ref{matrix_cr_rs}) can be diagonalised, 
as it is Hermitian by construction. Let $\mu^{(i)}$ be the real eigenvalues
and $e_r^{(i)}$ the associated orthonormal $d_S$ dimensional eigenvectors
\begin{eqnarray}
   \sum_{i=1}^{d_S} e^{(i)}_r e^{(i)\,*}_s  = \delta_{rs} \,, \qquad
   \sum_{r=1}^{d_S} e^{(i)\,*}_r e^{(j)}_r  = \delta^{ij} \,.
\end{eqnarray}
Thus we have
\begin{eqnarray}
   D_{rs} = \sum_{i=1}^{d_S} \mu^{(i)} e^{(i)}_r e^{(i)\,*}_s \,.
\label{D_spectral_decomp}
\end{eqnarray}
(Note that to find the eigenvalues we have to first solve a $d_S$-dimensional
polynomial.) So all together using this in eq.~(\ref{ME_1+D}) we find
the intermediate result
\begin{eqnarray}
   \langle B_r|e^{-(\hat{H}_0 + \lambda\hat{\tilde{{\cal O}}})t} | B_s\rangle
      =  \sum_{i=1}^{d_S} e^{(i)}_r \left[ 1 - t \mu^{(i)} 
                                \right] e^{(i)\,*}_s
          \times e^{-\bar{E}t} \,,
\label{matrix_rs}
\end{eqnarray}
which we now use to find the final form of the correlation function.


\subsection{The correlation function}
\label{matrix_diag}


\subsubsection{General result}


Finally, we re-exponentiate the first term in eq.~(\ref{matrix_rs})
and then substitute back into eq.~(\ref{2pt_correl_degen_final}) to give the 
leading term at large $t$ of
\begin{eqnarray}
   C_{\lambda\,B^\prime B}(t)
      = \sum_{i=1}^{d_S} A^{(i)}_{\lambda\,B^\prime B} \, e^{-E^{(i)}_\lambda t} \,,
\label{C_fo}
\end{eqnarray}
where the perturbed energies are given by
\begin{eqnarray}
   E_\lambda^{(i)} = \bar{E} + \mu^{(i)} \,,
\label{pert_degen_en}
\end{eqnarray}
and the amplitude
\begin{eqnarray}
   A_{\lambda\,{B^\prime B}}^{(i)} = w_{B^\prime}^{(i)} \bar{w}^{(i)}_B \,,
\end{eqnarray}
with
\begin{eqnarray}
   w_{B^\prime}^{(i)} = \sum_{r=1}^{d_S} Z^{B^\prime}_r e_r^{(i)}\,, 
   \quad \mbox{and} \quad
   \bar{w}_{B}^{(i)} = \sum_{s=1}^{d_S} \bar{Z}^B_s e_s^{(i)*} \,,
\label{w_def}
\end{eqnarray}
where the wavefunctions, or overlaps, are
\begin{eqnarray}
   Z^{B^\prime}_r = {}_\lambda\langle 0 | \hat{\tilde{B}}^\prime(\vec{p}^\prime)
                                      | B_r(\vec{p}_r) \rangle_\lambda
   \quad \mbox{and} \quad
   \bar{Z}^B_s = {}_\lambda\langle B_s(\vec{p}_s) | \hat{\bar{B}}(\vec{0}) 
                                                  |0 \rangle_\lambda \,.
\label{Z_def_gen}
\end{eqnarray}
Eqs.~(\ref{C_fo}) -- (\ref{Z_def_gen}) are the results that
we shall be using in the following.

In the final/initial baryon space, $\{B^\prime, B\}$, the determination 
of $E_\lambda^{(i)}$, $i = 1,\ldots,d_S$ is now equivalent
to a GEVP, where we diagonalise a matrix of correlation functions. 
To determine all the energies we thus require this to be at least 
a $d_S \times d_S$ matrix, so both the sets $\{B^\prime\}$ and $\{B\}$ 
must be at least $d_S$ dimensional.

If there were states $|Z\rangle$ with lower energy than the
quasi-degenerate energy states and hence less damped than these states then
the $\{B^\prime, B\}$ space must be increased and a larger GEVP applied.
We do not consider this lower energy case further here, 
and take the quasi-degenerate energy states to be the lowest states.
Additionally, if the higher energy states have not died away sufficiently
then a larger $\{B^\prime, B\}$ space could also be used.


\subsubsection{A simplification}
\label{simple}


The above result is true for general source and sink operators. 
If as mentioned before, we set $\hat{B}^\prime$ and $\hat{B}$ 
`close' to $\hat{B}_r$ and $\hat{B}_s$ respectively then the above 
expressions greatly simplify and we expect that eq.~(\ref{w_def}) reduces to
\begin{eqnarray}
   w_r^{(i)}=Z_r^{B_r} e_r^{(i)}\,, \quad \mbox{and} \quad
   \bar{w}_s^{(i)} = \bar{Z}_s^{B_s} e_s^{(i)*} \,.
\end{eqnarray}
In turn this means that the overlaps $Z_r^{B_r}$ and $\bar{Z}_s^{B_s}$
although defined using the perturbed states of eq.~(\ref{state_lambda}),
the $O(\lambda)$ terms have then little effect. For example for $Z_r^{B_r}$
using eq.~(\ref{state_lambda})%
\footnote{We shall assume that this also holds for the perturbed vacuum, 
$|0\rangle_\lambda$.}
to expand 
${}_\lambda\langle 0|\hat{\tilde{B}}_r(\vec{p}_r)|B_r(\vec{p}_r)\rangle_\lambda$
the $O(\lambda)$ terms which involve overlaps such as 
$\langle 0 |\hat{B}_r|Y\rangle$ or $\langle X|\hat{B}_r|B_r\rangle$ 
vanish or are small due to the orthogonality of the spectrum,
so the effect of the perturbation on the overlaps is higher order in $\lambda$. 
We thus have
\begin{eqnarray}
   Z^{B_r}_r = \langle 0 | \hat{B}_r(\vec{0})
                                   | B_r(\vec{p}_r) \rangle + \ldots \,
   \quad \mbox{and} \quad
   \bar{Z}^{B_s}_s = \langle B_s(\vec{p}_s) | \hat{\bar{B}}_s(\vec{0}) 
                                                   |0 \rangle + \ldots \,.
\label{Z_def}
\end{eqnarray}
where in addition for $Z^{B_r}_r$ we have also used
$\hat{B}(\vec{x}) = e^{-i\hat{\vec{p}}\cdot\vec{x}} \, \hat{B}(\vec{0}) \,
                          e^{i\hat{\vec{p}}\cdot\vec{x}}$ to re-write it in the
above form.


\subsection{The relation between the initial and final momenta}
\label{matrix_element}


While the equations in section~\ref{matrix_diag} are the basic results, 
this discussion is general and can be applied to many quantum systems. 
We shall now be more specific to the situation here.
However before considering some examples we shall first discuss some
properties of the matrix element appearing in eq.~(\ref{matrix_cr_rs}). 
Using $\hat{O}(\vec{x}) = e^{-i\hat{\vec{p}}\cdot\vec{x}} \, \hat{O}(\vec{0}) \,
       e^{i\hat{\vec{p}}\cdot\vec{x}}$
we soon find
\begin{eqnarray}
   \lefteqn{
     \langle B_r(\vec{p}_r)| \hat{\tilde{{\cal O}}}(\vec{q}) 
                          | B_s(\vec{p}_s) \rangle}
     \hspace*{0.50in}
    & &                                              \label{matrix_el} \\
    &=& \langle B_r(\vec{p}_r)| \hat{O}(\vec{0})| B_s(\vec{p}_s) \rangle
          \, \delta_{\vec{p}_r,\vec{p}_s+\vec{q}}
       + \langle B_r(\vec{p}_r)| \hat{O}^\dagger(\vec{0})
                                                    |B_s(\vec{p}_s) \rangle
        \, \delta_{\vec{p}_r,\vec{p}_s-\vec{q}} \,.
                                                          \nonumber
\end{eqnarray}
Thus the initial momentum, $\vec{p}_s$ either steps up or down 
by $\vec{q}$, i.e.\
\begin{eqnarray}
   \vec{p}_r = \vec{p}_s + \vec{q} \,, \quad \mbox{or} \quad
   \vec{p}_r = \vec{p}_s - \vec{q} \,,
\end{eqnarray}
and the quasi-degenerate states, as sketched in 
Fig.~\ref{sketch_energy_levels}, are mixed together.

As a simple example, to be discussed in some detail in 
section~\ref{quasi_degeb_N_energy_ela}, let us take the two dimensional 
quasi-degenerate state subspace as having momentum $\vec{p}$ and
$\vec{p}+\vec{q}$. Thus the final momentum $\vec{p}_r$ 
can be chosen to be either $\vec{p}_r = \vec{p}+\vec{q}$ with
the $+$ sign and $\vec{p}_s = \vec{p}$ (or $\vec{p}_r = \vec{p}-\vec{q}$ 
with the $-$ sign and $\vec{p}_s = \vec{p}$) to remain within 
this subspace. We shall use these results frequently 
in the coming presentation.

A corollary from eq.~(\ref{matrix_el}) is that for a non-zero 
momentum transfer, $\vec{q} \not= 0$, the diagonal matrix elements
$a_{rr}$ in eq.~(\ref{matrix_cr_rs}) are zero, so the $O(\lambda)$
terms vanish and hence $D$ becomes trivial. This was alluded to before: 
if we are investigating momentum transfer and form factors, 
then we are forced to consider the degenerate energy case 
to determine the matrix element \cite{Chambers:2017tuf}.
Non-zero off-diagonal matrix elements leads to the phenomenon of
avoided energy levels, as discussed later in 
section~\ref{quasi_degeb_N_energy_ela}.

As well as degeneracies between levels differing in momentum by $\pm\,\vec{q}$
there will also be cases where states differing by $\pm2\,\vec{q}$,
$\pm3\,\vec{q}$ etc.\ are nearly degenerate. Such degeneracies will be 
converted into avoided level crossings by the operator acting multiple times. 
(These are determined by higher orders in $\lambda$ of the Dyson expansion 
in eq.~(\ref{dyson_expansion}).) We have not investigated these 
higher order cases here.


\subsection{Incorporating the spin index}
\label{spin_index}


We now consider the complications caused by the spinor index and
the consequent spin-$1/2$ carried by the octet baryons. Until now we
have postponed this discusion, so strictly the previous results
correspond to spinless scalar particles. To incorporate the spin index, 
$\sigma$ and the corresponding Dirac index $\alpha$ we shall see that 
this involves an alternative approach to that usually used when 
computing $3$-point correlation functions. We first generalise 
eqs.~(\ref{2pt_correl_degen_final}) and (\ref{ME_1+D}) appropriately
and together with $\hat{B}^\prime \sim \hat{B}_r$ and
$\hat{B} \sim \hat{B}_s$ this gives
\begin{eqnarray}
   \lefteqn{C_{\lambda\,B_{r\alpha}\,B_{s\beta}}(t)} \hspace*{0.25in}
     & &                                                   \nonumber \\
     &=& \sum_{\sigma_r\sigma_s} \,
            {}_\lambda\langle 0 |
                 \hat{\tilde{B}}_{r\alpha}(\vec{p}_r) 
                           | B_r(\vec{p}_r,\sigma_r) \rangle_\lambda \times  
                                 \label{2pt_correl_degen_final_spin} \\
     & &  \hspace*{0.50in}                                                     
          \langle B_r(\vec{p}_r,\sigma_r)
                           |e^{-(\hat{H}_0 + \lambda\hat{\tilde{{\cal O}}})t} 
                              | B_s(\vec{p}_s,\sigma_s)\rangle \times
         {}_\lambda\langle B_s(\vec{p}_s,\sigma_s) 
                              | \hat{\bar{B}}_{s\beta}(\vec{0}) 
                              |0\rangle_\lambda \,,
                                                         \nonumber
\end{eqnarray}
where
\begin{eqnarray}
   \langle B_r(\vec{p}_r,\sigma_r)|e^{-(\hat{H}_0 + \lambda\hat{\tilde{{\cal O}}})t} 
                         | B_s(\vec{p}_s,\sigma_s)\rangle
      &=& (\delta_{\sigma_r\sigma_s}\delta_{rs} - tD_{\sigma_rr,\sigma_ss}) 
                            \times e^{-\bar{E}t} \,,
\label{BrOBs}
\end{eqnarray}
and
\begin{eqnarray}
   D_{\sigma_rr,\sigma_ss} = \epsilon_r \delta_{\sigma_r\sigma_s} \delta_{rs}
        + \lambda a_{\sigma_r r, \sigma_s s} \,,
\label{matrix_cr_rs_spin}
\end{eqnarray}
where we have now defined $a_{\sigma_r r, \sigma_s s}$ as the matrix element
\begin{eqnarray}
   a_{\sigma_r r, \sigma_s s}
      =  \langle B_r(\vec{p}_r,\sigma_r) | \hat{\tilde{{\cal O}}}(\vec{q}) 
                                        | B_s(\vec{p}_s,\sigma_s) \rangle \,.
\label{a_spin_def}
\end{eqnarray}
As the spin $\sigma_r = \pm$ the $D$ matrix is doubled in size, 
now being a $2d_S \times 2d_S$ matrix, i.e.\ the $r$ index
is interlaced in $\pm$ pairs.
The matrix element is defined with respect to $\hat{H}_0$ and
we expect that the energies corresponding to the spin states
$|B(\vec{p}, \sigma)\rangle$, with $\sigma = \pm$ are degenerate.
(This is a reflection of Kramers degeneracy.) 

We could continue as before with this enlarged matrix. 
However when we have only spin non-flip (the case considered here)
or spin-flip matrix elements, it is simplest to try to keep as close 
as possible to the previous results. We can achieve this by writing 
the overlaps as
\begin{eqnarray}
   {}_\lambda\langle 0 | \hat{B}_{r\,\alpha}(\vec{0}) 
             | B_r(\vec{p}_r,\sigma_r) \rangle_{\lambda\,{\rm rel}}
     &=& Z_r \, u^{(r)}_\alpha(\vec{p}_r,\sigma_r) + \ldots \,,
                                                         \nonumber  \\
   {}_\lambda\langle B_s(\vec{p}_s,\sigma_s) 
             | \hat{\bar{B}}_{s\,\beta}(\vec{0}) |0 \rangle_{\lambda\,{\rm rel}}
     &=& \bar{Z}_s \, \bar{u}_\beta^{(s)}(\vec{p}_s,\sigma_s) + \ldots \,,
\label{wf_def}
\end{eqnarray}
where $Z_r$ and $\bar{Z}_s$ are taken as scalars 
with $\bar{Z} = Z^*$.
Although the states here are the perturbed states, rather then the
unperturbed states, again we expect the effect of the perturbation 
to be small, as discussed in section~\ref{simple}.

Furthermore, although we could consider the Dirac indices as a GEVP 
it is more convenient to sum over them with some matrix, $\Gamma$. 
In this article we shall primarily consider the unpolarised case with
\begin{eqnarray}
   \Gamma^{\rm unpol} = (1+\gamma_4)/2 \,,
\end{eqnarray}
so
\begin{eqnarray}
   C_{\lambda\,rs}(t) = {\rm tr}\Gamma^{\rm unpol}C_{\lambda\,B_rB_s}(t) \,.
\label{C_unpol}
\end{eqnarray}
Using
\begin{eqnarray}
   \bar{u}^{(r)}(\vec{p}_r, \sigma_r) \Gamma^{\rm unpol} u^{(s)}(\vec{p}_s, \sigma_s)
      = \sqrt{(E_r + M_r)(E_s + M_s)} \, \delta_{\sigma_r\sigma_s} \,,
\label{ubar_u_unpol_explicit}
\end{eqnarray}
(see Appendix~\ref{spinor_results} and eq.~(\ref{ubar_u_unpol})) means that
due to the $\delta_{\sigma_r\sigma_s}$ term appearing there, the 
$\sigma_r$, $\sigma_s$ sums in  eq.~(\ref{2pt_correl_degen_final_spin}) 
become diagonal, and hence we just sum over them in eqs.~(\ref{BrOBs}), 
(\ref{matrix_cr_rs_spin}). This reduces $D$ to the previous $d_S\times d_S$ 
matrix as in section~\ref{matrix_diag}, where
\begin{eqnarray}
   D_{rs} = \epsilon_r \delta_{rs} + \lambda a_{rs} \,, 
   \quad \mbox{with} \quad
   a_{rs} = {1 \over 2} ( a_{+r,+s} + a_{-r,-s} ) \,.
\label{matrix_cr_rs_unpol}
\end{eqnarray}
This effectively is the same result as before, but we are now just averaging
over the diagonal spin terms. This gives finally
\begin{eqnarray}
   C_{\lambda\,rs}(t)
      = \sum_{i=1}^{d_S} w^{(i)}_r\bar{w}_s^{(i)} e^{-E_\lambda^{(i)}t} \,,
\label{C_rs_nodirac}
\end{eqnarray}
with%
\footnote{For notational simplicity we have have absorbed some factors
into a redefinition of the overlap definitions
\begin{eqnarray}
   \sqrt{ E_r + M_r \over E_r}\,Z_r \to Z_r\,, \quad \mbox{and} \quad
   \sqrt{ E_s + M_s \over E_s}\,\bar{Z}_s \to \bar{Z}_s \,.
                                                            \nonumber
\end{eqnarray}
This is due to to the relativistic normalisation, eq.~(\ref{foot_norm}),
used for the results in Appendix~\ref{spinor_results}, together with 
eq.~(\ref{ubar_u_unpol_explicit}) and a factor $2$ from the
averaging over polarisations. \label{redef_Z}}
\begin{eqnarray}
   w_r^{(i)}       = Z_r e_r^{(i)} \quad \mbox{and} \quad
   \bar{w}_s^{(i)} = \bar{Z}_s e_s^{(i)\,*} \,,
\label{w_wbar_nodirac}
\end{eqnarray} 
where the eigenvectors, $e_r^{(i)}$ are from the $D$ matrix in 
eq.~(\ref{matrix_cr_rs_unpol}) and the eigenvalues $\mu^{(i)}$ give
the energies $E_\lambda^{(i)} = \bar{E}+\mu^{(i)}$ as in eq.~(\ref{pert_degen_en}).

Another possibility is
$\Gamma^{\rm pol}_{\pm 3} = (1+\gamma_4)/2 \times (1 \pm i\gamma_5\gamma_3)$
(see Appendix~\ref{pol_unpol_gamma} and eq.~(\ref{ubar_u_pol3})) 
which again gives a reduced $D_{rs}$ together with $a_{rs} = a_{\pm r, \pm s}$. 
(Note that both these $\Gamma$-matrix forms are chosen so that the 
diagonal $\delta_{\sigma_r\sigma_s}\delta_{rs}$ term in eq.~(\ref{BrOBs}) 
remains as $\delta_{rs}$.) The choice of projection matrix, $\Gamma$, 
depends on the symmetry of the operator and picks out the 
relevant matrix element. So, as discussed here for an unpolarised
or spin-non-flip matrix element we would use $\Gamma^{\rm unpol}$ or
$\Gamma^{\rm pol}_{\pm 3}$.

In Appendix~\ref{euclid_FF} (together with Appendix~\ref{spinor_results})
we have investigated the phase factor relationship between $a_{-r,-s}$ 
with $a_{+r,+s}$ (or $a_{-r, +s}$ with $a_{+r, -s}$) for all possible local 
bilinear currents culminating in eq.~(\ref{me_reality}) and 
Table~\ref{table:etag}.

Furthermore in Appendix~\ref{alt_energy_states}, the general 
result for the $d_S = 2$ case is given.
Some comments are also made for the spin-flip case using for example
$\Gamma^{\rm pol}_\pm =(1+\gamma_4)/2\times i\gamma_5(\gamma_1\pm i\gamma_2)$
which cannot be put in the form discussed in this section 
(i.e.\ as an effective $D_{rs}$).


\section{Quasi-degenerate baryon energy states}
\label{quasi_degeb_N_energy_ela}


\subsection{Flavour diagonal matrix elements}
\label{flav_diag_ele}


The simplest example, as alluded to in section~\ref{matrix_element},
is to consider two close energy states for the
same baryon but with different momentum. Thus the possible operators
in eq.~(\ref{op_def}) must be flavour diagonal. (We shall consider
flavour changing, that is flavour off-diagonal matrix elements 
in the next section.) To be concrete we shall consider the nucleon, 
$B = N(uud)$ here, although the results hold for other octet (or decuplet)
particles. As an example, we may take the quark content of the operator to be
\begin{eqnarray}
   O(\vec{x}) 
      \sim (\bar{u}\gamma u)(\vec{x}) - (\bar{d}\gamma d)(\vec{x}) \,,
\end{eqnarray}
where $\gamma$ is an arbitrary Dirac gamma matrix.
As discussed previously we shall first consider the general structure
and then finally incorporate the spin index as in section~\ref{spin_index}.

Clearly we have a degeneracy or near degeneracy when $\vec{q}$ is chosen
such that we have the energy states with
$E_N(\vec{p}) \approx E_N(\vec{p}+\vec{q})$ (or alternatively
$E_N(\vec{p}) \approx E_N(\vec{p}-\vec{q})$). Let us now consider 
some possible solutions focusing on 
$E_N(\vec{p}) \approx E_N(\vec{p}+\vec{q})$. For clarity we first describe
this for the the non-interacting case, later we generalise to the 
interacting case, leading to an avoided level crossing.

\begin{figure}[!tb]
\begin{minipage}{0.45\textwidth}

   \includegraphics[width=6.50cm]{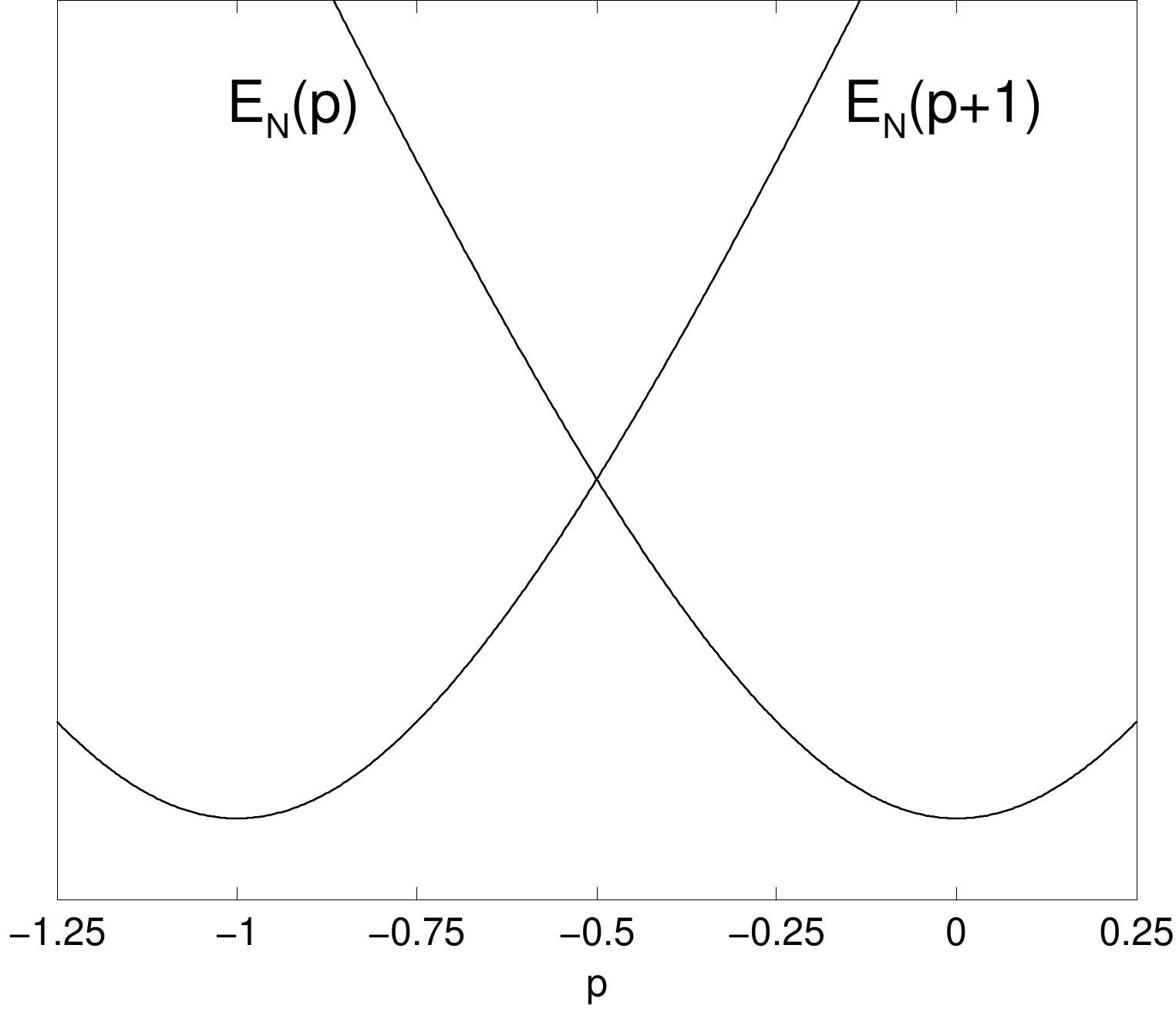}

\end{minipage}\hspace*{0.05\textwidth}
\begin{minipage}{0.45\textwidth}

   \includegraphics[width=6.50cm]{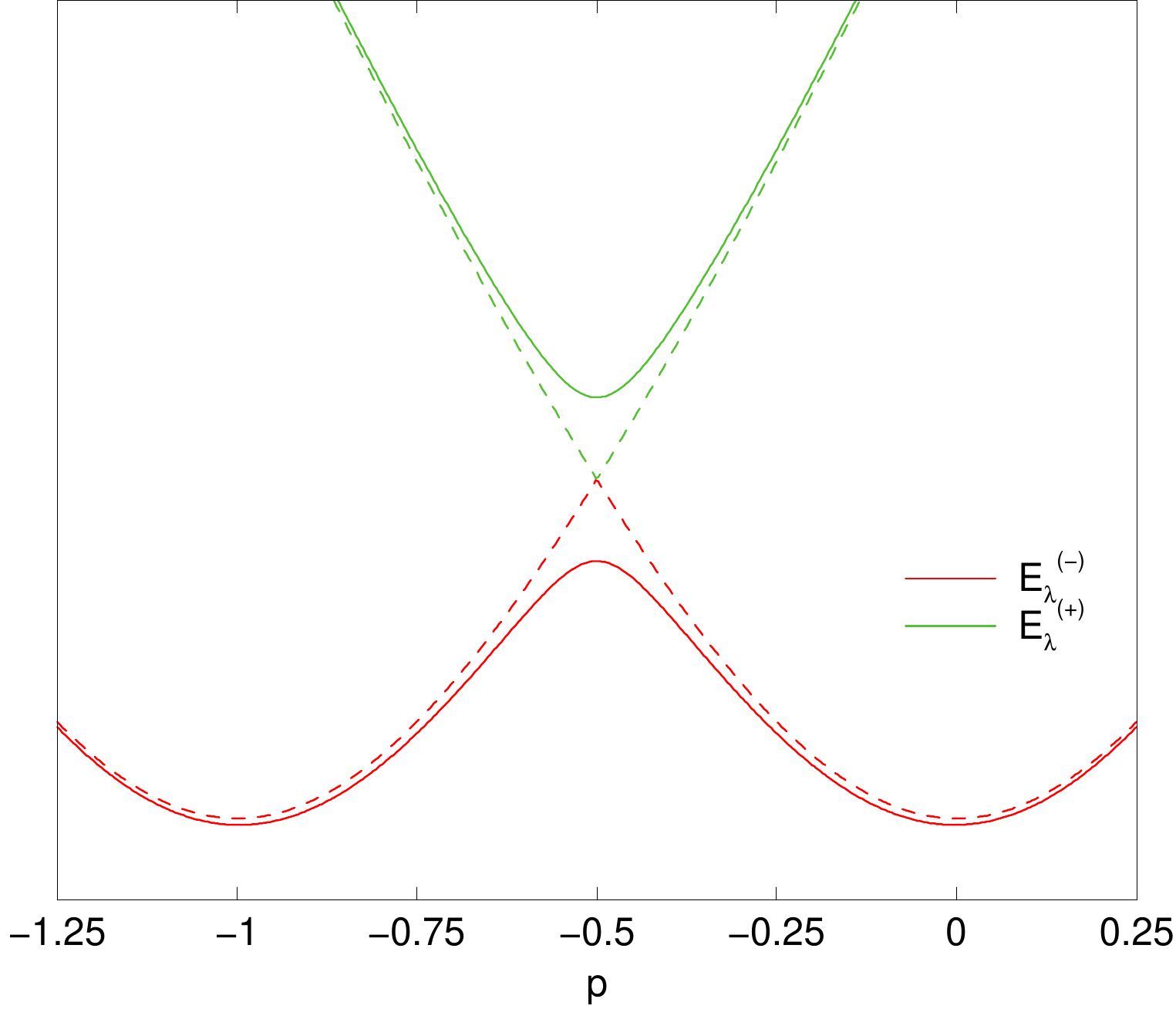}

\end{minipage}
\caption{LH panel: A sketch of the (unperturbed) energy states 
         $E_N(p)$, $E_N(p+q)$ versus $p$ in one dimension for fixed $q$
         using units where $q=1$. Using these units there is a degeneracy
         at $p=-1/2$.
         RH panel: An equivalent sketch of the perturbed
         energy states, $E^{(\pm)}$ based on eq.~(\protect\ref{E_pm}).
         The dashed lines are the free case.
         The sketch shows the avoided energy levels.}
\label{NN_avoided}
\end{figure}

Let us first consider the simpler $1$-dimensional case (for example
suppose that $\vec{q}$ is in the $z$-direction: $\vec{q} = (0, 0, q)$
and similarly for $\vec{p}$). There will now be a crossing at $p = - q/2$
where $p^2 = (p+q)^2$ and we would have a near-degenerate 
state close to these states. These are illustrated in the left hand, LH, 
panel in Fig.~\ref{NN_avoided}
where a sketch of the crossing is shown.
This is the region we wish to consider perturbation theory 
by applying the Feynman--Hellmann theorem -- well separated from other 
potential degeneracies.
In $3$-dimensions we have the corresponding simple solution 
$\vec{p} = - \vec{q}/2$. This possibility was 
considered in \cite{Chambers:2017tuf}.

In the following, we derive results close to (or at) the degeneracies.
We shall only consider $2$-fold degeneracies
as this means that $d_S =2$ and we have a quadratic eigenvalue equation 
to solve. (The doubling to include the spin index, as previously discussed in 
section~\ref{spin_index} is a simple generalisation and will be stated
at the end of this section.) While we can solve higher dimensional 
polynomials, they are likely to be less useful as the result will 
contain several different nucleon matrix elements, which are difficult
to disentangle. Note that this requirement becomes more difficult 
to achieve if $\vec{q}$ is too small as the $\lambda$ range 
where $D$ in eq.~(\ref{matrix_cr_rs}) takes the form of a $2\times 2$
matrix might become rather narrow, forcing the use of higher
dimensional $D$ matrices. 

After this general discussion let us take the two momenta to be $\vec{p}$ 
and $\vec{p}+\vec{q}$ and we consider the case where the two 
degenerate states form the subspace where 
$E_N(\vec{p}+\vec{q}) \approx E_N(\vec{p})$. So we set
\begin{eqnarray}
   |B_1(\vec{p}_1)\rangle = |N(\vec{p})\rangle \,, \qquad
   |B_2(\vec{p}_2)\rangle = |N(\vec{p}+\vec{q})\rangle \,,
\end{eqnarray}
with 
$E_{B_1}(\vec{p}_1) \equiv E_N(\vec{p}) = \bar{E} + \epsilon_1$ and
$E_{B_2}(\vec{p}_2) \equiv E_N(\vec{p}+\vec{q}) = \bar{E} + \epsilon_2$.
The geometry of $\vec{p}$ and $\vec{q}$ is chosen so that 
$E_N(\vec{p}+\vec{q}) \approx E_N(\vec{p})$ are the lowest energy
states in this sector, i.e.\ there is no state with a 
lower energy, as indicated in the LH panel of Fig.~\ref{NN_avoided}.
Momentum conservation, i.e.\ the step-up or step-down in $\vec{q}$ 
from eq.~(\ref{matrix_el}) gives the matrix of baryon matrix elements as
\begin{eqnarray}
   a_{rs} 
   = \langle B_r(\vec{p}_r) | \hat{\tilde{{\cal O}}}(\vec{q}) 
                           | B_s(\vec{p}_s) \rangle
   = \left( \begin{array}{cc}
               0  & a^*  \\
               a  & 0    \\
            \end{array}
     \right)_{rs} \,,
\label{NN_matrix}
\end{eqnarray}
where
\begin{eqnarray}
   a = \langle B_2(\vec{p}_2) |\hat{O}(\vec{0})| B_1(\vec{p}_1)\rangle \,.
\label{a_value_flav_diag}
\end{eqnarray}
To first find the eigenvalues of $D$ in eq.~(\ref{matrix_cr_rs})
we have to solve a quadratic equation. This gives
\begin{eqnarray}
   \mu^{(\pm)}
      = {1 \over 2}(\epsilon_1 + \epsilon_2) 
        \pm {1 \over 2}\sqrt{ (\epsilon_1-\epsilon_2)^2 + 4\lambda^2 |a|^2 } \,.
\label{eigenvals}
\end{eqnarray}
leading to the energies
\begin{eqnarray}
   E^{(\pm)}_\lambda
     = \bar{E} + \mu^{(\pm)} 
     = {1 \over 2}(E_1 + E_2) \pm {1 \over 2}\Delta E_\lambda \,,
\label{E_pm}
\end{eqnarray}
with
\begin{eqnarray}
   \Delta E_\lambda 
     = E^{(+)}_\lambda - E^{(-)}_\lambda
     = \sqrt{ (E_1 - E_2)^2 + 4\lambda^2 |a|^2 } \,.
\label{DeltaE}
\end{eqnarray}
We sketch these energy levels $E^{(\pm)}$ in the RH panel of 
Fig.~\ref{NN_avoided} and compare with the free case ($\lambda \to 0$),
dashed lines. We see that for $\lambda \not= 0$ then we have the
phenomenon of avoided energy levels for $E_\lambda^{(\pm)}$.

The eigenvectors $e^{(\pm)}_r$ are given by
\begin{eqnarray}
   e_r^{(\pm)} 
    = {1 \over \sqrt{\Delta E_\lambda}}
         \left( \begin{array}{c}
                   \sqrt{\kappa_{\pm}} \\[0.5em]
                   \pm \mbox{sgn}(\lambda) \sqrt{\kappa_{\mp}}\, {a \over |a|}
                \end{array}
          \right)_r \,,
   \quad \mbox{with} \quad
   \kappa_\pm = {1 \over 2}\Delta E_\lambda  \pm {1 \over 2}(E_1-E_2) \,,
\label{eigenvec}
\end{eqnarray}
where the normalisation factor has been chosen so that
$|e_1^{(\pm)}|^2 + |e_2^{(\pm)}|^2 = 1$. A useful relation is
$\kappa_+\kappa_- = \lambda^2|a|^2$. Note that the components of the 
eigenvectors are related: $e_2^{(-)} = - \mbox{sgn}(\lambda)a/|a|\,e_1^{(+)}$ 
and $e_2^{(+)} = \mbox{sgn}(\lambda)a/|a|\,e_1^{(-)}$. We also see that while
the Feynman--Hellmann approach cannot yield any information on the phase 
of the matrix element from the energy as it is the modulus, the phase
is however contained in the eigenvectors as $a = |a|\zeta_a$
(with $\zeta_a$ the phase of the matrix element).

This result of course includes the degenerate case when the nucleon 
$\vec{p}$, $\vec{q}$ momenta are arranged so that their energies 
are the same, $E_2 = E_1$ (the crossing point in the LH panel of 
Fig.~\ref{NN_avoided}. As discussed earlier, this requires the 
geometry of the $\vec{p}$ and $\vec{q}$ momenta to be chosen such that 
$\vec{q}^{\,2} = -2\vec{p}\cdot\vec{q}$ with a possible solution 
$\vec{p} = -\vec{q}/2$. In this case $\Delta E_\lambda = 2|\lambda||a|$
and eigenvectors $\vec{e}^{(\pm)} = (1,\pm\mbox{sgn}(\lambda) a/|a|)/\sqrt{2}$. 

Including the spin index, for the numerical case under consideration 
in section~\ref{lattice} where we set $\Gamma = \Gamma^{\rm unpol}$ is to simply 
average over the spins of the matrix element, $a \to (a_{++}+a_{--})/2$
as given in eq.~(\ref{matrix_cr_rs_unpol}). Relations between
$a_{--}$ and $a_{++}$ are given in Appendix~\ref{euclid_FF}
(together with Appendix~\ref{spinor_results}). The general result is
given in Appendix~\ref{alt_energy_states}.


\subsection{Flavour off-diagonal (transition) matrix elements}
\label{transition_els}


We shall now consider flavour off-diagonal, or transition, matrix elements 
taking for definiteness the $\Sigma^- \to n$ decay, or in the isospin limit 
considered here $\Sigma(sdd) \to N(udd)$ as our example, i.e.\ an $s \to u$ 
decay. We take the quark content of the operator as
\begin{eqnarray}
   O(\vec{x}) \sim (\bar{u} \gamma s)(\vec{x}) \,,
\label{trans_op}
\end{eqnarray}
thus the action is no longer diagonal in quark flavour space.
Let us consider the $|\Sigma\rangle$ and $|N\rangle$ as having 
nearly-degenerate energies (or quasi-degenerate energies) and apply the 
previous formalism, in particular eqs.~(\ref{C_fo}) and (\ref{pert_degen_en}).

\begin{figure}[!tb]
\begin{minipage}{0.45\textwidth}

   \begin{center}
      \includegraphics[width=6.45cm]{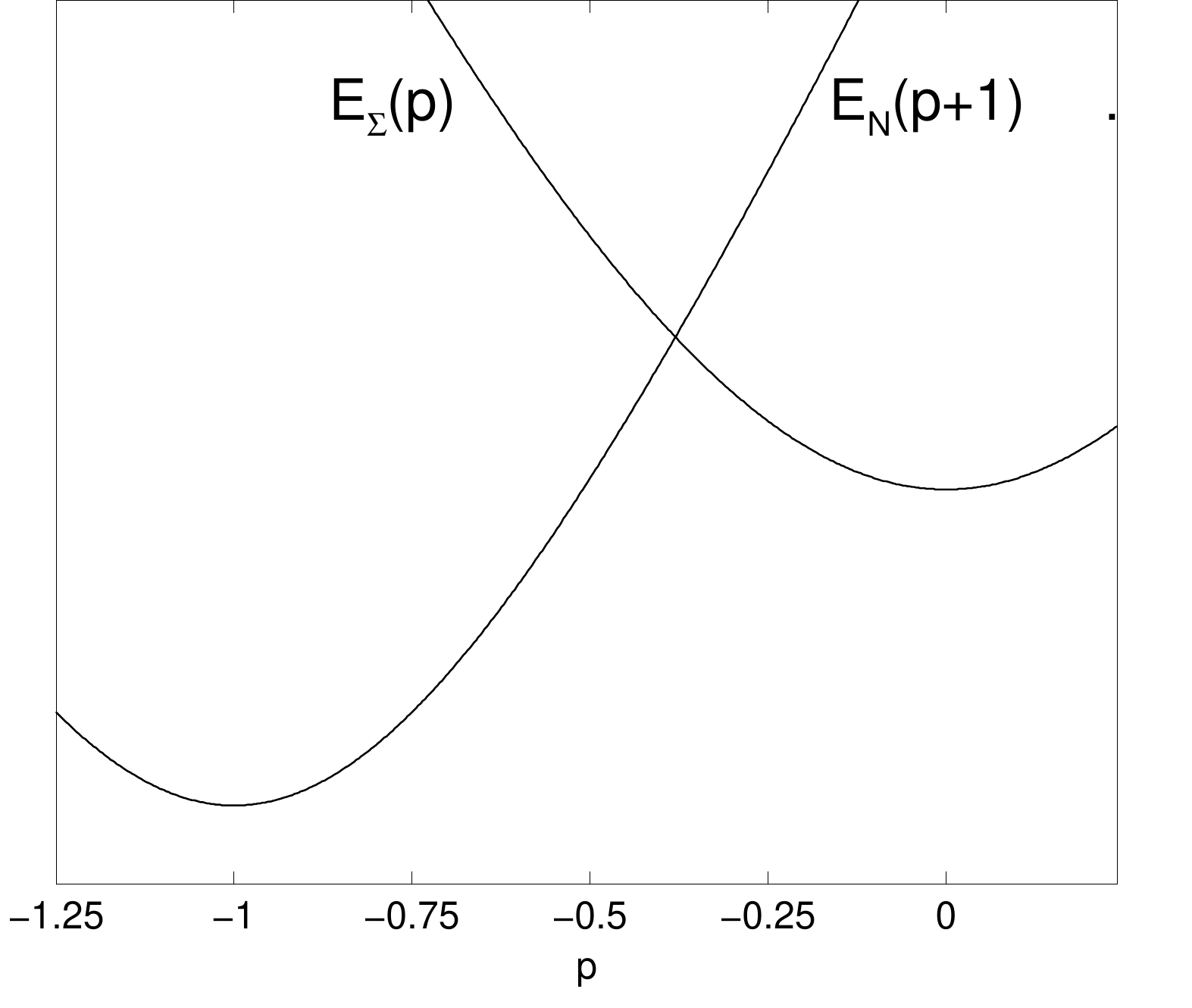}
   \end{center} 

\end{minipage}\hspace*{0.05\textwidth}
\begin{minipage}{0.45\textwidth}

   \begin{center}
      \includegraphics[width=6.60cm]{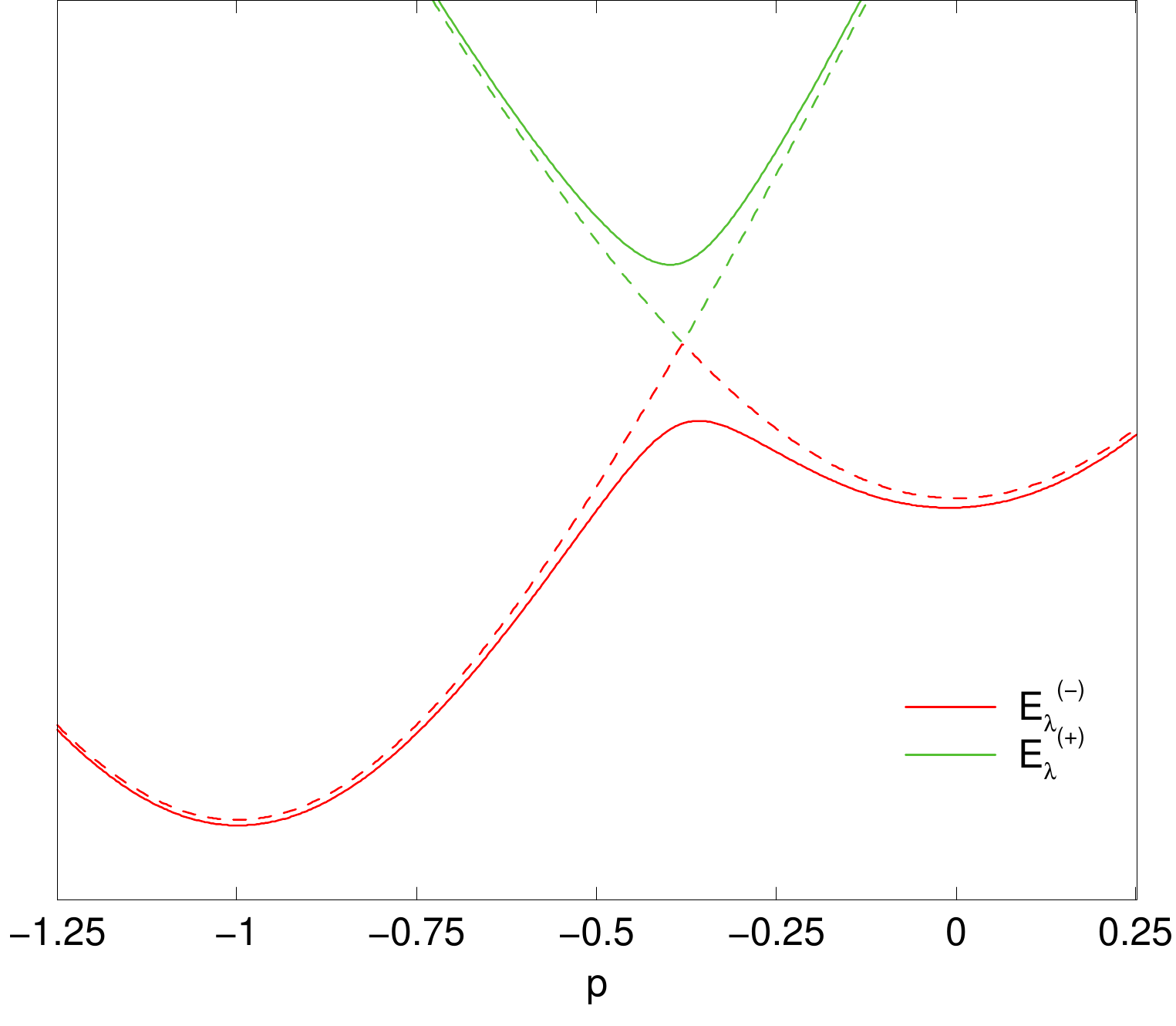}
   \end{center} 

\end{minipage}
\caption{LH panel: A sketch of the (unperturbed) energy states 
         $E_\Sigma(p)$, $E_N(p+q)$ versus $p$ in one dimension for fixed $q$
         using units where $q=1$.
         RH panel: An equivalent sketch of the perturbed
         energy states, $E^{(\pm)}$. The dashed lines are the free case.
         The sketch shows the avoided energy levels.}
\label{SigN_avoided}
\end{figure}

Following the discussion in section~\ref{flav_diag_ele}, let us
consider again $E_\Sigma(\vec{p}) \approx E_N(\vec{p}+\vec{q})$ 
the parallel case to that of the LH panel of Fig.~\ref{NN_avoided}
but now extended to the $\Sigma$ particle. In the LH panel of 
Fig.~\ref{SigN_avoided} we sketch this situation for the $1$-dimensional
example. As before we need 
to be in a region well separated from other degeneracies.
We now set 
\begin{eqnarray}
   |B_1(\vec{p}_1)\rangle = |\Sigma(\vec{p})\rangle \,, \qquad
   |B_2(\vec{p}_2)\rangle = |N(\vec{p}+\vec{q})\rangle \,.
\label{B_Sig_N}
\end{eqnarray}
Again from eq.~(\ref{almost_degen_energy}) let us write
$E_{B_1}(\vec{p}_1) = E_\Sigma(\vec{p}) = \bar{E}+\epsilon_1$ and
$E_{B_2}(\vec{p}_2) = E_N(\vec{p}+\vec{q}) = \bar{E}+\epsilon_2$.
We then find that 
$\langle B_r(\vec{p}_r) | \hat{\tilde{{\cal O}}}(\vec{q}) 
                       | B_s(\vec{p}_s) \rangle$ has the same structure
as in eq.~(\ref{NN_matrix}). So the results from section~\ref{flav_diag_ele}
from eq.~(\ref{NN_matrix}) -- eq.~(\ref{eigenvec}) remain unchanged.
In the RH panel of Fig.~\ref{SigN_avoided}, we show the interacting
(i.e.\ $\lambda \not= 0$) case from eq.~(\ref{E_pm}). Again we
now have an avoided level crossing. In comparison to the previous case, 
Fig.~\ref{NN_avoided}, while very similar, the degeneracy is now shifted
to a slightly smaller momentum value.

\begin{figure}[!tb]
\begin{minipage}{0.45\textwidth}

   \begin{center}
      \includegraphics[width=6.50cm]
                       {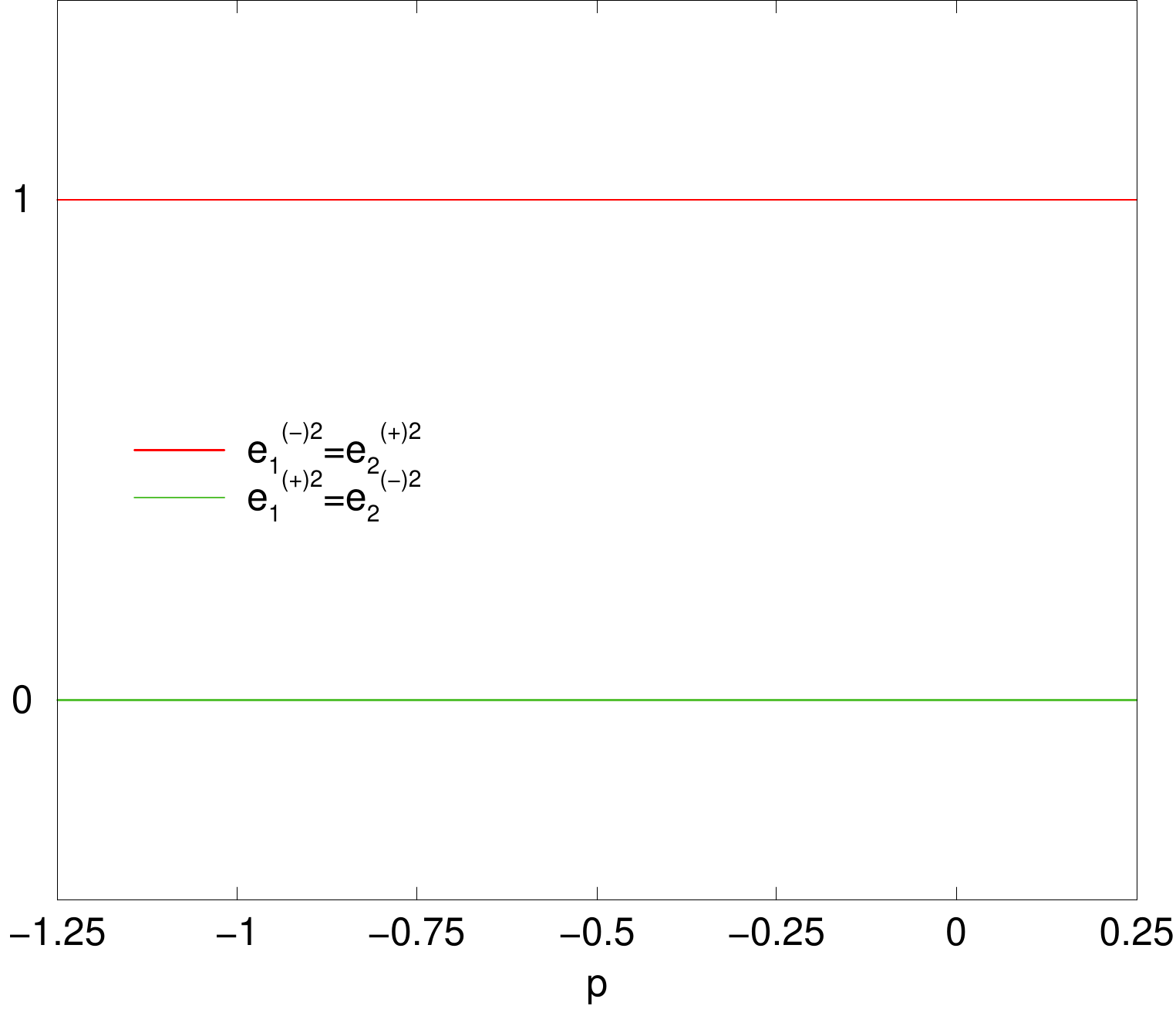}
   \end{center} 

\end{minipage}\hspace*{0.05\textwidth}
\begin{minipage}{0.45\textwidth}

   \begin{center}
      \includegraphics[width=6.50cm]
                       {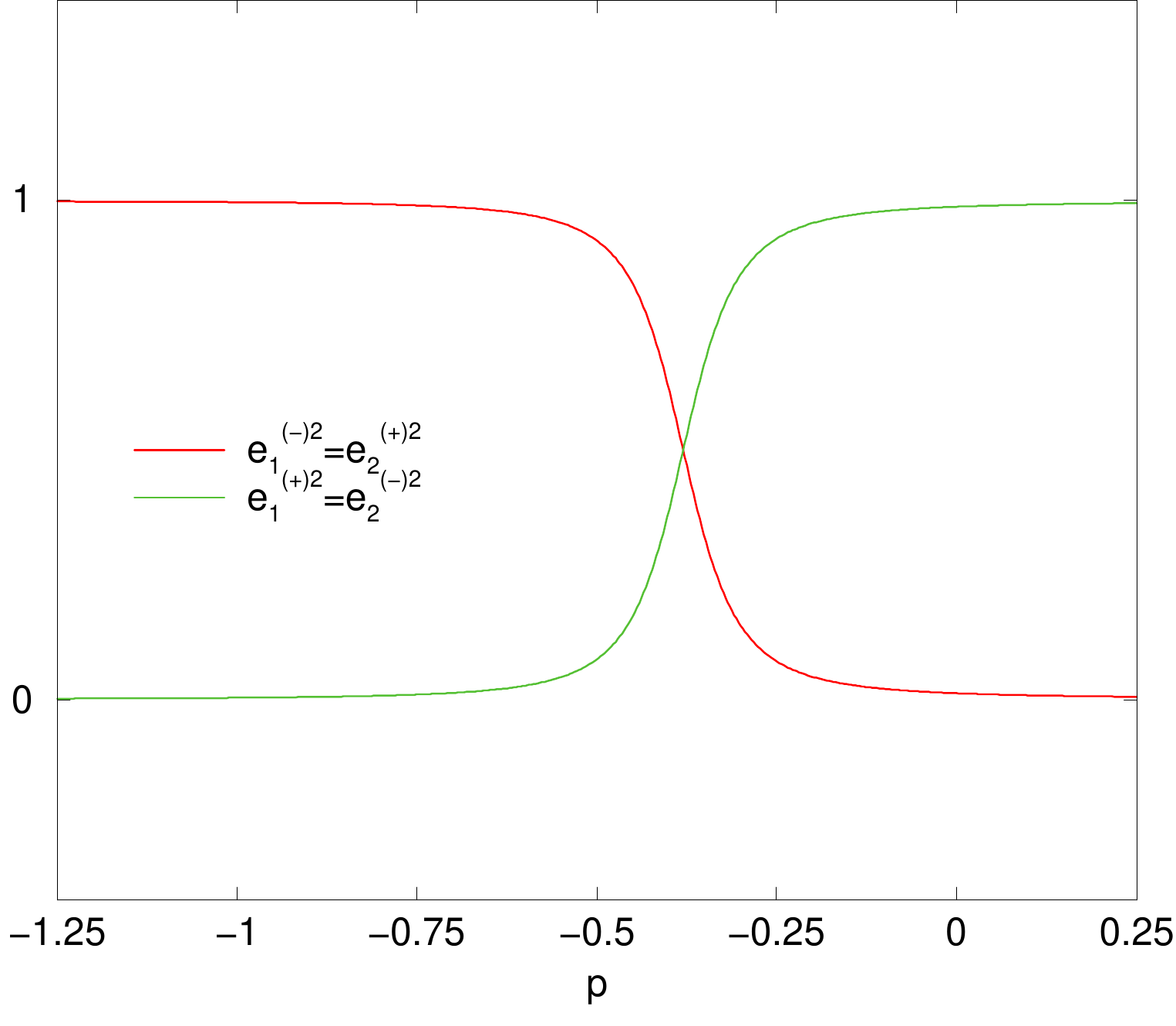}
   \end{center} 

\end{minipage}
\caption{Left panel: The free case where we have plotted $e_1^{(-)\,2}$
         and $e_2^{(-)\,2}$ against $p$, again taking units where $q=1$.
         Right panel: The interacting case showing the change of 
         state.}
\label{SigN_eigen2}
\end{figure}

In Fig.~\ref{SigN_eigen2} we sketch the corresponding eigenvectors to
the eigenvalues of Fig.~\ref{SigN_avoided}. Shown are
$e_1^{(-)\,2}$ and $e_2^{(-)\,2}$ against $p$ both for the free and 
interacting case. While in the free case the components of
$\vec{e}^{(\pm)}$ remain constant (left panel) for the interacting
case (right panel) they flip as the momentum $p$ changes.


\section{A lattice application for transition matrix elements}
\label{lattice}


As an example of this formalism, we shall now consider in more detail 
how the previous results can be applied to the $\Sigma \to N$ 
transition matrix element, i.e.\ decay $s \to u$ in the isospin limit 
as described in section~\ref{transition_els}. We first discuss 
in general the modifications to the action and the fermion 
inversion procedure before considering the specific numerical results.


\subsection{The fermion inversion and correlation functions}
\label{inversion}


To apply the results of section~\ref{transition_els} we need to consider
the action
\begin{eqnarray}
   S = S_g + \int_x \left(\bar{u}, \bar{s}\right) 
                               \left( \begin{array}{cc}
                                         D_u         & -\lambda\cal{T} \\
                                         -\lambda\cal{T}^\prime
                                                        & D_s          \\
                                      \end{array}
                               \right) \left( \begin{array}{c}
                                                 u  \\
                                                 s  \\
                                              \end{array}
                                       \right) + \int_x \bar{d}\,D_d\,d \,,
\label{lat_trans_act}
\end{eqnarray}
where $S_g$ is the gluon action and we shall now consider the fermionic
piece in more detail. For simplicity we absorb any clover terms into the
$D$s. We take the $u$ and $d$ quarks as mass degenerate
$m_u = m_d \equiv m_l$, with a common mass $m_l$.
(A more general situation would require a $3\times 3$ matrix, 
when the vector for $u$ and $s$ would be extended to $(u, d, s)$ 
with non-degenerate quark masses.)
For ${\cal T}$ we take the general local expression
\begin{eqnarray}
   {\cal T}(x,y;\vec{q}) = \gamma \, e^{i\vec{q}\cdot\vec{x}} \, \delta_{x,y} \,.
\end{eqnarray}
For $\gamma_5$-hermiticity for the matrix in eq.~(\ref{lat_trans_act}) 
we need ${\cal T}^\prime = \gamma_5 {\cal T}^\dagger \gamma_5$. From the action 
in eq.~(\ref{lat_trans_act}) we now define the larger flavour inverse 
propagator, $\cal M$, as
\begin{eqnarray}
   \cal M = \left( \begin{array}{cc}
                      D_u                   & -\lambda{\cal T} \\
                      -\lambda\gamma_5{\cal T}^\dagger \gamma_5 & D_s \\
                   \end{array}
            \right) 
     \equiv \left( \begin{array}{cc}
                      {\cal M}_{uu} & {\cal M}_{us} \\
                      {\cal M}_{su} & {\cal M}_{ss} \\
                   \end{array}
            \right) \,,
\label{M_def}
\end{eqnarray}
together with ${\cal M}_{dd} \equiv D_d$. 

We can generate correlation functions, $C_{\lambda\,rs}(t)$%
\footnote{While we could consider this as a 
$(2\times 4) \,\, \times \,\, (2\times 4)$ matrix, as in eq.~(\ref{C_unpol}),
we have projected each correlation function with $\Gamma$.}
for a fixed $\vec{p}$, $\vec{q}$ by choosing $B^\prime$ and $B$ to be either 
$B_r$ or $B_s$, as given in eq.~(\ref{B_Sig_N}). The correlation function 
matrix for a particular $\vec{p}$, $\vec{q}$ pair and suitable for a 
GEVP type procedure is thus given by
\begin{eqnarray}
   C_{\lambda\,rs}(t)
      = \left( \begin{array}{cc}
                  C_{\lambda\,\Sigma\Sigma}(t) &
                  C_{\lambda\,\Sigma N}(t)    \\
                  C_{\lambda\,N\Sigma}(t)     &
                  C_{\lambda\,NN}(t) 
               \end{array}
        \right)_{rs} \,,
\label{correl_fun_decay_mat}
\end{eqnarray}
(see Appendix~\ref{corr_fun} for more details).
The individual correlation functions in this equation are built from 
Green's functions given by
\begin{eqnarray}
   \left( \begin{array}{cc}
             G_{uu} & G_{us}   \\
             G_{su} & G_{ss}   \\
          \end{array}
   \right)
      = \left( \begin{array}{cc}
                 ({\cal M}^{-1})_{uu}  & ({\cal M}^{-1})_{us}   \\
                 ({\cal M}^{-1})_{su}  & ({\cal M}^{-1})_{ss}   \\
               \end{array}
        \right) \,.
\end{eqnarray}
The relations are standard between the correlation functions and Green's
functions, for completeness we give them in Appendix~\ref{corr_fun}.

We now need to invert the matrix ${\cal M}$ in eq.~(\ref{M_def}).
One possibility is to consider a fermion matrix twice the size to 
the standard single-flavour fermion matrix for the two flavours. 
Instead we shall consider here ${\cal M}$ as a $2\times 2$ block matrix
and invert that. This leads to 
\begin{eqnarray}
   G^{(uu)} &=& (1 - \lambda^2  D_u^{-1}{\cal T}
                    D_s^{-1}\gamma_5{\cal T}^\dagger\gamma_5)^{-1} D_u^{-1} \,,
                                                          \nonumber  \\
   G^{(ss)} &=& (1- \lambda^2 D_s^{-1}\gamma_5{\cal T}^\dagger\gamma_5 D_u^{-1} 
                             {\cal T})^{-1} D_s^{-1} \,,
\label{Guu+Gss_res}
\end{eqnarray}
and
\begin{eqnarray}
   G^{(us)} &=& \lambda D_u^{-1}{\cal T} G^{(ss)} \,,
                                                          \nonumber  \\
   G^{(su)} &=& \lambda D_s^{-1}\gamma_5{\cal T}^\dagger\gamma_5 G^{(uu)} \,.
\label{Gus+Gsu_res}
\end{eqnarray}
The problem with eq.~(\ref{Guu+Gss_res}) is that it involves an inversion
within an inversion, which computationally would be very expensive.
However for $\lambda$ small (the case considered here) it is sufficient
to expand to a low order in $\lambda$, especially as the expansion
parameter is $\lambda^2$. Thus given $G_{2n}^{(uu)}$, $G_{2n}^{(ss)}$ we have
\begin{eqnarray}
   G_{2n+2}^{(uu)} 
      &=&  D_u^{-1} + 
           \lambda^2 D_u^{-1}{\cal T}D_s^{-1}\gamma_5{\cal T}^\dagger \gamma_5 
                     G_{2n}^{(uu)} \,,
                                                          \nonumber  \\
   G_{2n+2}^{(ss)} 
      &=&  D_s^{-1} + 
           \lambda^2 D_s^{-1}\gamma_5{\cal T}^\dagger\gamma D_u^{-1}{\cal T}
                    G_{2n}^{(ss)} \,,
\label{iteration_lambda}
\end{eqnarray}
for $n = 0, 1, 2, \ldots$, the exact result being obtained for $n \to \infty$.
$G_{2n+1}^{(us)}$ and $G_{2n+1}^{(su)}$ for $n = 0, 1, 2,\ldots$ are then given 
from eq.~(\ref{Gus+Gsu_res}) again using $G_{2n}^{(uu)}$, $G_{2n}^{(ss)}$ as input.
Effectively each matrix inversion (either $D_u^{-1}$ or $D_s^{-1}$) is associated
with an additional power of $\lambda$. Even powers of $\lambda$ vanish for 
transition terms (and correspondingly odd powers of $\lambda$ vanish for 
the flavour-diagonal terms). Some more details are given in 
Appendix~\ref{fermion_inversion}. For example the leading order result 
($n = 0$) for both the diagonal and off-diagonal Green's functions are given by
\begin{eqnarray}
   G^{(uu)} &=& D_u^{-1} + O(\lambda^2)\,,
                                                          \nonumber  \\
   G^{(ss)} &=& D_s^{-1} + O(\lambda^2) \,,
                                                          \nonumber  \\
   G^{(us)} &=& \lambda D_u^{-1}{\cal T} D_s^{-1} + O(\lambda^3) \,,
                                                          \nonumber  \\
   G^{(su)} &=& \lambda D_s^{-1}\gamma_5{\cal T}^\dagger\gamma_5D_u^{-1} 
                                               + O(\lambda^3) \,,
\end{eqnarray}
which is possibly sufficient as there are no $O(\lambda^2)$ terms
so the validity of the linear term in $\lambda$ could extend further.
The off-diagonal correlation functions are now just like the usual 
$3$-point function integrated over the insertion time. 

To better justify the Feynman--Hellmann procedure, we shall 
consider higher order iterations to approximate the Green's functions
to within numerical accuracy. To build the Green's functions we use
$\delta_{\vec{x},\vec{0}}\delta_{t,0}$ as the initial source, and build the chain
using the previously calculated object as the new source as given 
in eq.~(\ref{iteration_lambda}). This has the advantage of producing 
the Green's function and hence correlation function as a continuous function
of $\lambda$ rather than needing a separate evaluation for each value 
of $\lambda$ chosen. Each subsequent insertion of the operator on the 
correlation function is constructed using a sequential source with the 
insertion time being summed over.

Note that for each different operator and momentum $\vec{q}$ we have to 
re-calculate everything. This is opposite to the usual common procedure
for three-point functions, where we calculate the second Green's function
from the sink to the operator (which allows many operators to be inserted
for one second inversion).


\subsection{The simulation}
\label{simulation}


\subsubsection{The decay matrix element and chosen kinematics}
\label{decay_me}


We shall consider in this article the vector matrix element $V_4$ for
$\Sigma \to N$ where the $\Sigma$ is stationary, i.e.\ $\vec{p} = \vec{0}$
(and $\vec{q} = \vec{p}^{\,\prime} - \vec{0}$). 
Then the (Euclidean) momentum transfer is given in this case by%
\footnote{Note that we have adopted the convention that $q$ is
positive for a scattering process where for the scattered baryon the 
momentum $q$ is added to the initial baryon momentum. This is opposite 
to the semi-leptonic case, where the lepton and neutrino carry
momentum $q$. This was reflected in the choice in
section~\ref{transition_els}. While here this convention does not
matter, when unified SU(3) flavour breaking expansions are considered,
\cite{Bickerton:2019nyz}, one specific $q$ convention has to be chosen
for all cases.} 
\begin{eqnarray}
   q = (i(M_\Sigma - E_N(\vec{q})), \vec{q})\,,
   \quad \mbox{or} \quad
   Q^2 = - (M_\Sigma - E_N(\vec{q}))^2 + \vec{q}^2 \,.
\label{q_Q2_def}
\end{eqnarray}
From eq.~(\ref{matrix_cr_rs_unpol}) we must average the matrix element
over the spin index. These can be computed using the results
given in Appendix~\ref{euclid_FF} together with those in
Appendix~\ref{spinor_results}. This gives%
\footnote{For simplicity we simply write 
$\hat{O}(\vec{0}) \to \bar{u}\gamma_4 s$.}
\begin{eqnarray}
   \lefteqn{\langle N(\vec{q},+)|\bar{u}\gamma_4 s
                                   |\Sigma(\vec{0},+)\rangle_{\rm rel}}
      & &                                  \label{V4_matrix_el} \\
      &=& \sqrt{2M_\Sigma(E_N(\vec{q})+M_N)}
                                                    \nonumber   \\
      & & \hspace*{0.25in} \times \left( f_1^{\Sigma N}(Q^2)
                 + {E_N(\vec{q})-M_N \over M_N+M_\Sigma} f_2^{\Sigma N}(Q^2)
                 + {E_N(\vec{q}) - M_\Sigma \over M_N + M_\Sigma} f_3^{\Sigma N}(Q^2)
          \right) \,.
                                                    \nonumber
\end{eqnarray}
This uses the relativistic normalisation, see eq.~(\ref{foot_norm}). 
(We emphasise this here with the subscript.)
Note that the matrix element in eq.~(\ref{V4_matrix_el}) can be considered
as a function of $Q^2$ as eq.~(\ref{q_Q2_def}) gives
$E_N(\vec{q}) = (Q^2+M_\Sigma^2+M_N^2)/(2M_\Sigma)$ which can be used to
eliminate $E_N(\vec{q})$ on the RHS of eq.~(\ref{V4_matrix_el}).
Denoting the various spin components by $a_{\pm\pm}$ then we also find
as expected $a_{--}=a_{++}$, $a_{+-}=0=a_{-+}$. (For this case, the
matrix element is real.) In the following for simplicity we will 
supress the spin index.

$\Delta E_\lambda$ from eq.~(\ref{DeltaE}) is given as the (positive) 
difference in the perturbed energies
\begin{eqnarray}
   \Delta E_\lambda
     = \sqrt{ ( M_{\Sigma} - E_{N}(\vec{q}))^2
               + 4\lambda^2 
                 \left( { \langle N(\vec{q})|\bar{u}\gamma_4 s
                                         |\Sigma(\vec{0})\rangle_{\rm rel}
                          \over
                          (2E_N(\vec{q}))(2M_\Sigma) }^2 \right) } \,.
\label{DeltaE_decay}
\end{eqnarray}
It is thus sufficient to construct just a matrix of correlation functions,
as given in eq.~(\ref{correl_fun_decay_mat}) and then apply the GEVP 
procedure to this.


\subsubsection{GEVP}
\label{gevp}


We apply the GEVP (Generalised Eigen-Value Problem) to the $2 \times 2$
correlator matrix $C_\lambda(t)$, eq.~(\ref{correl_fun_decay_mat}). 
The variation of the method we use here \cite{Owen:2012ts} is to first
determine the left $v^{(i)}$ and right $u^{(i)}$ eigenvectors and then 
project out the eigenvalues 
\begin{eqnarray}
   c^{(i)}(t) = e^{-E^{(i)}_\lambda t} \,,
\label{en_ev}
\end{eqnarray}
for $E^{(i)}_\lambda$, $i = \pm$ (see eq.~(\ref{E_pm})). To achieve this,
we consider $t_0$ and a further time $t_0 + \Delta t_0$ to construct 
the following eigenvalue equations
\begin{eqnarray}
   C_{\lambda}^{-1}(t_0) C_\lambda(t_0+\Delta t_0)u^{(i)}(t_0, \Delta t_0)
      &=& c^{(i)}(\Delta t_0) u^{(i)}(t_0, \Delta t_0) \,,
                                   \label{eigenvector_det_GEVP}     \\
   v^{(i)\dagger}(t_0, \Delta t_0) C_\lambda(t_0+\Delta t_0)C_{\lambda}^{-1}(t_0)
      &=& c^{(i)}(\Delta t_0) v^{(i)\dagger}(t_0, \Delta t_0) \,, 
                                                    \quad i = \pm \,.
                                                      \nonumber
\end{eqnarray}
Solving these equations will give the fixed eigenvectors $u$ and $v$ 
(i.e.\ independent of $t$) which can be combined with the correlator matrix
to construct a new correlation function
\begin{eqnarray}
   C_\lambda^{(i)}(t) = v^{(i)\,\dagger} C_\lambda(t)u^{(i)} \,, \quad i = \pm \,,
\end{eqnarray}
which projects out the eigenvalue $c^{(i)}(t)$, eq.~(\ref{en_ev}).
Using eqs.~(\ref{C_rs_nodirac}), (\ref{w_wbar_nodirac}) this means that
\begin{eqnarray}
   v_r^{(i)} = {N^{(i)} \over Z_r}\,e^{(i)}_r \,, \quad \mbox{and} \quad
   u_s^{(i)} = {\bar{N}^{(i)} \over \bar{Z}_s}\,e^{(i)}_s \,,
\label{u_v_e}
\end{eqnarray}
where $N^{(i)}$ and $\bar{N}^{(i)}$ are normalisation constants.
Essentially $v_r^{(i)*}$ measures the component of $B_r$ in the 
$i^{\rm th}$ eigenvector and similarly for $u_s^{(i)}$ and $\bar{B}_s$.
(The above statements and equations are not restricted to just the 
$d_s = 2$ case.)

These two correlators $C_\lambda^{(i)}(t)$, $i =\pm$ represent the two 
low-lying eigenstates of the system which of course includes the 
perturbation to the action. To relate this to the transition form factors
in eq.~(\ref{DeltaE_decay}) we thus require the energy splitting between 
these two states. To extract this energy splitting we construct 
the ratio of the correlators
\begin{eqnarray}
   R_\lambda(t)
      = { C_\lambda^{(+)}(t) \over C_\lambda^{(-)}(t) } 
      \,\, \stackrel{t\gg 0}{\propto} \,\, e^{-\Delta E_\lambda t} \,,
\label{ratio_R}
\end{eqnarray}
which in the large Euclidean time limit will behave like a 
single exponential function and  will show up in the effective energy 
as a plateau region. We thus use this effective energy to pick out 
a suitable plateau region and then fit a one-exponential function 
to the ratio. The two important parameters of the GEVP calculation are 
$t_0$ and $\Delta t_0$. Optimally the time range from $t_0$ 
and $t_0+\Delta t_0$ needs to be in a region where the ground state is 
saturated but the signal-to-noise ratio is still sufficiently high 
to exclude any effects from higher states.


\subsubsection{Lattice details}
\label{lattice_details}


Numerical simulations have been performed using $N_f = 2+1$ $O(a)$
improved clover Wilson fermions \cite{Cundy:2009yy} at $\beta = 5.50$ and
$(\kappa_l, \kappa_s) = (0.121040, 0.120620)$ on
a $N_s^3\times N_t = 32^3\times 64$ lattice size. More definitions and 
details are given in \cite{Bietenholz:2011qq}. We briefly mention here 
that our strategy is to keep the average bare quark mass constant from 
a value on the $SU(3)$ flavour symmetry line. The above 
$(\kappa_l, \kappa_s)$ have been chosen to correspond to 
$\kappa_l = \kappa_s \equiv \kappa_0 = 0.120900$ at the $SU(3)$ flavour
symmetric point. The `distance' in lattice units from the flavour 
symmetric line is given by $\delta m_l$ which is defined by
\begin{eqnarray}
   \delta m_l = {1 \over 2} \left( {1 \over \kappa_l} 
                                 - {1 \over \kappa_0} \right) \,,
\label{deltaml_def}
\end{eqnarray}
and here is $\sim -0.005$.
$SU(3)$ flavour breaking terms have been determined, which allows
an extrapolation to the physical point for both hadron masses, 
\cite{Bietenholz:2011qq} and matrix elements
\cite{Bickerton:2019nyz,Bickerton:2021yzn}. 
This situation corresponds to a lattice spacing of 
$a \sim 0.074\,\mbox{fm} \sim 1/(2.67\,\mbox{GeV})$ leading to a pion
mass of $\sim 330\,\mbox{MeV}$. Errors given in the following are 
primarily statistical (using $\sim O(500)$ configurations) using a 
bootstrap method.


\subsubsection{Twisted boundary conditions}
\label{chubby}


While the formalism developed here is designed to allow non-degenerate
energy states (quasi-degenerate energy states), it is clearly necessary
to keep the energy of the states close to each other. Spatial momentum 
on the lattice is discretised and given in each direction in steps of 
$2\pi/N_s$, which is coarse on this lattice size and makes this requirement
difficult to achieve. To obtain a finer energy level separation we use 
twisted boundary conditions, \cite{Bedaque:2004kc}, it being sufficient
to apply this just to the valence quarks, 
\cite{Sachrajda:2004mi,Bedaque:2004ax,Flynn:2007ess,Boyle:2008yd}.
In general we take for a quark, $q$,
\begin{eqnarray}
   q(\vec{x}+N_s\vec{e}_i,t) = e^{i\theta_i} q(\vec{x},t)\,, 
                                \quad i = 1, 2, 3 \,.
\end{eqnarray}
This is rather similar to the Feynman--Hellmann procedure
described earlier, and leads to a shift in the momentum in the Green's 
function by $\vec{\theta}/N_s$. 
Specifically we choose to compose $\vec{q}$ as a twist for the $u$ quark
in the $2$-direction. In other words we set the lattice momenta to
\begin{eqnarray}
   \vec{p} = \vec{0}\,, \qquad 
   \vec{q} = \left(0, {\theta_2\over N_s}, 0\right) \,,
\end{eqnarray}
and use the results of section~\ref{transition_els}.
For the runs and number of configurations used in this article, we have 
determined the masses (in lattice units) as $M_N = 0.424(11)$ and 
$M_\Sigma = 0.461(10)$ close to those given in \cite{Bietenholz:2011qq}
(the number of configurations used in this study is somewhat smaller).
In Table~\ref{twist_parameters} we give the $\theta_2$-parameter
\begin{table}[!htb]
   \begin{center}
   \begin{tabular}{cr|rrr}
      run \# & $\theta_2/\pi$ & $\vec{q}^{\,2}$ & $E_N$ &
      $M_\Sigma - E_N$  \\
      \hline
      1 & 0.0   & 0.0    & 0.424(11)  & 0.0366(33) \\
      2 & 0.448 & 0.0019 & 0.429(10)  & 0.0351(35) \\
      3 & 1     & 0.0096 & 0.437(10)  & 0.0301(42) \\
      4 & 1.6   & 0.0247 & 0.450(12)  & 0.0182(57) \\
      5 & 2.06  & 0.0408 & 0.462(12)  & 0.0030(69) \\
      6 & 2.25  & 0.0488 & 0.469(13)  &-0.0037(78) \\
   \end{tabular}
   \end{center}
   \caption{$\theta_2$-twist values, together with $\vec{q}^{\,2}$,
            $E_N(\vec{q})$, $M_\Sigma-E_N(\vec{q})$ in lattice units.
            In addition $M_N = 0.424(11)$ and $M_\Sigma = 0.461(11)$.}
\label{twist_parameters}
\end{table}
values that we have used in our investigation.
Run $\#1$ in the table corresponds to $\vec{q} = \vec{0}$ 
or ``$q_{\rm max}$'', run $\#2$ corresponds approximately to $Q^2 = 0$, 
while runs $\#5$, $\#6$ are the closest we have achieved to 
$E_N(\vec{0}+\vec{q}) = E_\Sigma(\vec{0}) = M_\Sigma$. 
In the table we also give $\vec{q}^{\,2}$, $E_N(\vec{q})$ and the difference
$M_\Sigma-E_N(\vec{q})$ (all in lattice units).
These are the measured values from the relevant $2$-point correlation 
functions.


\subsection{Tests}
\label{tests}


\subsubsection{Correlators and GEVP}
\label{correlator_fitting}

\begin{figure}[!tb]
\begin{minipage}{0.45\textwidth}

   \begin{center}
      \includegraphics[width=7.50cm]{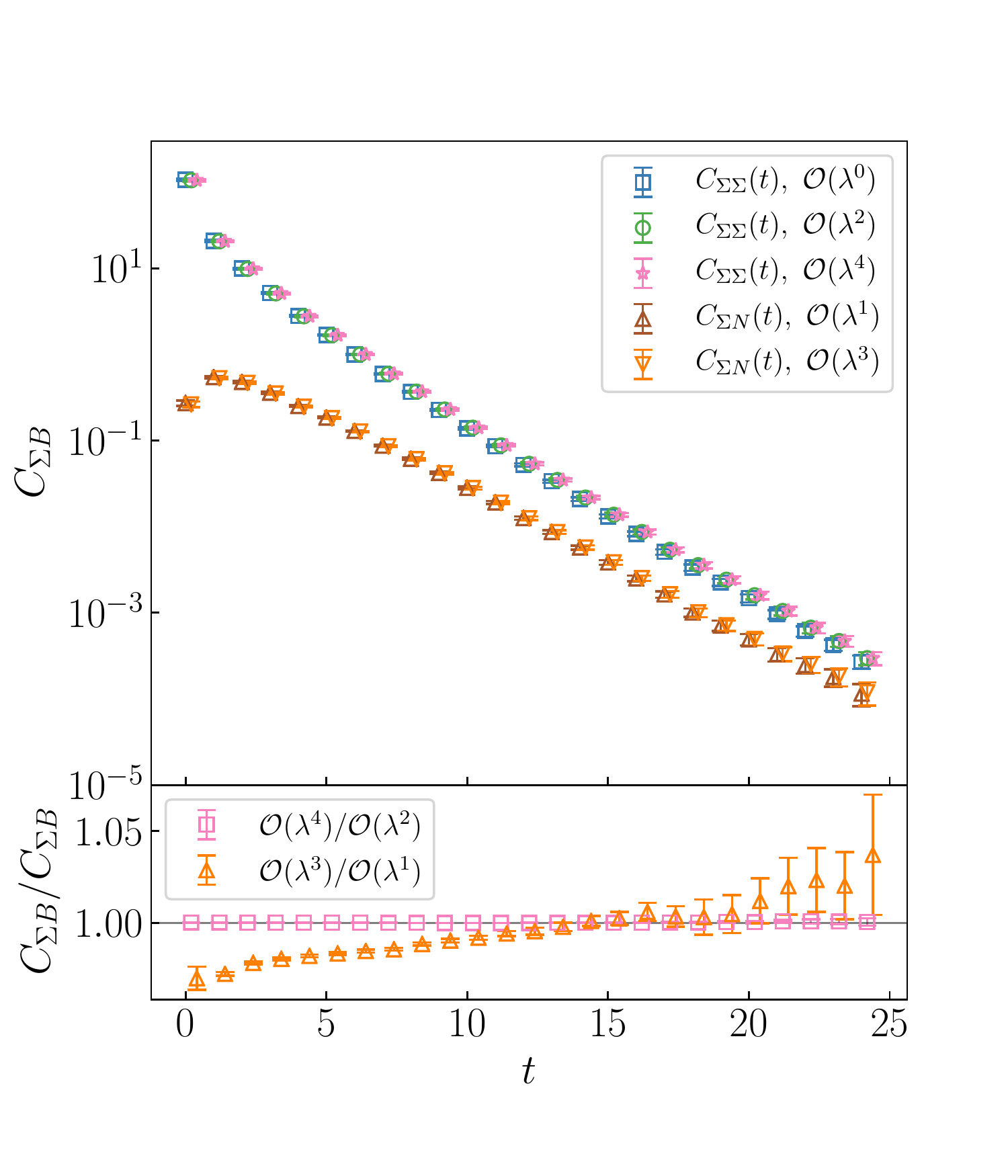}
   \end{center} 

\end{minipage}\hspace*{0.05\textwidth}
\begin{minipage}{0.45\textwidth}

   \begin{center}
      \includegraphics[width=7.50cm]{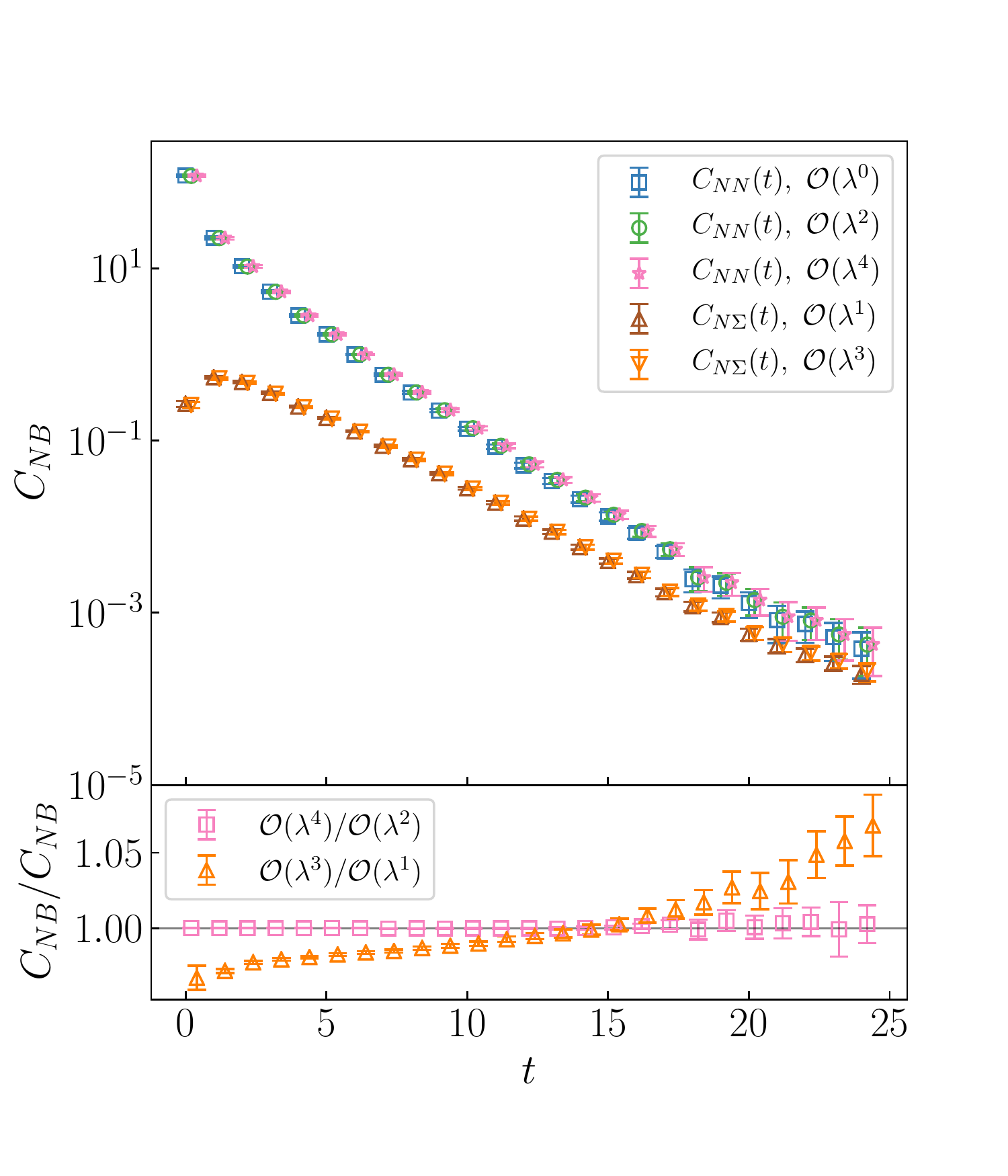}
   \end{center} 

\end{minipage}
\caption{LH panel: The $C_{\lambda\,\Sigma\Sigma}$ correlation 
         functions (at $O(1)$, $O(\lambda^2)$ and $O(\lambda^4)$, 
         squares, circles and stars respectively) and $C_{\lambda\,\Sigma N}$ 
         correlation functions (at $O(\lambda)$ and $O(\lambda^3)$, 
         upper triangles and lower triangles respectively) at 
         $\lambda = 0.025$ against $t$ for $t_0 = 6$ and $\Delta t_0 = 4$.
         The lower sub-plot shows the ratio of correlation functions
         $C_{\lambda\,\Sigma B}$ ($B = \Sigma$, squares and $B = N$, triangles)
         between the two highest orders of $\lambda$ available.
         RH panel: Similarly for $C_{\lambda\,NN}$ and $C_{\lambda\,N\Sigma}$.
         The points are slightly offset for visibility. Both results
         are for run \#5.}
\label{correlators_C}
\end{figure}

We first wish to determine the value of $n$ required for the expansions
in eq.~(\ref{iteration_lambda}) to provide a good approximation for 
the full Green's function of eq.~(\ref{Guu+Gss_res}).
In Fig.~\ref{correlators_C} values of the four correlators
for $t_0 = 6$ and $\Delta t_0 = 4$ where the Green's functions and hence 
correlation functions are computed to various orders of $\lambda$
by iterating eq.~(\ref{iteration_lambda})%
\footnote{By this we mean that at any order we include the appropriate
lower orders, so for example $O(\lambda^4)$ means we generate the
$O(\lambda^0)+O(\lambda^2)+O(\lambda^4)$ terms iterating
eq.~(\ref{iteration_lambda}) for the diagonal Green's functions.}.
The LH panel shows the $C_{\lambda\,\Sigma\Sigma}$ and $C_{\lambda\,\Sigma N}$ 
correlation functions with $\lambda = 0.025$. The lower sub-plot 
shows the ratio of the correlation functions between the two highest 
orders of $\lambda$ available, to give an impression of the convergence
of the series. For the diagonal correlators we see that the change is
negligible, while for the off-diagonal correlation functions the change
is at most a few $\%$ and in the region where the fits are made
(see the following Fig.~\ref{deltaE_expansion}) at most $\sim 1\%$. 
The RH panel shows the $C_{\lambda\,NN}$ and $C_{\lambda\,N\Sigma}$ correlators 
also at $\lambda = 0.025$. A similar discussion and conclusion holds 
as for the LH panel.

Applying the GEVP to the $2\times 2$ matrix of correlation functions
$\Delta E_\lambda$ is calculated from eq.~(\ref{ratio_R}).
The results for $\Delta E_\lambda$ are dependent on $\lambda$, 
eq.~(\ref{DeltaE_decay}), so as $\lambda$ increases, the resulting 
correlation functions will have increasing linear-in-time contributions 
that become dominant. In Fig.~\ref{deltaE_expansion} we investigate this 
\begin{figure}[!tb]
\begin{minipage}{0.45\textwidth}

   \begin{center}
      \includegraphics[width=7.50cm]{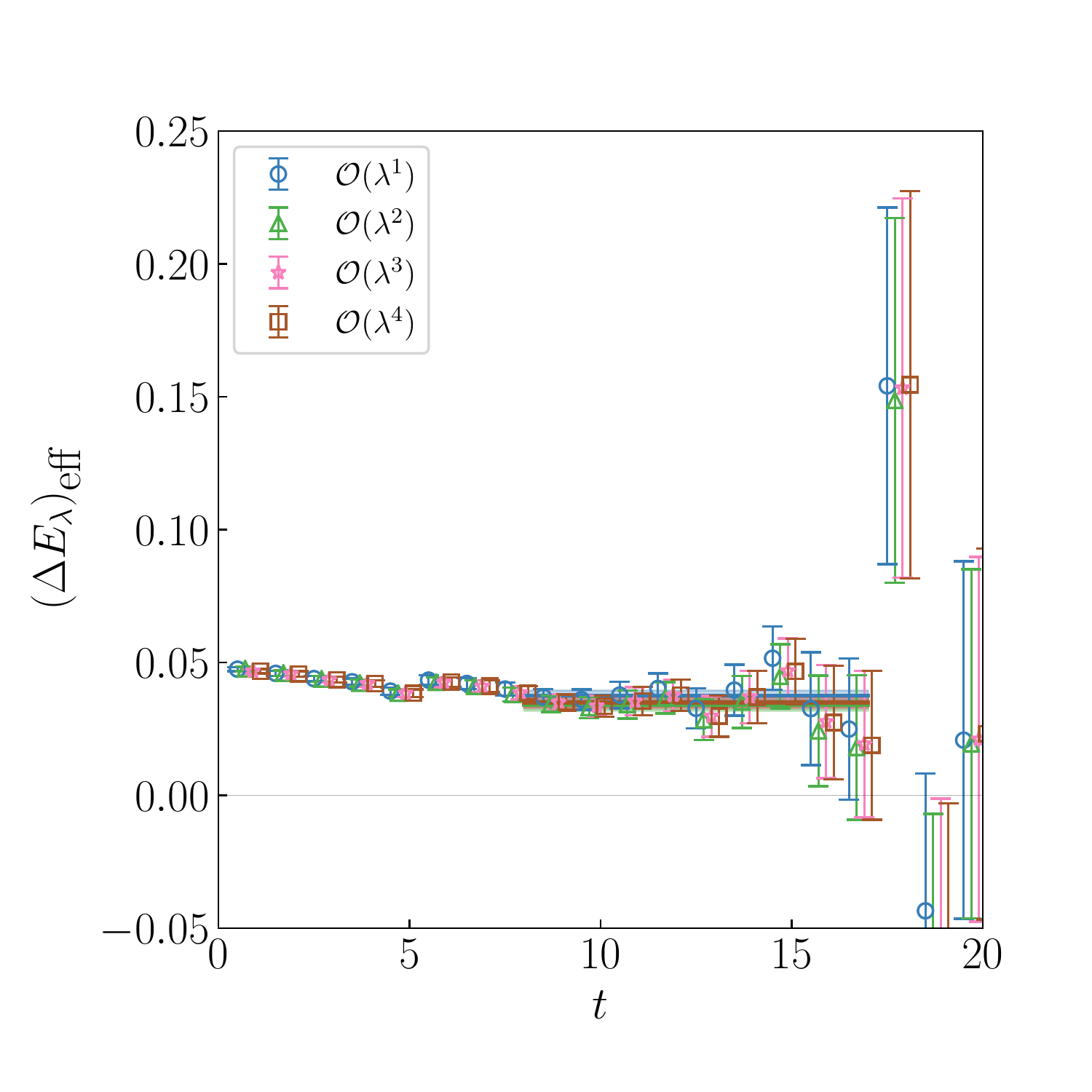}
   \end{center} 

\end{minipage}\hspace*{0.05\textwidth}
\begin{minipage}{0.45\textwidth}

   \begin{center}
      \includegraphics[width=7.50cm]{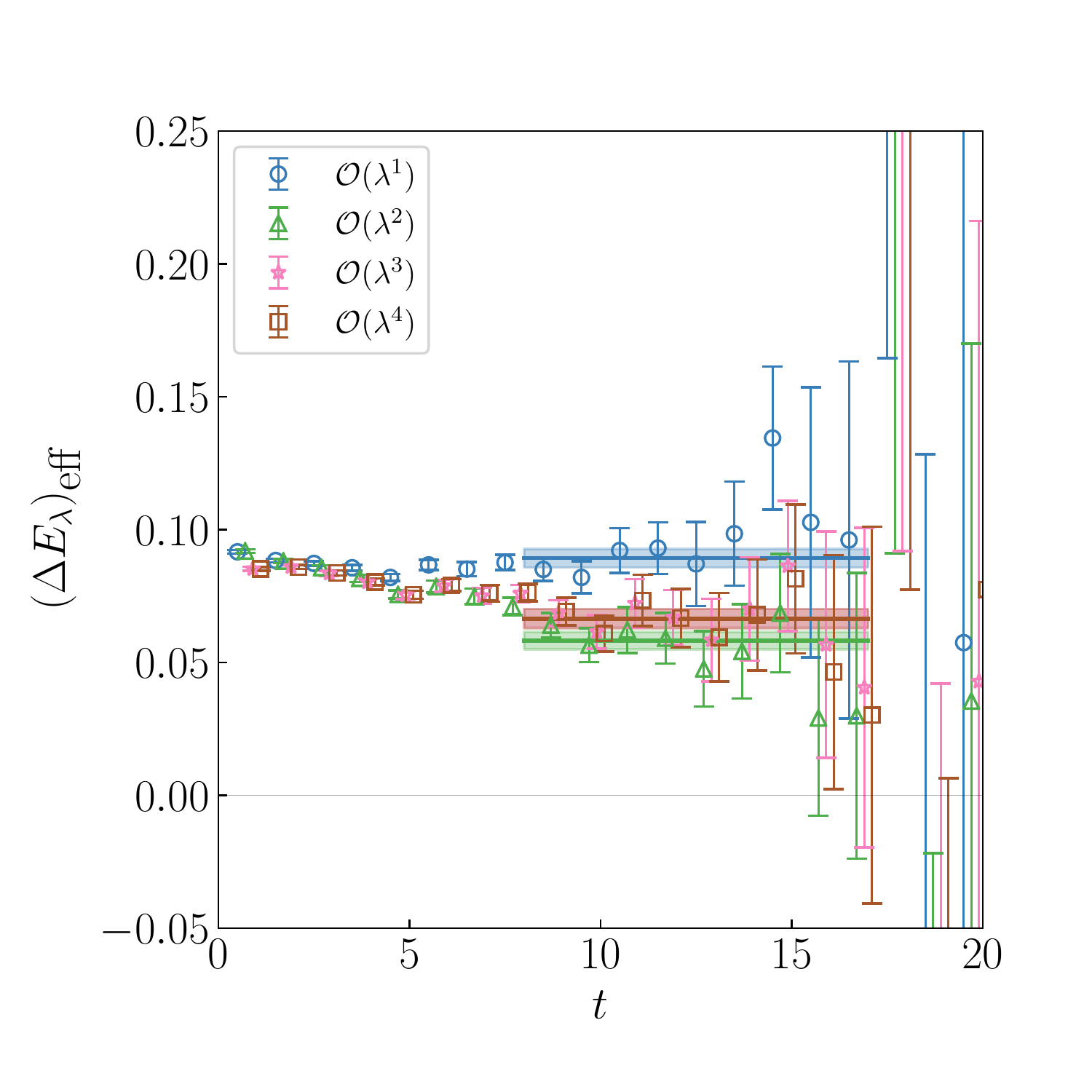}
   \end{center} 

\end{minipage}

\begin{minipage}{0.45\textwidth}

   \begin{center}
      \includegraphics[width=7.50cm]{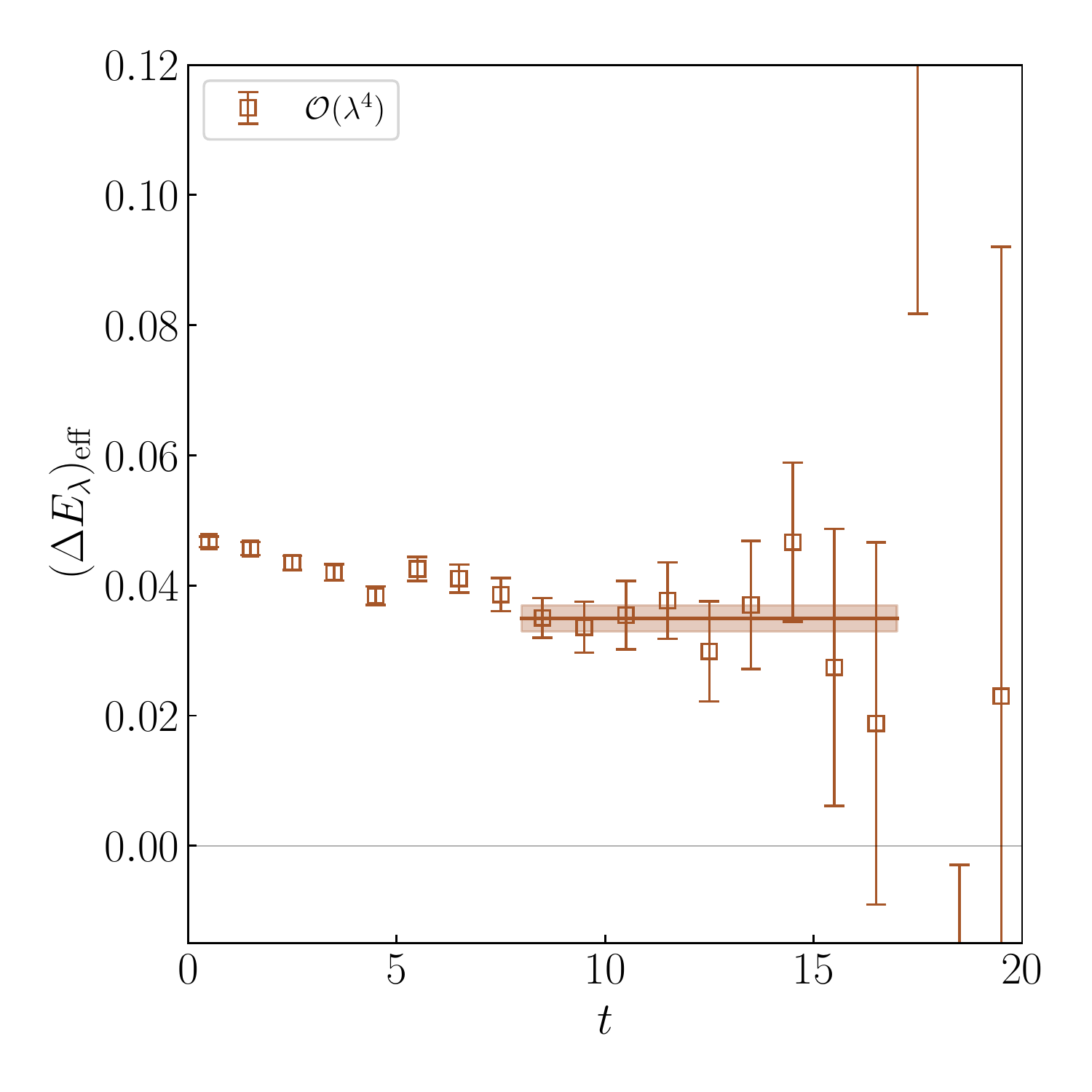}
   \end{center} 

\end{minipage}\hspace*{0.05\textwidth}
\begin{minipage}{0.45\textwidth}

   \begin{center}
      \includegraphics[width=7.50cm]{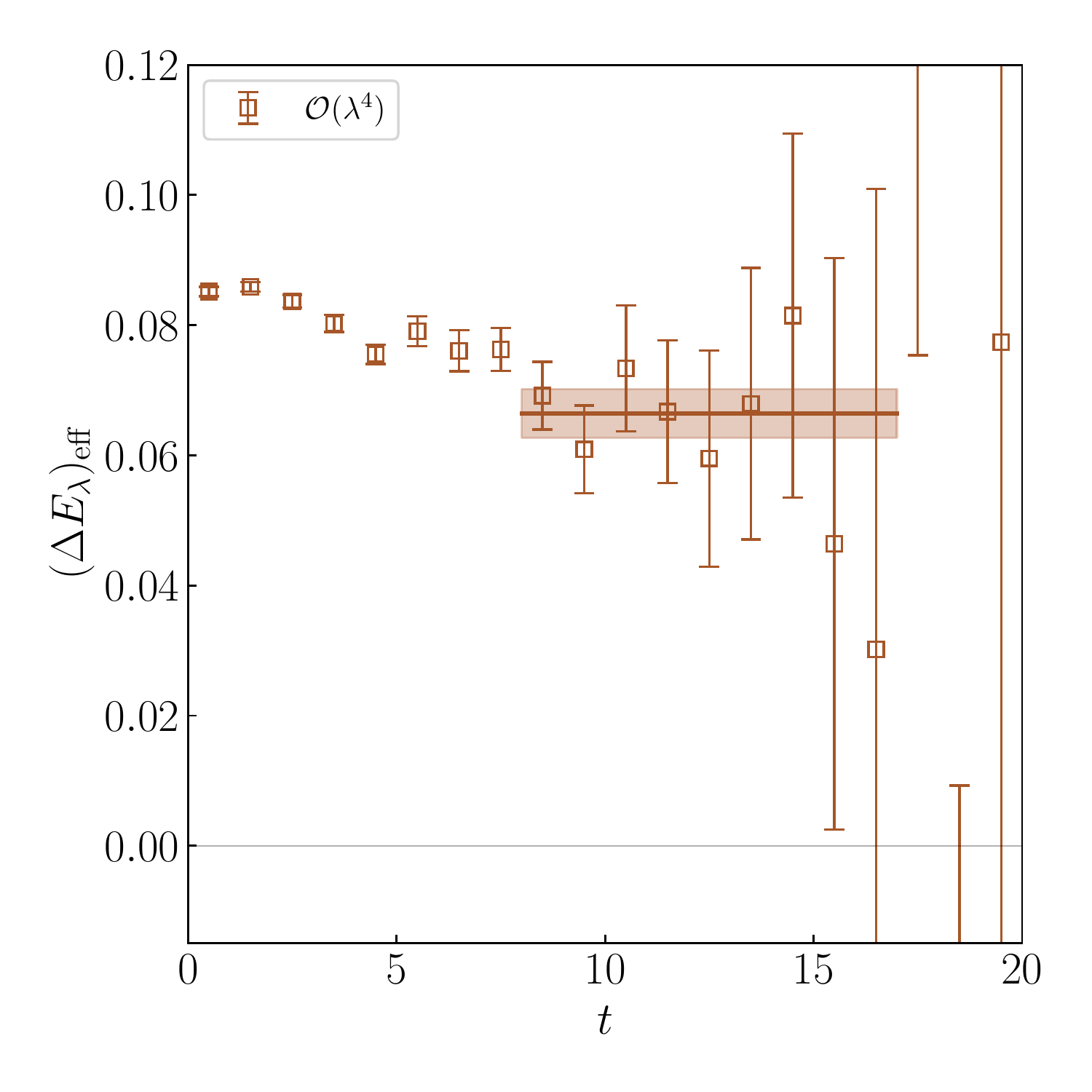}
   \end{center} 

\end{minipage}
\caption{Upper LH panel: $(\Delta E_\lambda)_{\rm eff}$ versus $t$ for 
         $\lambda = 0.025$ at $O(\lambda)$ (circles), 
         $O(\lambda^2)$ (triangles), $O(\lambda^3)$ (stars), 
         $O(\lambda^4)$ (squares) for run $\#5$.
         Upper RH panel: Similarly for $\lambda = 0.05$.
         The points are slightly offset for visibility.
         Also shown is the fit interval used and fit using
         eq.~(\protect\ref{ratio_R}).
         Lower LH plot: An expanded plot of the upper LH plot
         at $O(\lambda^4)$ (squares).
         Lower RH plot: Similarly for the upper RH plot.}
\label{deltaE_expansion}
\end{figure}
by showing the energy difference $(\Delta E_\lambda)_{\rm eff}$ versus $t$
where using eq.~(\ref{ratio_R}) we have 
$(\Delta E_\lambda)_{\rm eff} = - \ln (R_\lambda(t+1)/R_\lambda(t))$.
Again in the upper two plots the various orders in $\lambda$ are shown: 
$O(\lambda)$, $O(\lambda^2)$, $O(\lambda^3)$ and $O(\lambda^4)$. 
The upper LH panel is with $\lambda = 0.025$, while the upper RH panel is 
for $\lambda = 0.05$. It can be seen that the correlator at $O(\lambda)$ 
starts to drift up at the higher value of $\lambda$, 
however $\Delta E_\lambda$ for $O(\lambda^4)$ still shows a plateau for 
this value of $\lambda$. Again, as discussed previously for
the correlation functions in Fig.~\ref{correlators_C} this gives an
impression of the convergence of the Green's functions in 
eq.~(\ref{iteration_lambda}) and it's effect on the determined energies.
In the lower two plots we use an expanded scale for the $O(\lambda^4)$ 
results. 

We need to check that the parameters used in the GEVP are appropriate 
and give reliable results. This becomes more of an issue as the energies 
of the two states come closer together. We will use some criteria to 
determine an optimal set of parameters \cite{Yoon:2016dij},
\begin{itemize}

   \item The correlation functions should have a good statistical 
         signal over the range spanned by $t_0$ and $\Delta t_0$,

   \item The estimate of the energy difference from $c^{(i)}$ should 
         be close to the final estimate of the energy difference.

\end{itemize}
The energies can also be estimated directly from the eigenvalues $c^{(i)}$ 
by using $E_\lambda(c^{(i)}) = -\ln c^{(i)}/\Delta t_0$. Since we are interested 
in the energy difference between the two states, we will consider 
$\Delta E_\lambda(c^{+}, c^{-}) = -\ln (c^{(+)}/c^{(-)})/\Delta t_0$.
This alternative estimation will then be compared to the energy shift 
from fitting to the (diagonal) ratio of correlators as described 
in eq.~(\ref{ratio_R}).

\begin{figure}[!tb]
   \begin{center}
      \includegraphics[width=0.8\textwidth]
                        {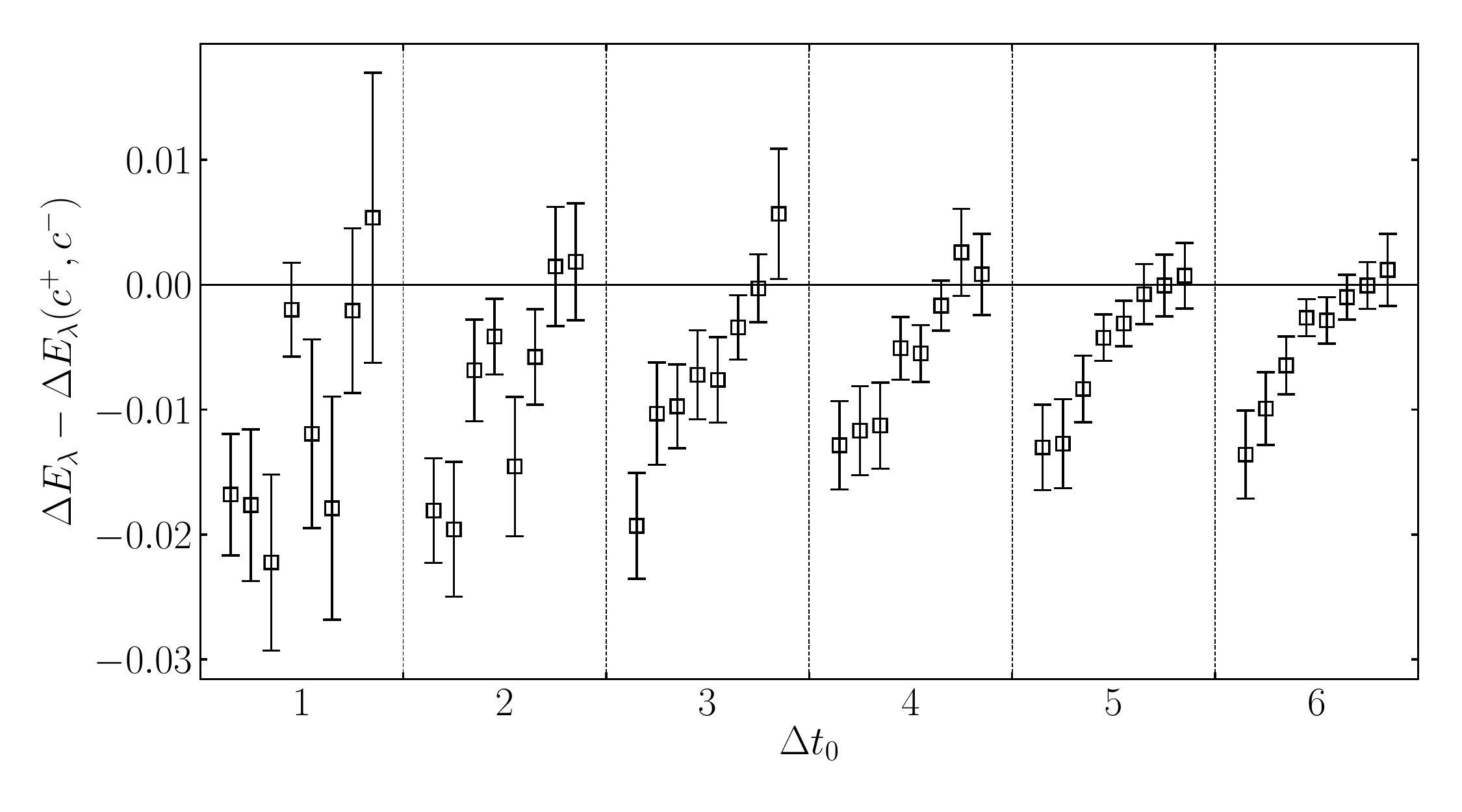}
   \end{center}
\caption{The difference between two estimates of $\Delta E_\lambda$, 
         one calculated from the eigenvalues of the GEVP and the 
         other from a fit to the ratio of correlators in 
         eq.~(\protect\ref{ratio_R}). The difference is shown 
         as a function of both $t_0$ and $\Delta t_0$.
         For each value of $\Delta t_0$ it is shown for the values 
         $t_0 = 1$ -- $8$, where the dashed lines separate the values of 
         $\Delta t_0$. These results are from run \#5. The uncertainties 
         are reduced for $\Delta t_0 \ge 4$ and they start agreeing 
         with zero for $t_0 \ge 6$.}
\label{fig:gevp-slice-run5}
\end{figure}

Fig.~\ref{fig:gevp-slice-run5} shows the difference between 
these two estimates of the energies as a function of both $t_0$ and
$\Delta t_0$ for run \#5. For $\Delta t_0 \ge 4$
the uncertainty in the difference is reduced and for $t_0 \ge 6$ 
the difference starts to agree with zero. Therefore we will choose 
$t_0=6$, $\Delta t_0=4$ as the parameters for the GEVP in runs 
\#4, \#5 and \#6. For the first three runs the difference between 
the energies of the nucleon and $\Sigma$ is large enough that the GEVP 
gives consistent results for smaller parameters and so we choose 
$t_0=4$, $\Delta t_0 = 2$ for those runs.

With this preliminary background we now discuss the energy shifts 
and state mixing.


\subsubsection{Energy shifts}

\begin{figure}[!tb]
\begin{minipage}{0.45\textwidth}

   \begin{center}
      \includegraphics[width=7.50cm]{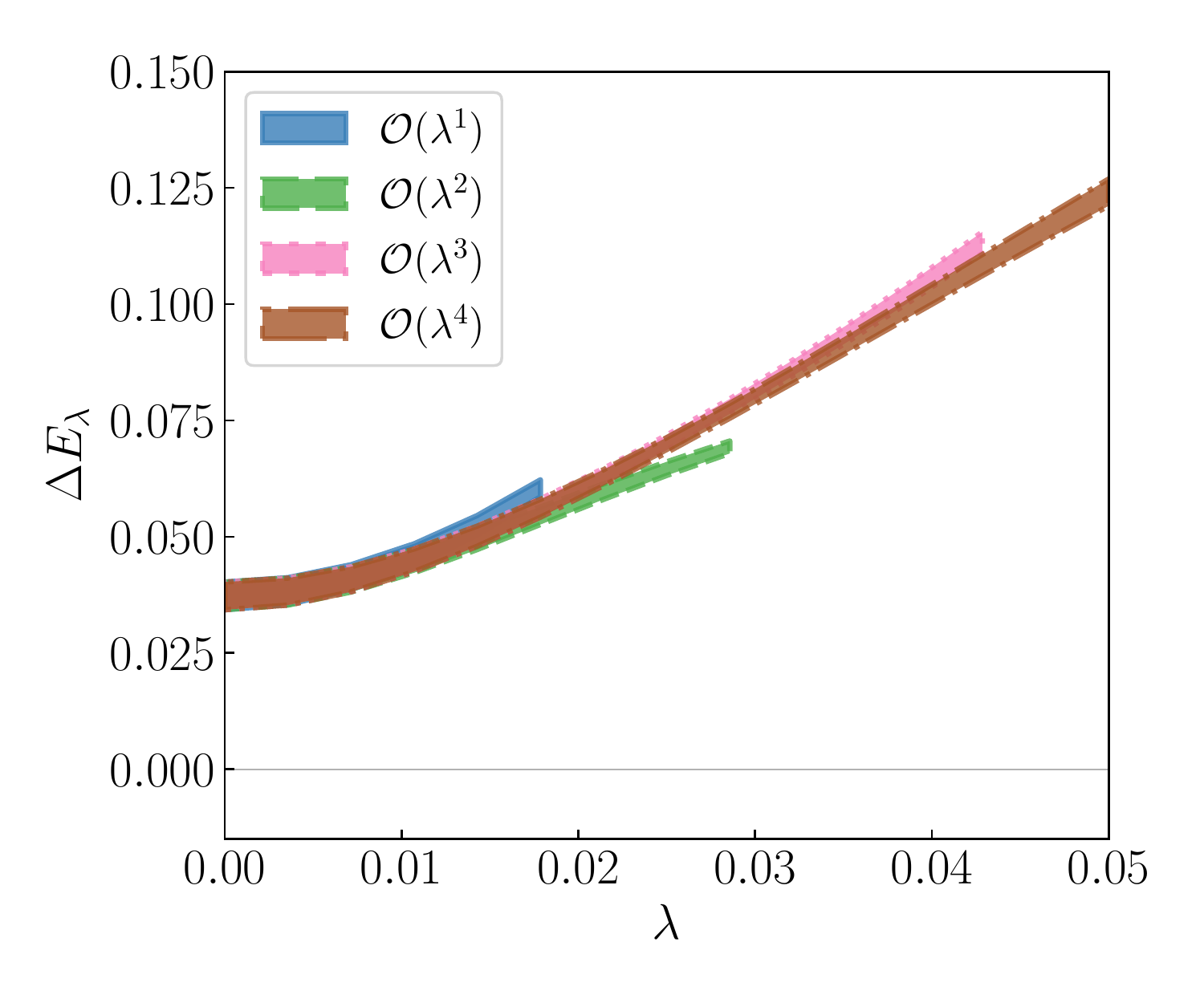}
   \end{center} 

\end{minipage}\hspace*{0.05\textwidth}
\begin{minipage}{0.45\textwidth}

   \begin{center}
      \includegraphics[width=7.50cm]{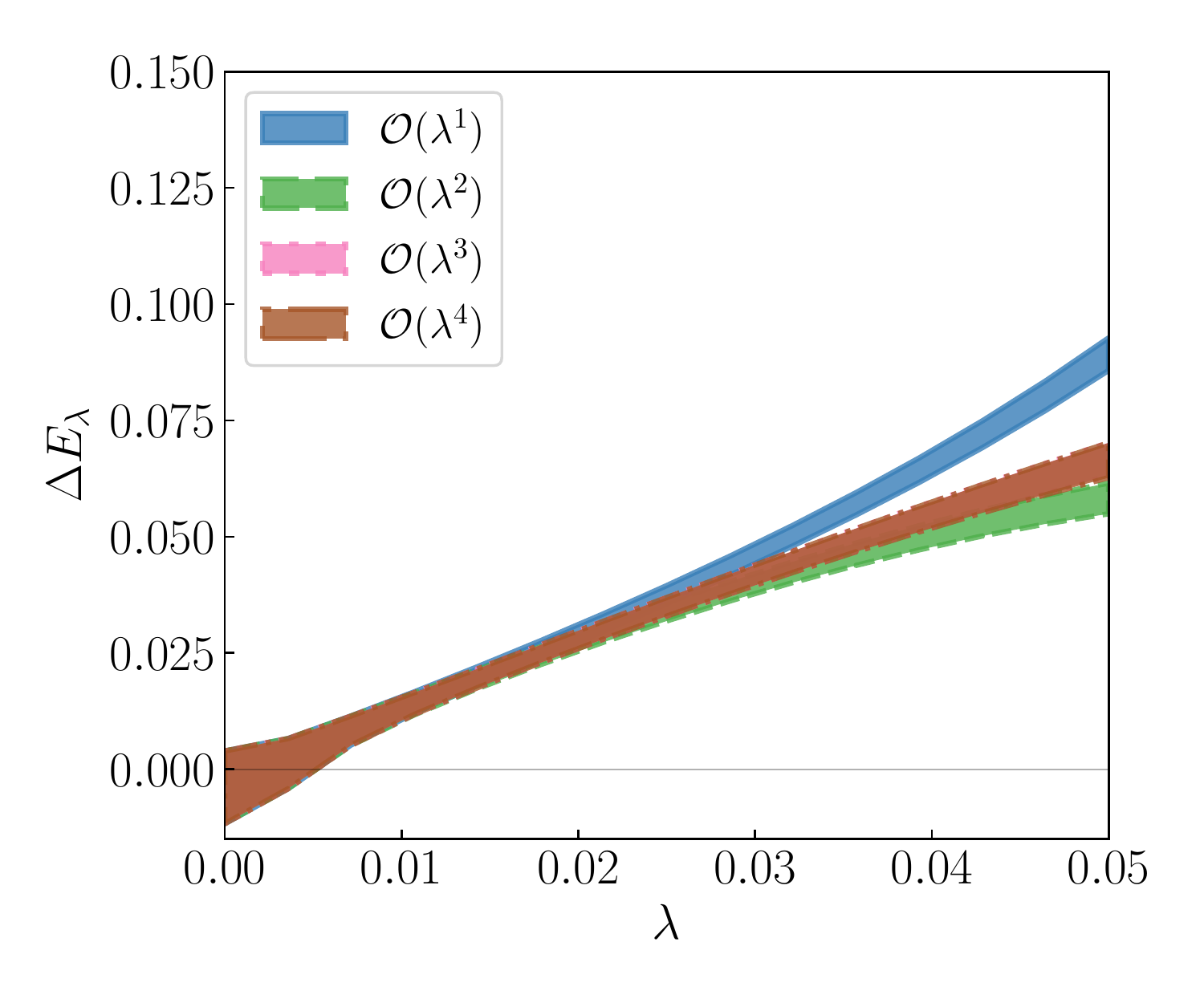}
   \end{center} 

\end{minipage}
\caption{LH panel: The $\lambda$-dependence for run $\#1$ for 
         $\Delta E_\lambda$. The numerical results for each
         order in $\lambda$ ($O(\lambda)$, $O(\lambda^2)$, $O(\lambda^3)$
         and $O(\lambda^4)$) are given as bands.
         RH panel: Similarly for run $\#5$.}
\label{energy_shift_run1+run5}
\end{figure}

We now consider the dependence of $\Delta E_\lambda$ with $\lambda$.
In Fig.~\ref{energy_shift_run1+run5} we show the $\lambda$ dependence
for run $\#1$ (left panel) and $\#5$ (right panel). As the numerical 
results for the correlation functions are coefficients of a polynomial 
in $\lambda$ to $\lambda^4$ we are able to give the results for 
$\Delta E_\lambda$ as a continuous function of $\lambda$. 
This allows a comparison of the numerical results for the various 
orders in $\lambda$. Following this we take the range 
of $\lambda$ to be determined when the last iteration in $\lambda$ 
produces little perceptible numerical effect and we have confidence
in the order of approximation of the Green's function in
eq.~(\ref{iteration_lambda}). From the plots in
Fig.~\ref{energy_shift_run1+run5} between the $O(\lambda^3)$ 
and $O(\lambda^4)$ terms this is the case for the 
range for $\lambda$ of $0 \le \lambda \lsim 0.04$.


\subsubsection{State mixing}
\label{mixing_states}


The eigenvectors which resulted from the GEVP calculation give insight 
into how much mixing is occuring between the two states at the given 
$\lambda$ value. We expect there to be minimal mixing for the data at 
momentum values which are far removed from the crossover point of 
the nucleon energy and the sigma mass, and more mixing for momentum values 
near the crossover point. To show how the mixing changes, we now consider 
the eigenvectors.

\begin{figure}[!tb]
\begin{minipage}{0.45\textwidth}

   \begin{center}
      \includegraphics[width=7.50cm]
                        {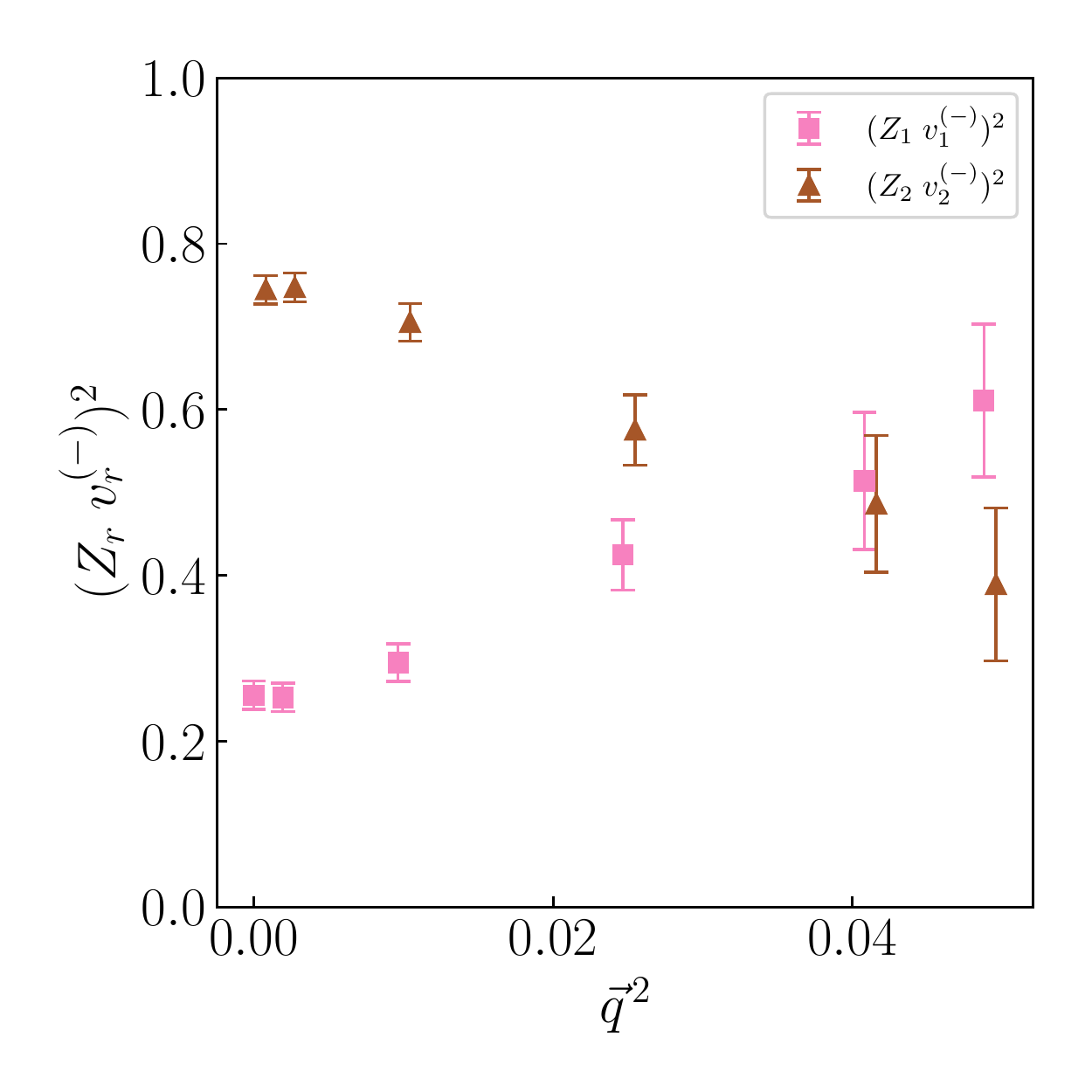}
   \end{center} 

\end{minipage}\hspace*{0.05\textwidth}
\begin{minipage}{0.45\textwidth}

   \begin{center}
      \includegraphics[width=7.50cm]
                       {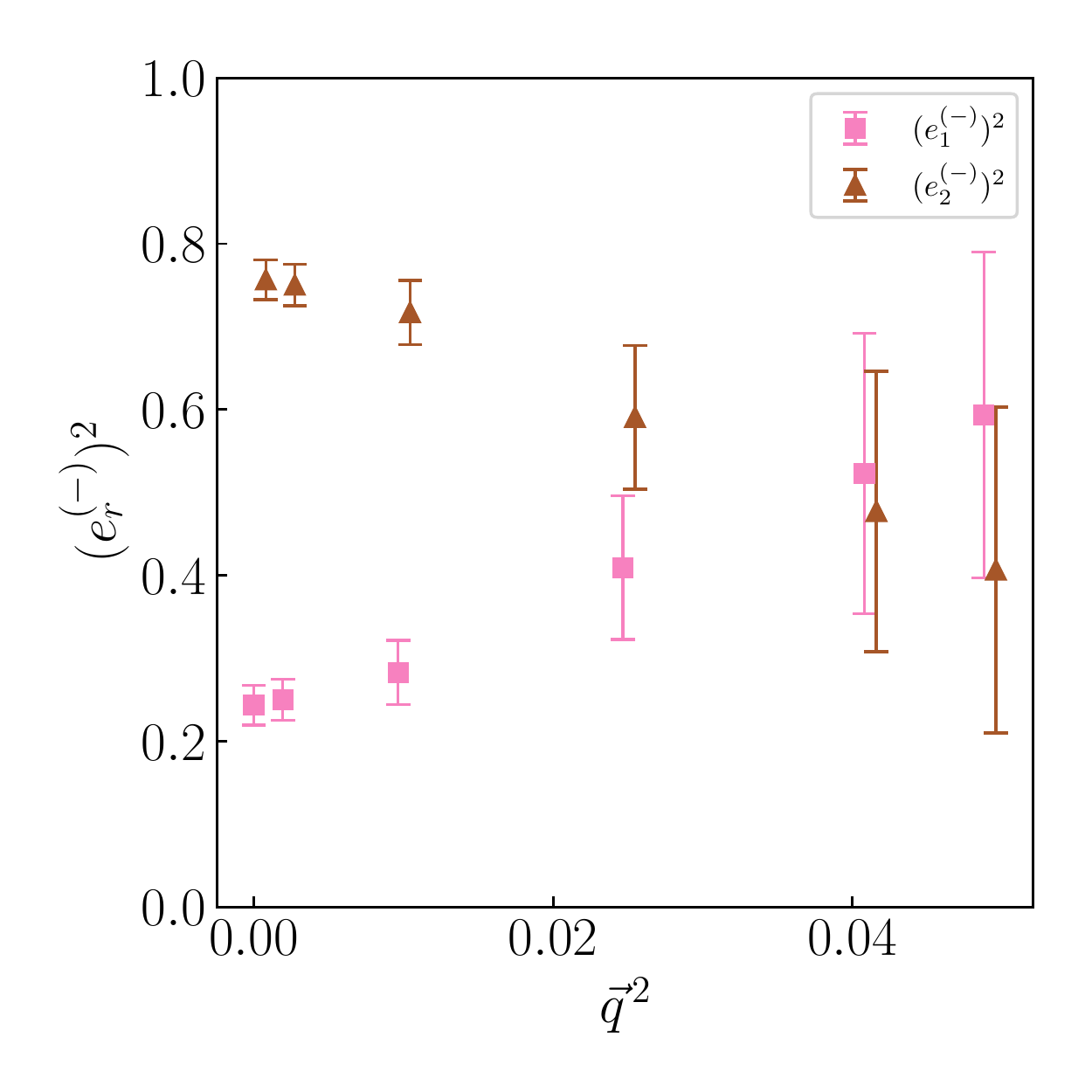}
    \end{center} 

\end{minipage}

\caption{LH panel:
         Normalised $Z_rv_r^{(-)\,2}$ ($r=1$ squares, $r=2$ triangles) 
         against $\vec{q}^{\,2}$. From eq.~(\ref{u_v_e}) we see that these 
         are equal to $e_r^{(-)\,2}$. 
         RH panel: 
         $e_r^{(-)\,2}$, ($r=1$ squares, $r=2$ triangles) from
         eq.~(\protect\ref{eigenvec}) against momentum $\vec{q}^{\,2}$.
         Both plots are for $\lambda=0.025$.} 
\label{mixing_v1p}
\end{figure}

For each eigenvector the square of each component separately 
will then give an indication of how the mixing changes with respect 
to the momentum (the normalisation of each eigenvector being $1$). 
This can be seen in Fig.~\ref{mixing_v1p} where in the LH panel
we plot the normalised $(Z_rv_r^{(-)})^2$ for $r = 1$ (squares) and
$2$ (triangles) versus $\vec{q}^{\,2}$, where $v_r^{(-)}$ is determined 
by the GEVP procedure. From eq.~(\ref{u_v_e}) we see that these are equal
to $e_r^{(-)\,2}$. As a check we also show in the RH panel of the figure 
$e_r^{(-)\,2}$, $r = 1$, $2$ directly computed from eq.~(\ref{eigenvec}) 
using the previously determined fit values from the energies. 
(As also discussed there the $e_r^{(+)\,2}$ are related to the $e_r^{(-)\,2}$
by an interchange.) Values near zero or one indicate minimal state mixing
and values near $1/2$ indicate a high amount of mixing between the states. 
Mixing occurs after run $\#4$ where $E_N \approx M_\Sigma$.

We shall consider avoided energy level mixing in more detail in the 
next section, section~\ref{avoided_energy_levels}.


\section{Results}
\label{results}


\subsection{Energy level comparison}
\label{energy_levels_comparison}


In the RH panel of Fig.~\ref{energy_shift_run1+run5} it can be seen 
that it is possible for the energy shift to be negative for small 
values of $\lambda$. This is due to the ordering of the states being 
difficult to determine at these values of $\lambda$. 
Since the fitting function in 
eq.~(\ref{DeltaE_decay}) is strictly positive, it will not produce a good 
fit for the runs where $\Delta E_\lambda$ gets close to zero. To solve this, 
we square the data and fit to the square of the function in 
eq.~(\ref{DeltaE_decay}). We will also predetermine the value of the energy
shift for the unperturbed two-point function 
$\Delta E_0 = |E_N(\vec{q})-M_\Sigma| $ (i.e.\ $\lambda=0$) 
and fix this in the fitting function. 
The matrix element is now the only free parameter.

\begin{figure}[!h]
\begin{minipage}{0.45\textwidth}

   \begin{center}
      \includegraphics[width=7.50cm]
                         {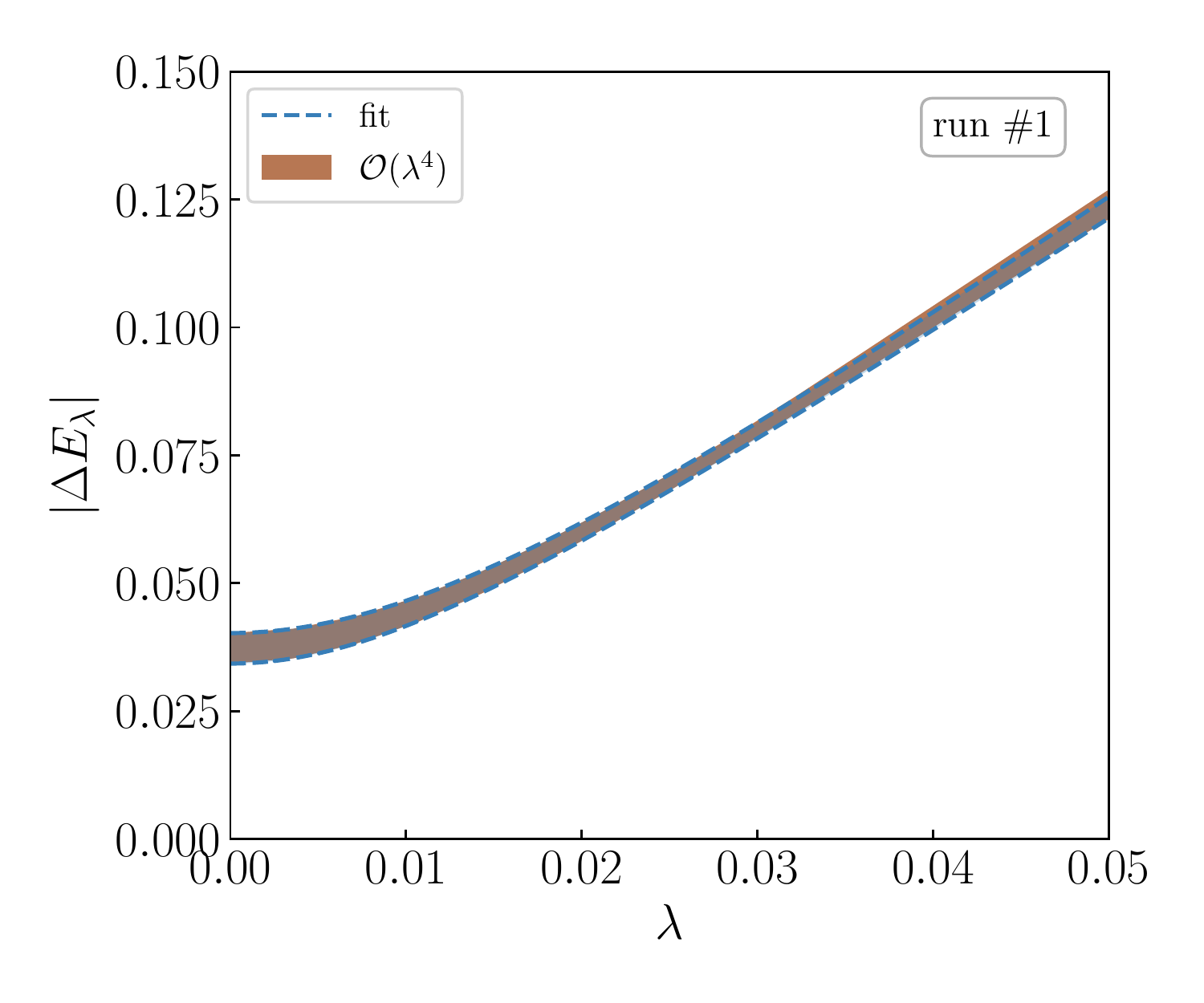}
   \end{center} 

\end{minipage}\hspace*{0.05\textwidth}
\begin{minipage}{0.45\textwidth}

   \begin{center}
      \includegraphics[width=7.50cm]
                         {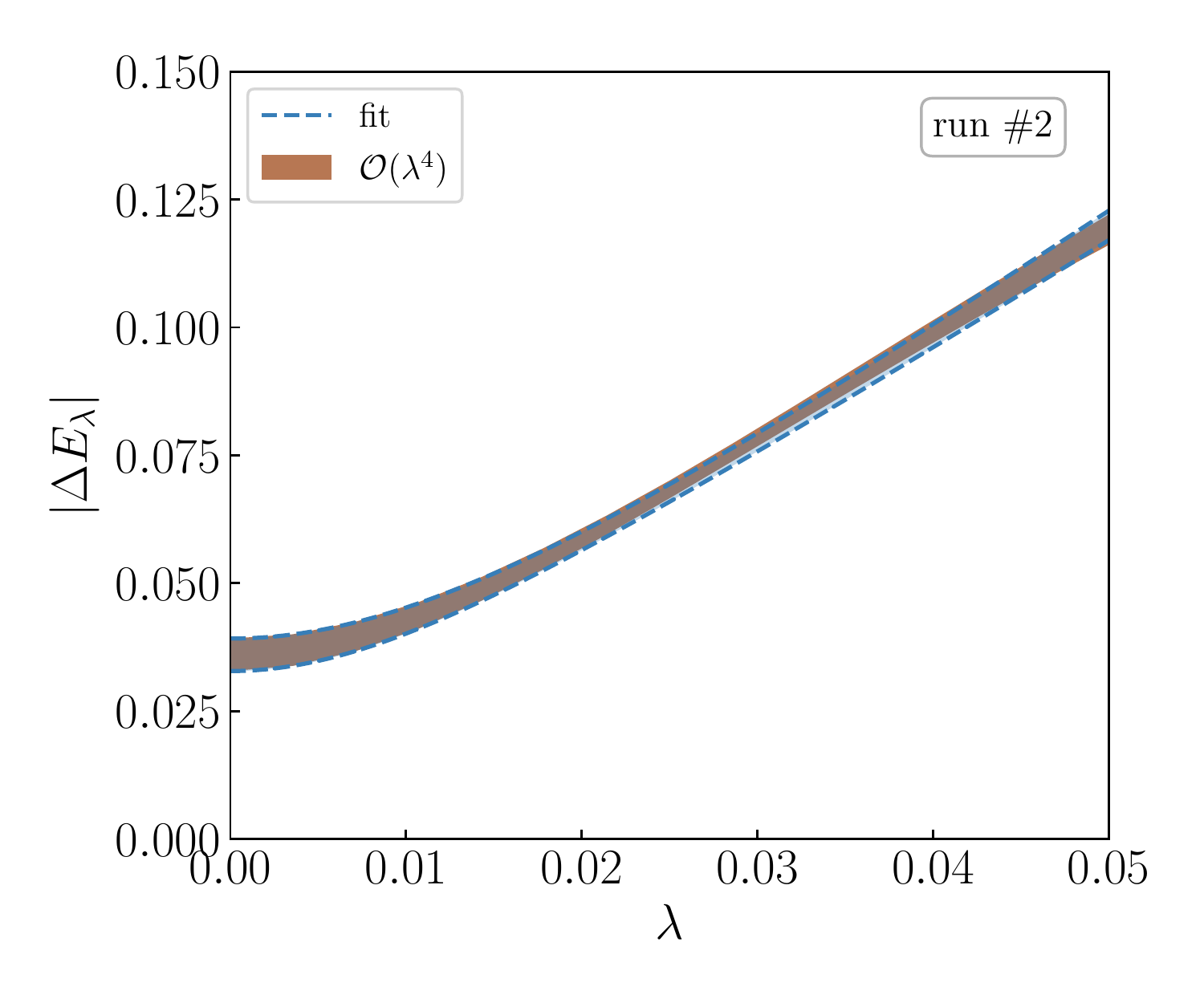}
   \end{center} 

\end{minipage}

\begin{minipage}{0.45\textwidth}

   \begin{center}
      \includegraphics[width=7.50cm]
                         {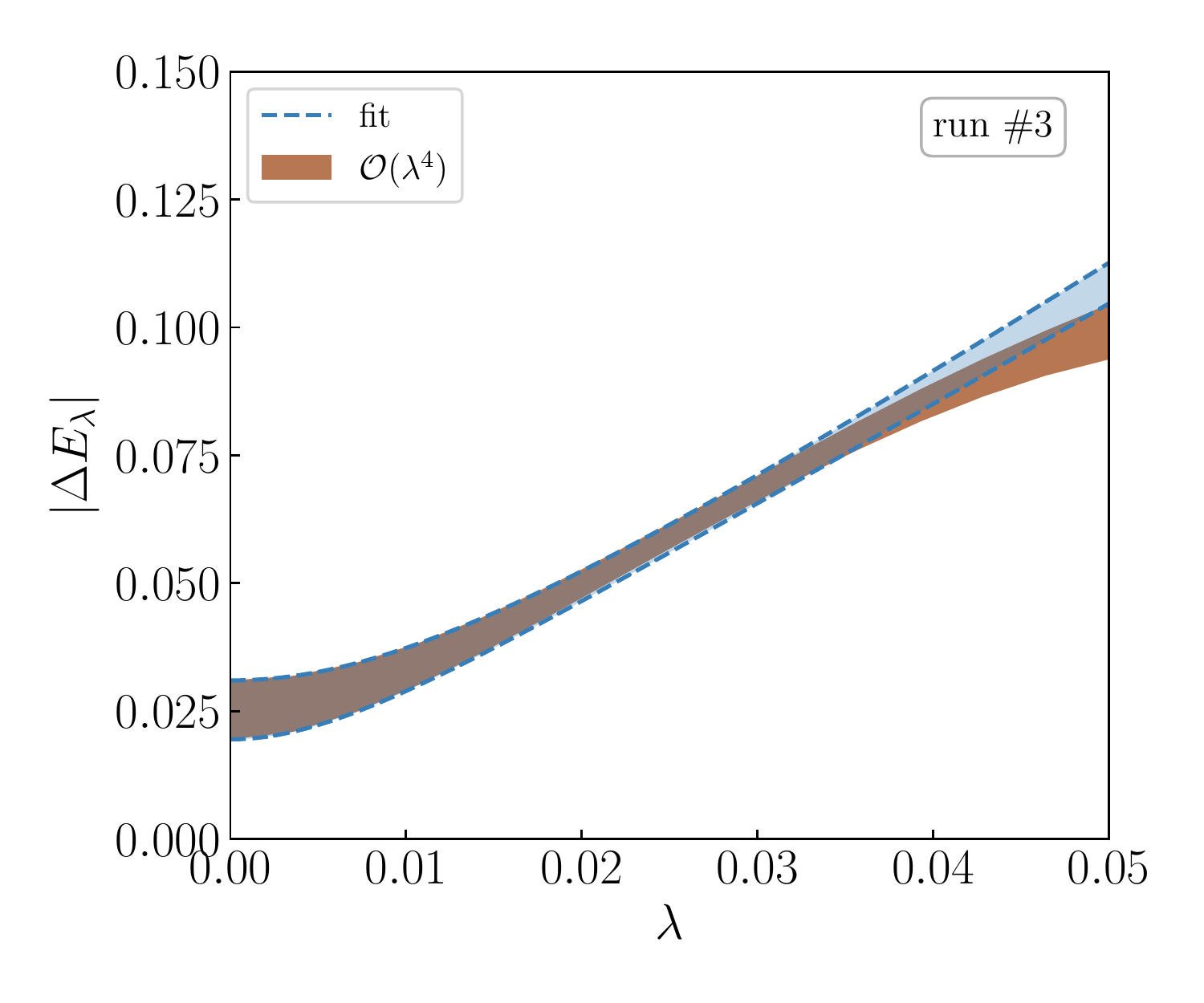}
   \end{center} 

\end{minipage}\hspace*{0.05\textwidth}
\begin{minipage}{0.45\textwidth}

   \begin{center}
      \includegraphics[width=7.50cm]
                         {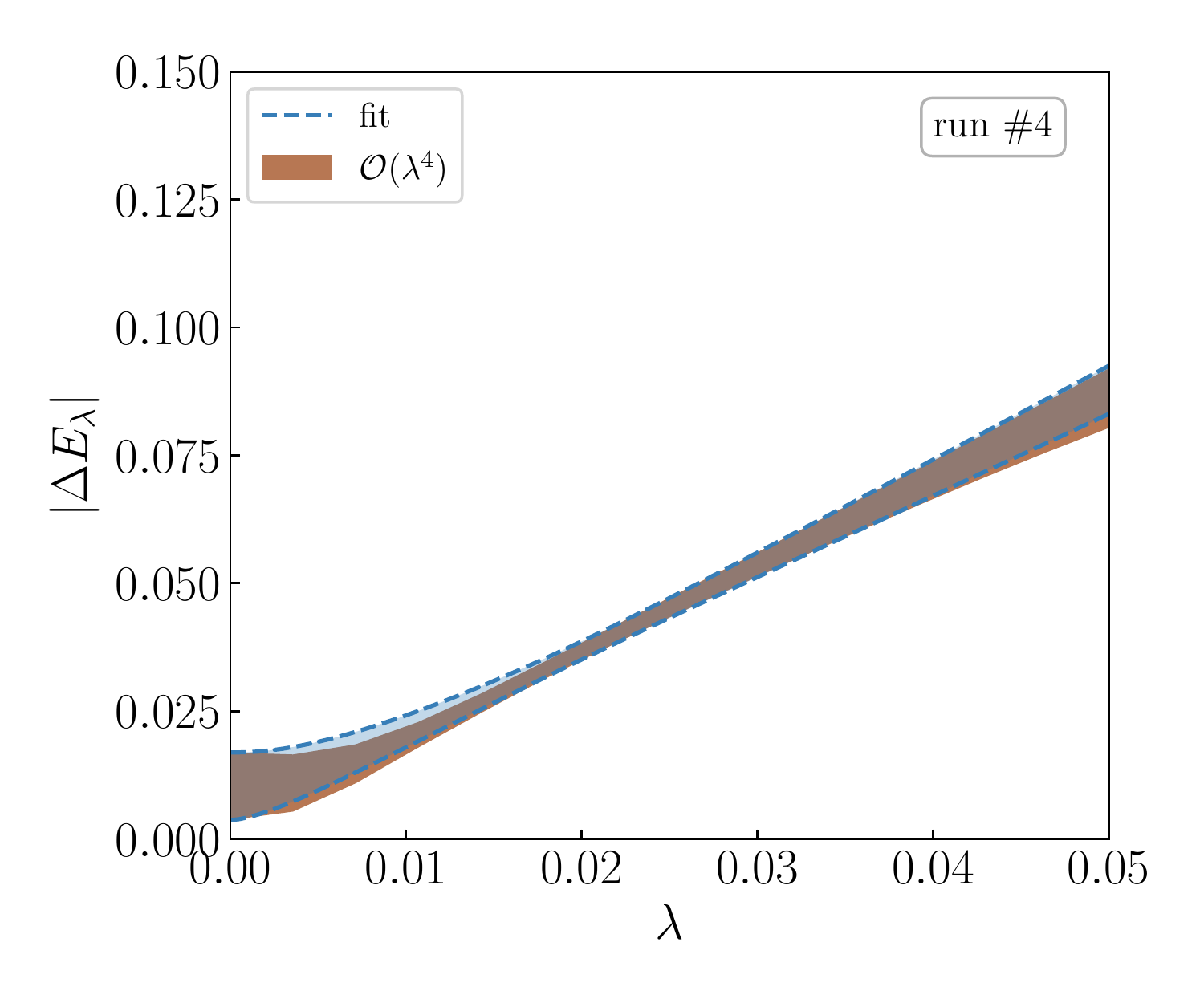}
   \end{center} 

\end{minipage}

\begin{minipage}{0.45\textwidth}

   \begin{center}
      \includegraphics[width=7.50cm]
                         {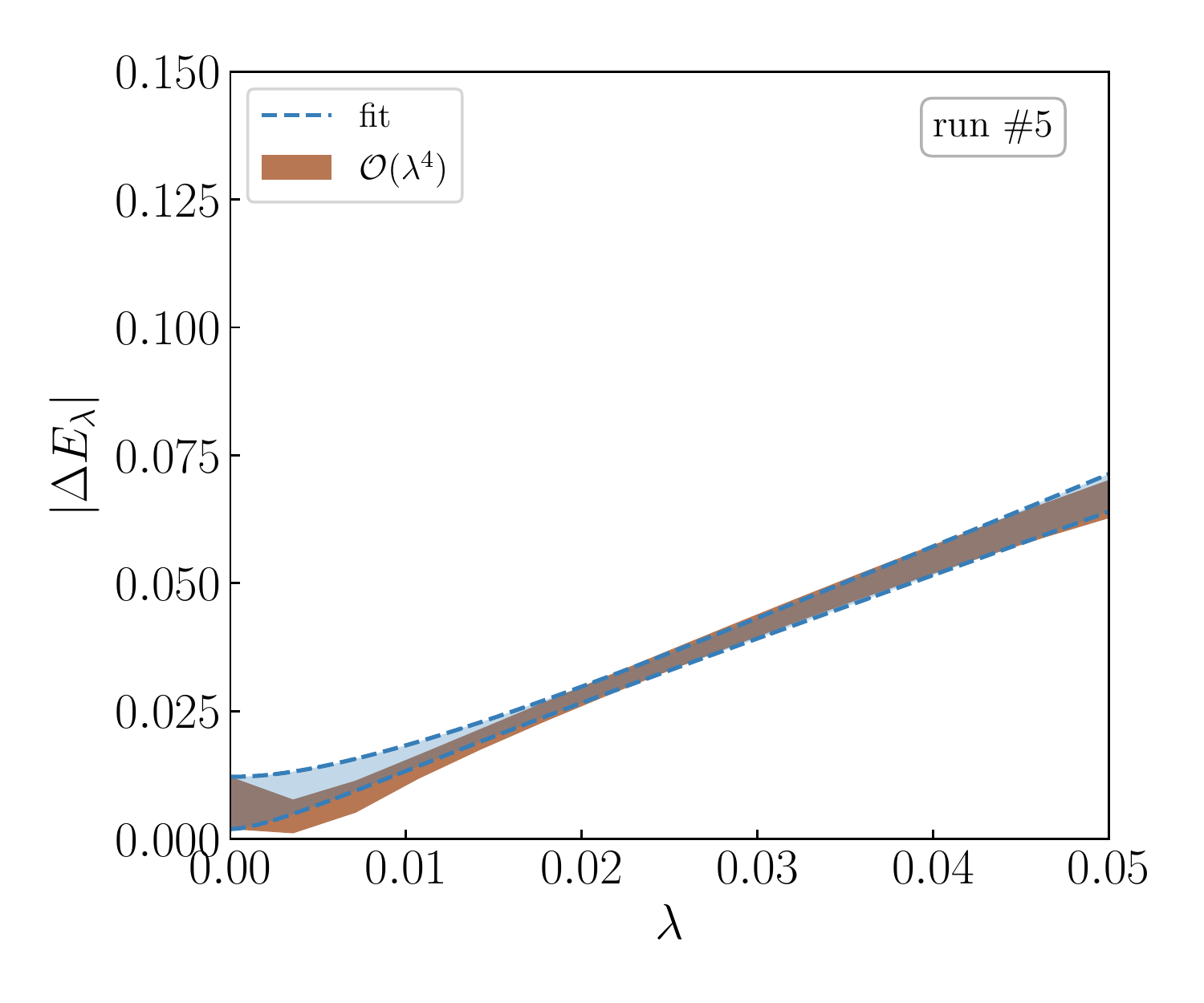}
   \end{center} 

\end{minipage}\hspace*{0.05\textwidth}
\begin{minipage}{0.45\textwidth}

   \begin{center}
      \includegraphics[width=7.50cm]
                         {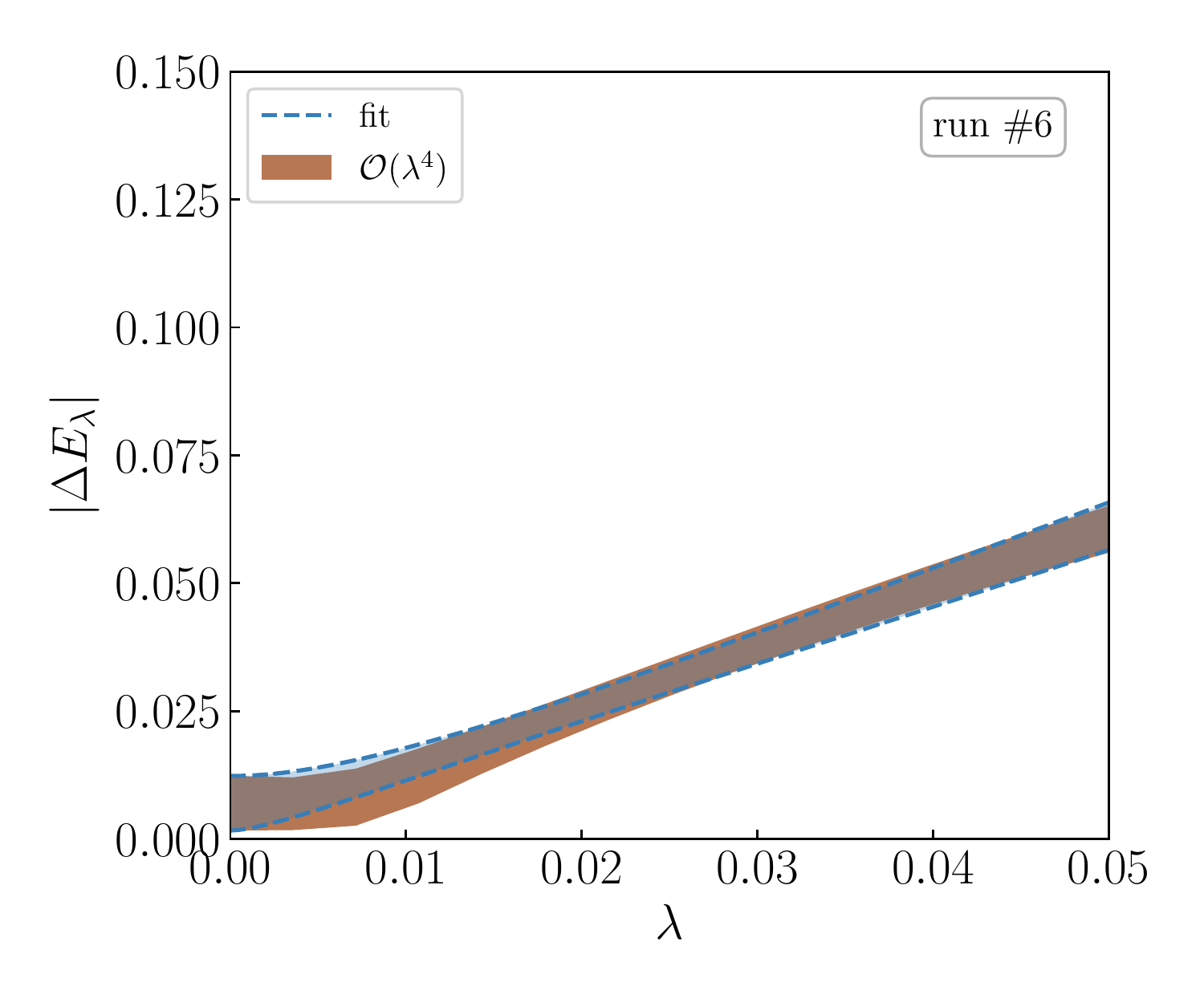}
   \end{center} 

\end{minipage}
\caption{The $\lambda$-dependence for runs $\#1$ (top left),
         $\#2$ (top right), $\#3$ (centre left), $\#4$ (centre right)
         $\#5$ (bottom left) and $\#6$ (bottom right) for 
         $\Delta E_\lambda$. The numerical results at $O(\lambda^4)$
         are given as bands. The fits are given from using
         the square of eq.~(\protect\ref{DeltaE_decay}) 
         as further discussed in the text.}
\label{energy_shift}
\end{figure}

In Fig.~\ref{energy_shift} we show the $\lambda$ dependence of the energy
shifts, $\Delta E_\lambda$, for runs $\#1$ -- $\#6$ at 
$O(\lambda^4)$. These, together with their associated errors, are shown 
as bands in the figures. 
A fit is made by using the square of eq.~(\ref{DeltaE_decay}).
We clearly see in the series of plots that when the quasi-degenerate 
states become simply degenerate states i.e.\ if $E_N(\vec{q}) \approx M_\Sigma$ 
(runs $\#5$, $\#6$) then we have approximate linear behaviour in 
$\lambda$ through the origin.


\subsection{Avoided energy level crossing}
\label{avoided_energy_levels}

\begin{figure}[htb]
\begin{minipage}{0.45\textwidth}

   \begin{center}
      \includegraphics[width=7.50cm]{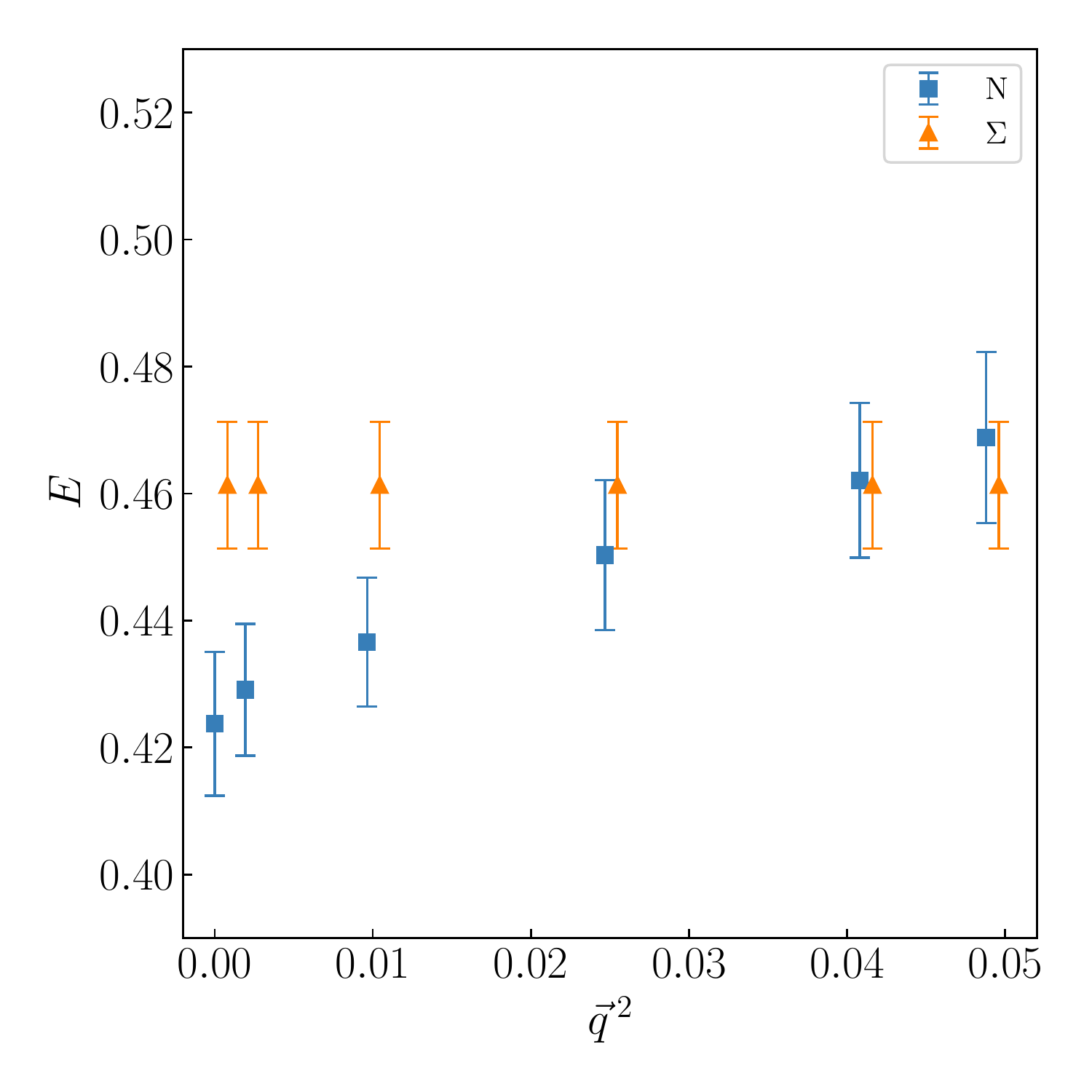}
   \end{center} 

\end{minipage}\hspace*{0.05\textwidth}
\begin{minipage}{0.45\textwidth}

   \begin{center}
      \includegraphics[width=7.50cm]{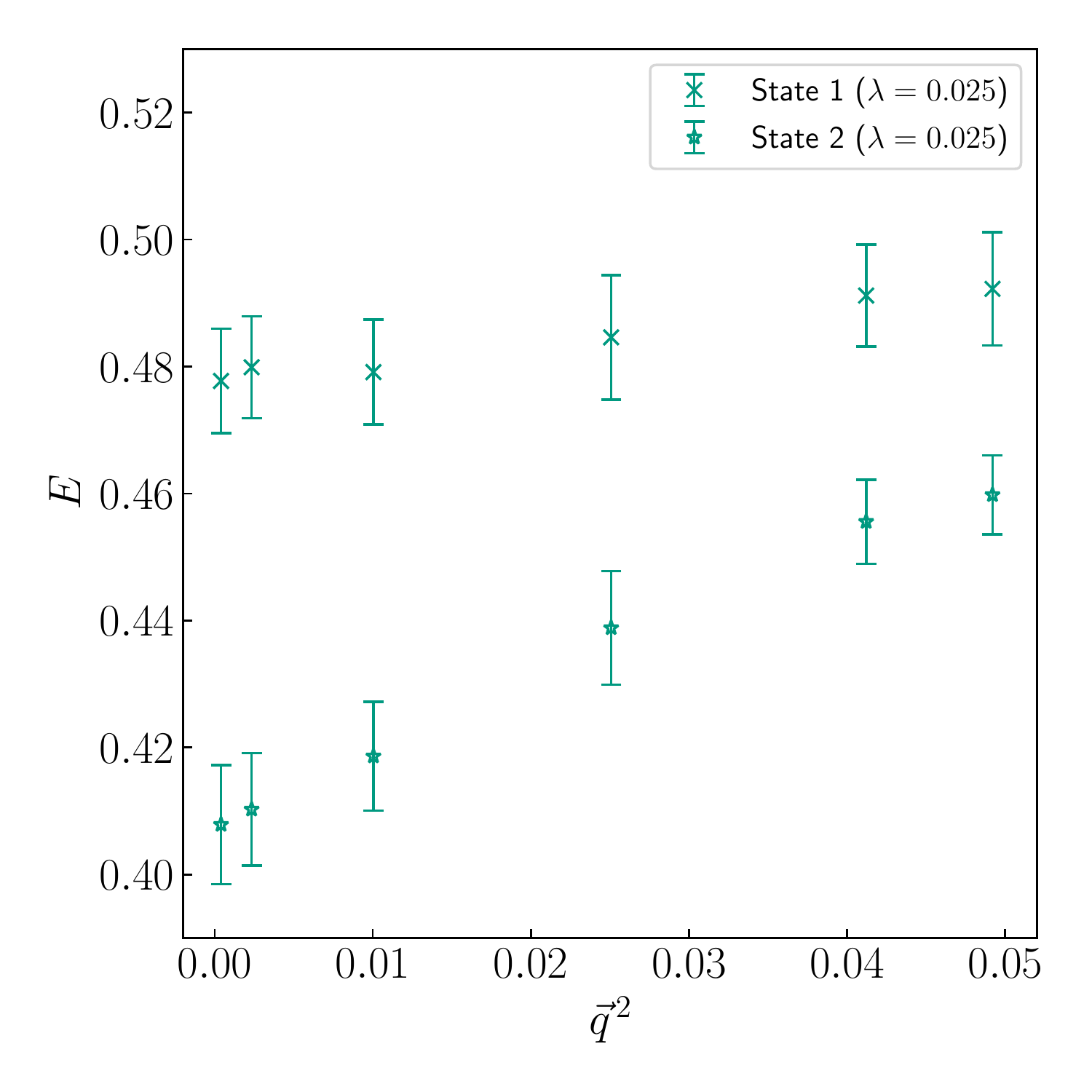}
   \end{center} 

\end{minipage}
\caption{LH panel: The non-interacting $\Sigma$ mass (filled triangles) 
         and $N$ energy states (filled squares) as a function of $\vec{q}^2$
         using the results of Table~\ref{twist_parameters}.
         RH panel: The mixed states $E^{(-)}_\lambda$ shown as crosses
         and $E^{(+)}_\lambda$ shown as stars.}
\label{avoided_level_crossing}
\end{figure}

We now investigate avoided energy level crossing.
In the LH plot of Fig.~\ref{SigN_avoided} we sketched the 
non-interacting case. In the interacting case (RH plot of 
Fig.~\ref{SigN_avoided}) the quasi-energy levels will avoid each other. 
While previously we only considered $\Delta E_\lambda$, we now consider 
each energy level, $E_\lambda^{(\pm)}$, separately.
We compute these from eq.~(\ref{E_pm}) by using the previously 
determined $\Delta E_\lambda$ together with 
$E_N(\vec{q})$ and $M_\Sigma$. In the LH panel of 
Fig.~\ref{avoided_level_crossing} we plot 
non-interacting stationary $\Sigma$ and the measured $E_N(\vec{q})$ against 
$\vec{q}^{\,2}$ using the $\lambda = 0.025$ results. 
We see that the energies cross for runs \#5 and \#6.
In the RH panel of Fig.~\ref{avoided_level_crossing} we show the
interacting case, where we now see avoided level crossing 
of the two energy states. This is similar to the case discussed previously
in section~\ref{transition_els}.


\subsection{Result comparison}


The results shown in section~\ref{energy_levels_comparison}
are for the `bare' matrix element. We take the renormalisation
constant $Z_V = 0.863(4)$, \cite{Bickerton:2019nyz}. This is determined 
from quark-counting for the flavour diagonal matrix elements at 
zero $3$-momentum. Practically this means for this transition
matrix element that $f_1^{\Sigma N}$ is renormalised. For from
eq.~(\ref{V4_matrix_el}) at $\vec{q} = \vec{0}$, the coefficient of
the $f_2^{\Sigma N}$ term vanishes, the coefficient of the
$f_3^{\Sigma N}$ term is $O(M_\Sigma-M_N) \sim O(\delta m_l)$, 
while $f_3^{\Sigma N}$ is also $O(\delta m_l)$ and hence this term is
$O(\delta m_l)^2$ and so is negligible. ($\delta m_l$ the `distance' from
the flavour symmetric line is given in eq.~(\ref{deltaml_def}).) 
For the matrix element expansions in $\delta m_l$ see, for example, 
\cite{Bickerton:2019nyz,Bickerton:2021yzn}.

We first wish to compare our results with other derivations using 
the standard approach, e.g.\ \cite{Gockeler:2003ay}, by computing 
$3$-point correlation functions. Briefly, for completeness, 
defining an (unpolarised) $3$-point correlation function
\begin{eqnarray}
   C_{NV_4\Sigma}(t, \tau; \vec{q}, \vec{0})
     = \mbox{tr}\Gamma^{\rm unpol} 
          \langle \hat{\tilde{N}}(t;\vec{q}) \hat{V}_4(\tau)
                           \hat{\bar{\Sigma}}(0,\vec{0}) \rangle \,,
\label{3pt_corr_fun}
\end{eqnarray}
analogously to the $2$-point correlation function of eq.~(\ref{baryon_2pt})
and eq.~(\ref{C_unpol}) and applying the same techniques as described earlier
and results from section~\ref{bilin_gen_res} we look for a plateau in the ratio 
$R(t, \tau; \vec{0}, \vec{p})$ defined as
\begin{eqnarray}
   R(t, \tau; \vec{q}, \vec{0})
     &=& {C_{NV_4\Sigma}(t, \tau; \vec{q}, \vec{0} )
          \over C_{\Sigma\Sigma}(t; \vec{0})} 
       \sqrt{ { C_{\Sigma\Sigma}(\tau; \vec{0})C_{\Sigma\Sigma}(t; \vec{0})
               C_{NN}(t-\tau; \vec{q}) \over
               C_{NN}(\tau; \vec{q}) C_{NN}(t; \vec{q}) 
               C_{\Sigma\Sigma}(t-\tau; \vec{0})} }
                                                       \nonumber    \\
      &=& {1 \over \sqrt{ 2E_N(\vec{q}) 2M_\Sigma }}
                 \langle N(\vec{q})|\bar{u}\gamma_4 s
                                         |\Sigma(\vec{0})\rangle_{\rm rel} \,.
\label{3-pt_ratio}
\end{eqnarray}
A similar result holds for $C_{\Sigma V^\dagger_4 N}(t, \tau; \vec{0}, \vec{q})$
by swapping $\Sigma \leftrightarrow N$ and considering the inverse process.
At $\vec{q} = \vec{0}$, the `double ratio' method,
e.g.\ \cite{Flynn:2007ess} is employed
\begin{eqnarray}
   R(t, \tau; \vec{0}, \vec{0})
      &=& \sqrt{ { C_{NV_4\Sigma}(t,\tau;\vec{0},\vec{0}) 
                 C_{\Sigma V_4^\dagger N}(t,\tau;\vec{0},\vec{0}) \over
                 C_{\Sigma\Sigma}(t;\vec{0})C_{NN}(t;\vec{0}) } }
                                                       \nonumber    \\
      &=& {1 \over \sqrt{ 2M_N 2M_\Sigma }}
                 \langle N(\vec{q})|\bar{u}\gamma_4 s
                                         |\Sigma(\vec{0})\rangle_{\rm rel} \,.
\label{double_ratio}
\end{eqnarray}   
For this case this gives reduced error bars and a more prominent plateau.

\begin{figure}[!h]
   \begin{center}
     \includegraphics[width=15.00cm]
                      {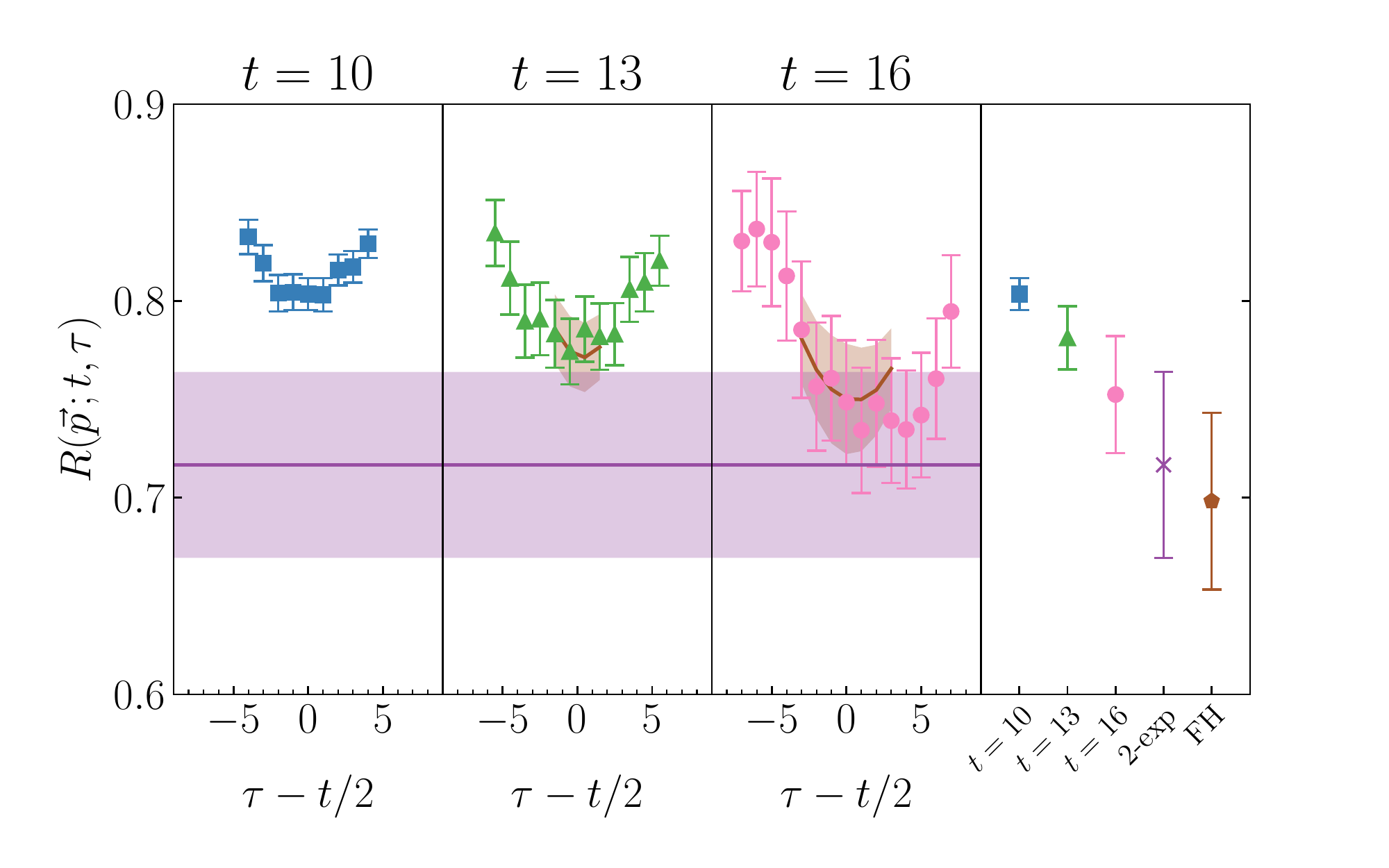}
   \end{center}
\caption{Comparing the three-point correlation function results to the 
         Feynman-Hellmann results. Left to right the first three plots
         for the three-point function ratios for sink--source separation 
         $t = 10$, $13$ and $16$ (filled squares, diamonds and crosses
         respectively). Global fits including a single additional
         excited state are also shown. The horizontal band 
         shows the global fit value for the matrix element.
         The fourth RH plot shows these results for the three $t$ values
         together with their extrapolated value (cross). For comparison we
         also show the closest Feynman-Hellmann result, filled upper triangle
         for run $\#5$.}
\label{correlation_fun_comparison}
\end{figure}

In Fig.~\ref{correlation_fun_comparison} various comparison ratios are shown
for $Q^2 \sim 0.27\,\mbox{GeV}^2$ for the $3$-point correlation function
approach, using eq.~(\ref{3-pt_ratio}). The individual results for a 
given $t$ (i.e.\ difference between baryon source and sink times) 
have smaller error bars, but due to excited states in the $3$-point 
correlation functions we have to perform measurements for various $t$ 
and extrapolate. An excited state can be accounted for by expanding 
the $2$- and $3$-point correlation functions to include contributions
from such an excited state and globally fitting
for various $t$ values, here $t = 10$, $13$ and $16$, simultaneously
(following for example \cite{Dragos:2016rtx}). The masses (including the
excited state masses) have been previously determined from $2$-point 
correlation function. This gives the various fits in 
Fig.~\ref{correlation_fun_comparison}. The constant in the fit then
gives the relevant matrix element as in eq.~(\ref{3-pt_ratio}).
Again all calculations are performed on the same set of gauge 
configurations with $500$ configurations used for each source-sink separation.

The Feynman-Hellmann approach has larger error bars, but as a $2$-point
correlation function measurement we largely avoid this extrapolation.
A comparison with the result of run $\#5$ ($Q^2 \sim 0.29\,\mbox{GeV}^2$) 
is also given in the figure both for the various $t_{\rm sep}$ and the 
extrapolated value. The results are compatible for the different approaches.
 
\begin{figure}[!htbp]
   \begin{center}
     \includegraphics[width=10.00cm]{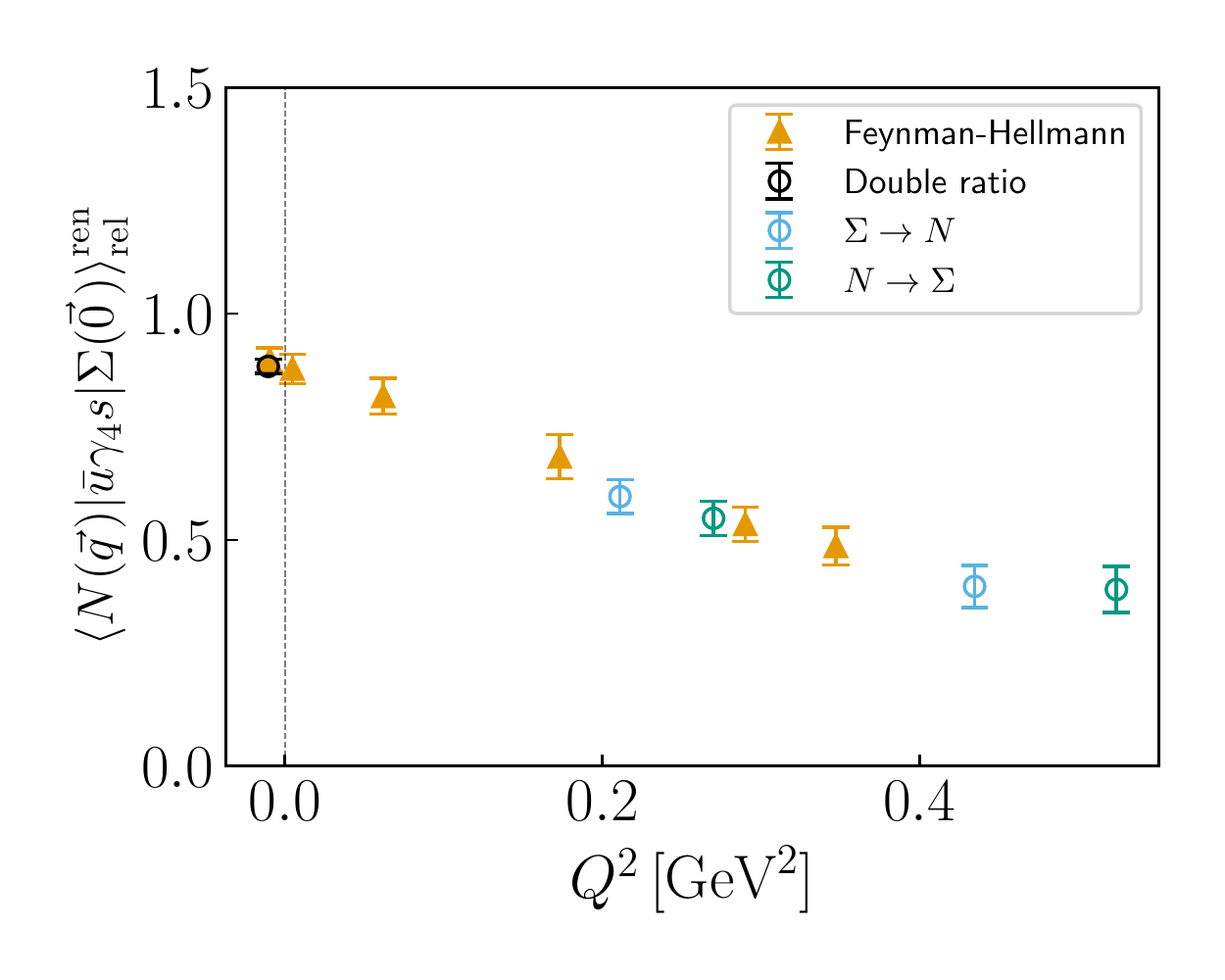}
   \end{center}
\caption{Results for 
         $\langle N(\vec{q})|\bar{u}\gamma_4 s
                              |\Sigma(\vec{0})\rangle_{\rm rel}^{\rm ren}$
         versus $Q^2$. Runs $\#1$ -- $\#6$ are given as (filled)
         triangles. We also make a comparison for this result with 
         results using standard approaches to the computation. 
         The (open) circles are results obtained from the $3$-point 
         correlation function, eq.~(\protect\ref{3-pt_ratio})
         or the double ratio, eq.~(\protect\ref{double_ratio}).}
\label{comparison}
\end{figure}

The results are given in Table~\ref{matrix_element_results}
\begin{table}[!htb]
   \begin{center}
   \begin{tabular}{r|rc}
      run \# & $Q^2$ [GeV$^2$]
             & $\langle N(\vec{q})|\bar{u}\gamma_4 s
                          |\Sigma(\vec{0})\rangle_{\rm rel}^{\rm ren}$  \\
    \hline
    1 & -0.0095 & 0.897(27) \\
    2 & 0.0048  & 0.878(32) \\
    3 & 0.062   & 0.817(40) \\
    4 & 0.17    & 0.684(49) \\
    5 & 0.29    & 0.535(38) \\
    6 & 0.35    & 0.486(42) \\
    \hline
    a & -0.01   & 0.883(16) \\
    b & 0.21    & 0.596(37) \\
    c & 0.27    & 0.548(38) \\
    d & 0.43    & 0.397(47) \\
    e & 0.52    & 0.390(51) \\
   \end{tabular}
   \end{center}
   \caption{The renormalised matrix element,
            $\langle N(\vec{q})|\bar{u}\gamma_4 s
                          |\Sigma(\vec{0})\rangle_{\rm rel}^{\rm ren}$
            against $Q^2$ in $\mbox{GeV}^2$ for the six runs. We also give
            five additional $Q^2$ results: runs $\#a$ -- $\#e$ using the
            methods described in eqs.~(\protect\ref{3pt_corr_fun}) --
            (\protect\ref{double_ratio}).}
\label{matrix_element_results}
\end{table}
and in Fig.~\ref{comparison} we plot
$\langle N(\vec{q})|\bar{u}\gamma_4 s|\Sigma(\vec{0})\rangle_{\rm rel}^{\rm ren}$
for runs $\#1$ -- $\#6$ against $Q^2$. 
There is good overall agreement between the two methods and in
particular confirm the values obtained from the approach using the 
Feynman--Hellmann theorem.

These results show that the Feynman-Hellmann theorem can be used for the 
calculation of transition form factors using two-point functions. 
This opens the way for more extensive calculations which can make use of 
the many tools and techniques available for controlling the contamination 
due to excited states in two-point functions.


\section{Conclusions}
\label{conclusions}


In this article we have extended the use of the Feynman--Hellmann
theorem in calculating (nucleon) matrix elements with momentum transfer
to situations where the relevant energy levels are not degenerate, 
but nearly degenerate or quasi-denerate as sketched in 
Fig.~\ref{sketch_energy_levels}. While for flavour-diagonal matrix elements 
this increases the scope of application of the Feynman--Hellmann theorem, 
as the associated energy levels now do not have to be exactly degenerate,
it now additionally allows for the computation of transition matrix 
elements. These latter matrix elements are naturally derived using
quasi-degenerate energy states.

In section~\ref{dyson_series}, using the Dyson expansion in the
Hamiltonian formalism, where the relevant operator is regarded as a 
perturbation in an expansion parameter $\lambda$, see eq.~(\ref{pert_ham}),
we gave a derivation of the basic result leading to
eq.~(\ref{C_fo}). In section~\ref{quasi_degeb_N_energy_ela} several
examples are discussed, first for flavour diagonal matrix elements
and then followed by flavour off-diagonal or transition matrix elements.

An example of the $\Sigma \to N$ decay (i.e.\ $s \to u$ transition)
for the vector current is considered. This necessitates the generalisation
of the action to include flavour non-diagonal terms. To minimise
numerical cost this is expanded to a sufficiently high order in $\lambda$
and then the two-point correlation function is reconstructed allowing
a range of $\lambda$ to be considered. Numerical results are then
discussed in section~\ref{lattice}. Avoided level crossing is demonstrated
for the quasi-degenerate enery states. A comparison is made with
results for the matrix element using the conventional $3$-point 
correlation function approach. Although in this article we only
consider the vector current transition matrix element, in the
Appendices, for completeness, we give the results for all possible Dirac
structures of the currents.

The availability of a large range of source-sink separations and the 
fact that there is only one exponentially decreasing set of excited 
states to deal with reduces the complexity of this task compared 
to the equivalent three-point function calculations. 
Additionally, the expansion of the matrix inversion in terms 
of the perturbation parameter used in this method presents a convenient way
to extend the applicable range of this method. Further calculations will be
required to determine whether it remains viable when approaching the 
physical quark masses when there are larger mass splittings.


\section*{Acknowledgements}


The numerical configuration generation (using the BQCD lattice QCD 
program \cite{Haar:2017ubh}) and data analysis (using the CHROMA software 
library \cite{Edwards:2004sx}) was carried out on the
DiRAC Blue Gene Q and Extreme Scaling Service
(Edinburgh Centre for Parallel Computing (EPCC), Edinburgh, UK), 
the Data Intensive Service 
(Cambridge Service for Data-Driven Discovery, CSD3, Cambridge, UK),
the Gauss Centre for Supercomputing (GCS) supercomputers JUQUEEN and JUWELS 
(John von Neumann Institute for Computing, NIC, Jülich, Germany) 
and resources provided by the North-German Supercomputer Alliance
(HLRN), the National Computer Infrastructure 
(NCI National Facility in Canberra, Australia supported by the
Australian Commonwealth Government) 
and the Phoenix HPC service (University of Adelaide). 
K.U.C. is supported by the ARC Grant No. DP220103098. 
R.H. is supported in part by the STFC Grant No. ST/P000630/1. 
P.E.L.R. is supported in part by the STFC Grant No. ST/G00062X/1.
G.S. is supported by DFG Grant No. SCHI 179/8-1.
R.D.Y. and J.M.Z. are supported by the ARC Grants
No. DP190100298 and No. DP220103098. For the purpose
of open access, the authors have applied a Creative
Commons Attribution (CC BY) licence to any author
accepted manuscript version arising from this submission.


\appendix

\section*{Appendix}


\section{Euclidean matrix elements}
\label{euclid_FF}


We take the Minkowski generalised currents to be given by
\begin{eqnarray}
   J^F(x) = (\bar{q}F\gamma q)(x)
      \equiv \sum_{f_1,f_2=1}^3 F_{f_1,f_2}(\bar{q}_{f_1}\gamma q_{f_2})(x) \,,
\label{JF}
\end{eqnarray}
where $q$ is a flavour vector, $q = (u,d,s)^T$, $F$ is a flavour matrix
and with $\gamma = \gamma_\mu$, $\gamma_\mu\gamma_5$, $I$, $i\gamma_5$, 
$\sigma_{\mu\nu} = i/2[\gamma_\mu,\gamma_\nu]$ for the vector, axial, scalar, 
pseudoscalar and tensor currents. (This ensures that $J^{F\,\dagger} = J^{F^T}$.)
The further generalisation to operators including covariant derivatives 
is straightforward.
We then take the Minkowski form factors as given in \cite{Bickerton:2019nyz}.

The Euclidean matrix elements are given by%
\footnote{We use in this Appendix the relativistic normalisation, see 
eqs.~(\ref{normalisation}) and (\ref{foot_norm}).}
\begin{eqnarray}
   \langle B^\prime(\vec{p}^\prime,\sigma^\prime)| J^F(\vec{q}) 
                                             | B(\vec{p},\sigma) \rangle
      = \bar{u}_{B^\prime}(\vec{p}^\prime,\sigma^\prime){\cal J}^F(q) 
           u_B(\vec{p},\sigma) \,,
\label{me_ubar_gam_u}
\end{eqnarray}
where the generalised currents $J^F(\vec{q})\,$%
\footnote{Again we are simplifying the notation, from eq.~(\ref{op_def})
we have $\hat{O}(\vec{x}) \to J^F(\vec{x})$ and
$\hat{\tilde{O}}(\vec{q}) \to J^F(\vec{q})$.}
also have the same flavour structure as defined by eq.~(\ref{JF}) 
but now using the conventions in \cite{Best:1997qp} with the Euclidean 
gamma matrices $\gamma = \gamma_\mu$, $i\gamma_\mu\gamma_5$, $I$, 
$\gamma_5 =\gamma_1\gamma_2\gamma_3\gamma_4$, 
$\sigma_{\mu\nu} = i/2[\gamma_\mu,\gamma_\nu]$ for vector $V_\mu$, axial $A_\mu$,
scalar, $S$, pseudoscalar, $P$, and tensor, $T_{\mu\nu}$, respectively. 
The Euclidean gamma matrices, $\gamma_\mu$, are now all Hermitian, 
$\gamma_\mu^\dagger = \gamma_\mu$. The relation between the Euclidean 
formulation of the action and Hamiltonian
(and hence also for matrix elements) is discussed in 
\cite{Creutz:1976ch,montvay+munster}.
Suppressing the flavour matrix, the ${\cal J}^F(q)$ are given by
\begin{eqnarray}
   {\cal V}_\mu(q) 
     &=& \gamma_\mu f_1^{\ind{BB^\prime}}(Q^2) 
            + \sigma_{\mu\nu}q_v {f_2^{\ind{BB^\prime}}(Q^2) \over M_{B^\prime}+M_B}
            - i q_\mu  {f_3^{\ind{BB^\prime}}(Q^2) \over M_{B^\prime}+M_B} \,,
                                                            \nonumber  \\
   {\cal A}_\mu(q) 
     &=& i\gamma_\mu\gamma_5 g_1^{\ind{BB^\prime}}(Q^2) 
             + i\sigma_{\mu\nu}\gamma_5q_v {g_2^{\ind{BB^\prime}}(Q^2)
                                        \over M_{B^\prime}+M_B}
             + q_\mu\gamma_5  {g_3^{\ind{BB^\prime}}(Q^2) \over M_{B^\prime}+M_B} \,,
                                                            \nonumber  \\
   {\cal S}(q) &=& g^{\ind{BB^\prime}}_S(Q^2) \,,
                                                            \nonumber  \\
   {\cal P}(q) &=& g^{\ind{BB^\prime}}_P(Q^2) \,,
                                                            \nonumber  \\
   {\cal T}_{\mu\nu}(q)
     &=& \sigma_{\mu\nu} h^{\ind{BB^\prime}}_1(Q^2)
                  + (q_\mu \gamma_\nu - q_\nu \gamma_\mu)
                         {h^{\ind{BB^\prime}}_2(Q^2)\over M_{B^\prime}+M_B}
                                                            \nonumber  \\
     & &         - (\sigma_{\mu\lambda}q_\nu - \sigma_{\nu\lambda}q_\mu)q_\lambda
                     {h^{\ind{BB^\prime}}_3(Q^2)\over (M_{B^\prime}+M_B)^2}
                 + 2\epsilon_{\mu\nu\rho\sigma}q_\rho\gamma_\sigma\gamma_5
                   {h^{\ind{BB^\prime}}_4(Q^2)\over M_{B^\prime}+M_B} \,.
\label{lorentz_decomp}
\end{eqnarray}
As we are using the conventions of \cite{Bickerton:2019nyz} then we have
taken in these expressions
\begin{eqnarray}
   q = p^\prime - p = (i(E_{B^\prime}-E_B), \vec{p}^\prime - \vec{p}) \,.
\end{eqnarray}
This is the convention for scattering processes, rather than the natural 
convention for decay processes where the opposite holds. However for 
consistency we remain with the above.

We have also manipulated the tensor results from the expressions given in
\cite{Bickerton:2019nyz}. For completeness, we briefly describe this here.
First to give them in a form as a function of $q$, only we use a 
Gordan identity which in Euclidean form is
\begin{eqnarray}
   p_{\mu} \pm p^{\prime}_{\mu}
      = \gamma_{\mu}\slashed{p}
          \pm \slashed{p}^{\prime} \gamma_{\mu}
          + i\sigma_{\mu\nu}(p_\nu \mp p_\nu^\prime)  \,,
\end{eqnarray}
together with $\slashed{p}u=iMu$ and $\bar{u}\slashed{p}=iM\bar{u}$.
This means that $h_2$ is replaced by $h_2 + h_3$ and there is now a
new structure $(\sigma_{\mu\lambda}q_\nu - \sigma_{\nu\lambda}q_\mu)q_\lambda$.
Secondly we use the antisymmetric piece of the identity
\begin{eqnarray}
   \gamma_\mu \gamma_\rho \gamma_\nu
      = \gamma_\mu\delta_{\rho\nu} - \gamma_\rho\delta_{\mu\nu}
                                + \gamma_\nu\delta_{\mu\rho}
        - \epsilon_{\mu\rho\nu\sigma} \gamma_\sigma\gamma_5 \,,
\end{eqnarray}
with $\epsilon_{1234} = +1$ to modify the $h_4$ structure. Finally for
the axial current for the $g_2$ term we can use
\begin{eqnarray}
   \sigma_{\mu\nu}\gamma_5 
      = - {1 \over 2}\epsilon_{\mu\nu\rho\lambda}\sigma_{\rho\lambda} \,.
\label{sigma_gamma5}
\end{eqnarray}
With these additional manipulations all the terms in the matrix
element decomposition are functions of $q$ and also all the Dirac
structure is in the standard gamma-matrix basis.

From the direct computation of the spinor bilinears, as detailed
in Appendix~\ref{spinor_results} and then using eq.~(\ref{me_ubar_gam_u})
together with eq.~(\ref{lorentz_decomp}) we find
\begin{eqnarray}
   \langle B^\prime(\vec{p}^\prime, -)| J^F | B(\vec{p}, -) \rangle
      &=& \eta_\gamma \langle B^\prime(\vec{p}^\prime, +)| J^F 
                                                | B(\vec{p}, +) \rangle^* \,,
                                                             \nonumber  \\
   \langle B^\prime(\vec{p}^\prime, -)| J^F | B(\vec{p}, +) \rangle
      &=& - \eta_\gamma \langle B^\prime(\vec{p}^\prime, +)| J^F 
                                                | B(\vec{p}, -) \rangle^* \,,
\label{me_reality}
\end{eqnarray}
where $\eta_\gamma = \pm$. Explicitly we have the results as given in
Table~\ref{table:etag}.
\begin{table}[htb]
   \begin{center}
   \begin{tabular}{c|cccccccccc}
      $\gamma$ & $\gamma_4$ & $\gamma_i$ &  $i\gamma_4\gamma_5$ 
        & $i\gamma_i\gamma_5$ & $I$ & $\gamma_5$ & $\sigma_{i4}$ & $\sigma_{ij}$ 
        & $\sigma_{i4}\gamma_5$ & $\sigma_{ij}\gamma_5$ \\
   $\eta_\gamma$ & $+$ & $-$ & $+$ & $-$ & + & $-$ & $+$ & $-$ & $-$ & $+$ \\
   \end{tabular}
   \end{center}
   \caption{The $\eta_\gamma$ factors.}
\label{table:etag}
\end{table}
These can be helpful in determining whether the computed matrix element is
real or imaginary.


\section{Spinor bilinear results}
\label{spinor_results}


The spinor bilinear forms are the most general possible, so to deal with
this we shall consider a specific representation -- the Dirac representation.
Some more general results are given for example in 
\cite{Wightman:2001zj,Pal:2007dc,Lorce:2017isp,Olpak:2019cth}.
Again we shall be in Euclidean space.


\subsection{General}


Sigma matrices
\begin{eqnarray}
   \sigma_1 = \left( \begin{array}{rr}
                        0 & 1 \\
                        1 & 0 
                     \end{array}
              \right) \,, \qquad
   \sigma_2 = \left( \begin{array}{rr}
                        0 & -i \\
                        i & 0 
                     \end{array}
              \right) \,, \qquad
   \sigma_3 = \left( \begin{array}{rr}
                        1 & 0 \\
                        0 & -1
                     \end{array}
              \right) \,,
\end{eqnarray}
where
\begin{eqnarray}
   \sigma_i\sigma_j = \delta_{ij} + i \epsilon_{ijk}\sigma_k \,, \qquad
   \sigma_i^\dagger = \sigma_i \,.
\end{eqnarray}
Gamma matrices
\begin{eqnarray}
   \gamma_i = \left( \begin{array}{rr}
                        0 & -i\sigma_i \\
                        i\sigma_i & 0 
                     \end{array}
              \right) \,, \qquad
   \gamma_4 = \left( \begin{array}{rr}
                        1 & 0 \\
                        0 & -1
                     \end{array}
              \right) \,, \qquad
   \gamma_5 = \gamma_1\gamma_2\gamma_3\gamma_4
            = \left( \begin{array}{rr}
                       0  & -1 \\
                       -1 & 0 
                     \end{array}
              \right) \,.
\end{eqnarray}


\subsection{$u$-spinors}


Solving the (free) Dirac equation gives for the $+ve$ energy spinors
\begin{eqnarray}
   u(\vec{p},\sigma) 
      = s \left( \begin{array}{c}
                             \chi^{(\sigma)} \\[0.5em]
                             {\vec{\sigma}\cdot\vec{p} \over s^2}\, 
                                                         \chi^{(\sigma)}
                          \end{array}
                   \right) \,.
\label{u_dirac}
\end{eqnarray}
where it is convenient to define in the following
\begin{eqnarray}
   s(\vec{p}) = \sqrt{E(\vec{p})+M} \,.
\end{eqnarray}
The spin is quantised along the $3^{\rm rd}$ direction (due to the
nature of the $\sigma_3$ matrix in particular) so
$\sigma = \pm$ and the $2$ component spinors at rest are given by
\begin{eqnarray}
   \chi^{(+)} = \left( \begin{array}{r}
                         1  \\
                         0 
                      \end{array} 
               \right) \,, \qquad
   \chi^{(-)} = \left( \begin{array}{r}
                         0  \\
                         1 
                      \end{array} 
               \right) \,,
\end{eqnarray}
or in components
\begin{eqnarray}
   \chi^{(\sigma)}_{\sigma_r} = \delta_{\sigma\sigma_r} \,, \quad
                                      \sigma = \pm\,,\,\, \sigma_r = \pm \,.
\end{eqnarray}
We also have
\begin{eqnarray}
   \bar{u}(\vec{p},\sigma) 
      = s \left( \chi^{(\sigma)T}\,,\, -\chi^{(\sigma)T}\,
                             {\vec{\sigma}\cdot\vec{p} \over s^2}
          \right) \,.
\label{ubar_dirac}
\end{eqnarray}
As a check we have $\slashed{p}\,u = iMu$, $\bar{u}\,\slashed{p} = iM\bar{u}$
as expected, as the Minkowski free Dirac equation $(\slashed{p}-m)u=0$
and upon Euclideanisation $\slashed{p} \to -i\slashed{p}$ where $p_4$
is imaginary.

$\chi^{(\sigma)}$ has the (obvious) property
$\chi^{(\sigma^\prime) T}\chi^{(\sigma)} = \delta_{\sigma^\prime \sigma}$
which from eqs.~(\ref{u_dirac}), (\ref{ubar_dirac})
gives the standard normalisation of
\begin{eqnarray}
   \bar{u}(\vec{p}, \sigma^\prime)\, u(\vec{p}, \sigma)
      = 2M \, \chi^{(\sigma^\prime) T}\chi^{(\sigma)}
      = 2M \, \delta_{\sigma^\prime \sigma} \,.
\end{eqnarray}
As $\chi^{(\sigma^\prime)}$, $\chi^{(\sigma)}$ just pick out the components of 
$\sigma_k$ in $\chi^{(\sigma^\prime)\,T}\sigma_k\chi^{(\sigma)}$ then we have
\begin{eqnarray}
   \chi^{(\sigma^\prime)\,T}\sigma_k\chi^{(\sigma)}
      = (\sigma_k)_{\sigma^\prime\sigma}
      = \sigma \delta_{k3}\delta_{\sigma^\prime \sigma}
       + (\delta_{k1} + i\sigma\delta_{k2})\delta_{\sigma^\prime, -\sigma} \,,
\label{sigma_explicit}
\end{eqnarray}
or in vector form
\begin{eqnarray}
    (\vec{\sigma})_{\sigma^\prime,\sigma}
      = \sigma \vec{e}_3\,\delta_{\sigma^\prime \sigma}
         + (\vec{e}_1 + i\sigma\vec{e}_2)\,\delta_{\sigma^\prime, -\sigma} \,.
\label{sigma_explicit_index}
\end{eqnarray}
In the following we will find
\begin{eqnarray}
   \bar{u}^\prime(\vec{p}^\prime, -) \gamma u(\vec{p}, -)
      &=& \eta_\gamma [\bar{u}^\prime(\vec{p}^\prime, +) \gamma u(\vec{p}, +)]^* \,,
                                                           \nonumber   \\
   \bar{u}^\prime(\vec{p}^\prime, -) \gamma u(\vec{p}, +)
      &=& -\eta_\gamma[\bar{u}^\prime(\vec{p}^\prime, +) \gamma u(\vec{p}, -)]^* \,,
\label{ubar_gam_u_spin_rel}
\end{eqnarray}
where $\eta_\gamma = \pm$. These will be the same factors as given in 
Table~\ref{table:etag}.


\subsubsection{Bilinears -- general case}
\label{bilin_gen_res}


\subsubsection*{Vector: $\gamma_4$, $\gamma_i$}

\vspace*{-0.15in}


\begin{eqnarray}
   \bar{u}^\prime(\vec{p}^\prime, \cdot) \gamma_4 u(\vec{p}, \cdot)
      &=& \left(s^\prime s + {\vec{p}^\prime\cdot\vec{p} \over s^\prime s} \right)I
         + {i \over s^\prime s}(\vec{p}^\prime\times\vec{p})\cdot\vec{\sigma} \,,
                                                            \nonumber   \\
   \bar{u}^\prime(\vec{p}^\prime, \cdot) \gamma_i u(\vec{p}, \cdot)
      &=& -i\left( {s^\prime \over s}\vec{p} +  {s \over s^\prime}\vec{p}^\prime
            \right)_i I
          + \left(\left( {s^\prime \over s}\vec{p}
                   - {s \over s^\prime}\vec{p}^\prime
            \right) \times\vec{\sigma}\right)_i \,,
\label{ubar_gam_mu_u}
\end{eqnarray}
where we have suppressed the spin $\sigma$ index, to have a matrix equation
in spin space (e.g.\ the $\sigma^\prime,\sigma$ components of
$\bar{u}^\prime(\vec{p}^\prime, \cdot) \gamma_4 u(\vec{p}, \cdot)$ are
$\bar{u}^\prime(\vec{p}^\prime, \sigma^\prime) \gamma_4 u(\vec{p}, \sigma)$).
The above equation (eq.~(\ref{ubar_gam_mu_u})) is written in a compact form.
This can be `undone' by using eq.~(\ref{sigma_explicit}) (or 
eq.~(\ref{sigma_explicit_index})) for $\vec{\sigma}_{\sigma^\prime\sigma}$.
This then allows the spin relation in eq.~(\ref{ubar_gam_u_spin_rel})
to be shown with, for this case, $\eta_{\gamma_4} = +$ and $\eta_{\gamma_i} = -$.
As expected this result is independent of the kinematic factors $\vec{p}^\prime$
and $\vec{p}$ and just depends on the factors: $i$, $\sigma$ and 
the combination $i\sigma$. The other cases follow a similar pattern.


\subsubsection*{Axial: $i\gamma_4\gamma_5$, $i\gamma_i\gamma_5$}

\vspace*{-0.15in}


\begin{eqnarray}
   \bar{u}^\prime(\vec{p}^\prime, \cdot) i\gamma_4\gamma_5 u(\vec{p}, \cdot)
      &=& - i\left( {s^\prime \over s}\vec{p} + {s \over s^\prime}\vec{p}^\prime
            \right)\cdot\vec{\sigma} \,,
                                                                        \\
   \bar{u}^\prime(\vec{p}^\prime, \cdot) i\gamma_i\gamma_5 u(\vec{p}, \cdot)
      &=& {i \over s^\prime s}(\vec{p}^\prime\times\vec{p})_i I
        - \left(s^\prime s \delta_{ij}
          + {1 \over s^\prime s}(p_i^\prime p_j + p_j^\prime p_i
                                 - \vec{p}^\prime\cdot\vec{p}\,\delta_{ij})
         \right)\sigma_j \,,
                                                            \nonumber
\end{eqnarray}
with $\eta_{i\gamma_4\gamma_5} = +$ and $\eta_{i\gamma_i\gamma_5} = -$.


\subsubsection*{Scalar: $I$}

\vspace*{-0.15in}


\begin{eqnarray}
   \bar{u}^\prime(\vec{p}^\prime, \cdot) I u(\vec{p}, \cdot)
      = \left(s^\prime s - {\vec{p}^\prime\cdot\vec{p} \over s^\prime s} \right) I
         - {i \over s^\prime s}(\vec{p}^\prime\times\vec{p})\cdot\vec{\sigma} \,,
\end{eqnarray}
with $\eta_I = +$.


\subsubsection*{Pseudoscalar: $\gamma_5$}

\vspace*{-0.15in}


\begin{eqnarray}
   \bar{u}^\prime(\vec{p}^\prime, \cdot) \gamma_5 u(\vec{p}, \cdot)
      = - \left( {s^\prime \over s}\vec{p} - {s \over s^\prime}\vec{p}^\prime
          \right)\cdot\vec{\sigma} \,,
\end{eqnarray}
with $\eta_{\gamma_5} = -$.


\subsubsection*{Tensor:}


\begin{itemize}

\item $\sigma_{i4} = i\gamma_i\gamma_4$, 
                        $\sigma_{ij} = i\gamma_i\gamma_j\,(i \neq j)$

\vspace*{-0.15in}

\begin{eqnarray}
   \bar{u}^\prime(\vec{p}^\prime, \cdot) \sigma_{i4} u(\vec{p}, \cdot)
      &=& -\left({s^\prime \over s}\vec{p} - {s \over s^\prime}\vec{p}^\prime
          \right)_iI
          - i\left(
             \left({s^\prime \over s}\vec{p} + {s \over s^\prime}\vec{p}^\prime
             \right)\times \vec{\sigma} \right)_i \,,
                                                           \nonumber   \\
   \bar{u}^\prime(\vec{p}^\prime, \cdot) \sigma_{ij} u(\vec{p}, \cdot)
     &=& -{i \over s^\prime s}(p_i^\prime p_j - p^\prime_j p_i)I
                                                    \label{sigma_ij}   \\
     & & \hspace*{0.250in}
         + \epsilon_{ijk} \left( -s^\prime s \,\delta_{kl}
                                +{1 \over s^\prime s} \left(
                                   p_k p_l^\prime + p_k^\prime p_l
                                   - \vec{p}^\prime\cdot\vec{p}\,\delta_{kl}
                                                   \right)
                          \right)\sigma_l \,,
                                                           \nonumber
\end{eqnarray}
with $\eta_{\sigma_{i4}} = +$ and $\eta_{\sigma_{ij}} = -$.

\item An alternative tensor form for $\sigma_{\mu\nu}\gamma_5$ and using the 
identity of eq.~(\ref{sigma_gamma5}) is
\begin{eqnarray}
   \bar{u}^\prime(\vec{p}^\prime, \sigma^\prime) 
               \sigma_{i4}\gamma_5 u(\vec{p}, \sigma)
      = -{1 \over 2}\epsilon_{ikl} \,
           \bar{u}^\prime(\vec{p}^\prime, \sigma^\prime) 
               \sigma_{kl} u(\vec{p}, \sigma) \,,
\end{eqnarray}
with $\eta_{\sigma_{i4}\gamma_5} = -$. A more explicit expression can then be 
given using eq.~(\ref{sigma_ij}). Similarly
\begin{eqnarray}
   \bar{u}^\prime(\vec{p}^\prime, \sigma^\prime) 
               \sigma_{ij}\gamma_5 u(\vec{p}, \sigma)
      = -\epsilon_{ijk} \,
           \bar{u}^\prime(\vec{p}^\prime, \sigma^\prime) 
               \sigma_{k4} u(\vec{p}, \sigma) \,,
\end{eqnarray}
with $\eta_{\sigma_{ij}\gamma_5} = +$.

\end{itemize}


\subsubsection{Bilinears -- unpolarised/polarised cases}
\label{pol_unpol_gamma}


Useful combinations discussed here are
\begin{itemize}

   \item $\Gamma^{\rm unpol} = (1+\gamma_4)/2$ giving
         \begin{eqnarray}
            \bar{u}^\prime(\vec{p}^\prime, \sigma^\prime)
                           \Gamma^{\rm unpol} u(\vec{p}, \sigma)
            = s^\prime s \, \delta_{\sigma^\prime\sigma} \,,
         \label{ubar_u_unpol}
         \end{eqnarray}

   \item $\Gamma^{\rm pol}_{\pm 3} 
            = (1+\gamma_4)/2 \times (1 \pm i\gamma_5\gamma_3)$ giving
         \begin{eqnarray}
            \bar{u}^\prime(\vec{p}^\prime, \sigma^\prime)
                           \Gamma^{\rm pol}_{\pm 3} u(\vec{p}, \sigma)
            = s^\prime s \, (1 \pm \sigma) \, \delta_{\sigma^\prime\sigma} \,,
         \label{ubar_u_pol3}
         \end{eqnarray}

   \item $\Gamma^{\rm pol}_{\pm} 
            = (1+\gamma_4)/2 \times i\gamma_5(\gamma_1 \pm i\gamma_2)$
         giving
         \begin{eqnarray}
            \bar{u}^\prime(\vec{p}^\prime, \sigma^\prime)
                           \Gamma^{\rm pol}_{\pm} u(\vec{p}, \sigma)
            = s^\prime s \, (1 \mp \sigma) \, \delta_{\sigma^\prime,-\sigma} \,.
         \label{ubar_u_polpm}
         \end{eqnarray}

\end{itemize}


\section{General derivation of energy states for the $d_S = 2$ case}
\label{alt_energy_states}


We give here an alternative derivation of the energy states in 
section~\ref{quasi_degeb_N_energy_ela} and in particular
eqs.~(\ref{E_pm}), (\ref{DeltaE}) which does not depend on the choice
of a particular $\Gamma$ matrix choice in section~\ref{spin_index}.
In eq.~(\ref{matrix_cr_rs_spin}) we now have a $(2\times 2)\times(2\times 2)$
(i.e.\ $d_S = 2$) matrix to diagonalise. 

Including the spin index we now have
\begin{eqnarray}
   \langle B_r(\vec{p}_r,\sigma_r) | \hat{\tilde{{\cal O}}}(\vec{q}) 
                           | B_s(\vec{p}_s,\sigma_s) \rangle
   = \left( \begin{array}{cc}
               0  & a^\dagger  \\
               a  & 0        \\
            \end{array}
     \right)_{\sigma_rr,\sigma_ss} \,,
\end{eqnarray}
where $a$ is replaced by a $2\times 2$ matrix (and the complex conjugate 
$a^*$ by the Hermitian conjugate $a^\dagger$). Thus, using 
eq.~(\ref{me_reality}) we replace
\begin{eqnarray}
   a \to \left( \begin{array}{rr}
                   a_{++}    & a_{+-}  \\
                   a_{+-}    & a_{--}
                \end{array}
         \right) \,,
\label{a_with_spin}
\end{eqnarray}
with $a_{-+} = -\eta a^*_{+-}$ and $a_{--} = \eta a_{++}^*$ where $\eta$ is
given in Table~\ref{table:etag}.
The eigenvalues of the resulting enlarged $D$ matrix in 
eq.~(\ref{matrix_cr_rs_spin}) are easily found, by first writing $D$
as a product of $2\times 2$ sub-matrices as in eq.~(\ref{M_splitting})
and then taking the determinant with the identification 
$A \to (\epsilon_1 - \mu)I$,
$B \to (\epsilon_2 - \mu)I$, $C \to a^\dagger$ and $D \to a$. 
We now have to solve the eigenvalue equation
\begin{eqnarray}
     \det\left((\epsilon_1-\mu)(\epsilon_2-\mu)I - \lambda^2 aa^\dagger\right) 
                  = 0 \,,
\end{eqnarray}
for $\mu$. Furthermore note that
\begin{eqnarray}
   aa^\dagger = (|a_{++}|^2+|a_{+-}|^2)I = |\det a|I \,,
\end{eqnarray}
which is diagonal. So this means that each eigenvalue is doubly 
degenerate as expected (the double energy degeneracy)  
and leads to the replacement in eqs.~(\ref{eigenvals}), (\ref{DeltaE}) of
\begin{eqnarray}
   |a| \to |\det a|^{1/2}\,,
\label{deta}
\end{eqnarray}
(or $|a| \to (|a_{++}|^2+|a_{+-}|^2)^{1/2}$), together with the appropriate
change in $\kappa_\pm$.

Matrix elements are either unpolarised or polarised (including spin flip)
and either real or imaginary. But one of these corresponds to a matrix 
element picked out by a $\Gamma$ matrix in section~\ref{spin_index}. 
For example for $\Gamma^{\rm unpol}$ from eq.~(\ref{matrix_cr_rs_unpol}) where
we have the replacement $a \to (a_{++}+a_{--})/2$ which also gives one 
matrix element for $|a|^2$ (i.e.\ $|a_{++}|^2$). Thus eq.~(\ref{deta}) 
may be considered the general result. Additionally the eigenvectors 
are found to be
\begin{eqnarray}
   e^{(\pm\sigma)}_{\sigma_r r} = {1 \over \sqrt{\Delta E_\lambda}}
       \left( \begin{array}{c}
                 \sqrt{\kappa_\pm} \chi^{(\sigma)}_{\sigma_r} \\[0.5em]
                  \pm\,\mbox{sgn}(\lambda)
                  {\sqrt{\kappa_\mp} \over \sqrt{|\det a|}} \, 
                    (a\chi^{(\sigma)})_{\sigma_r}
               \end{array}
        \right)_r \,.
\label{eigenv_dirac}
\end{eqnarray}
Parallel to eq.~(\ref{C_rs_nodirac}) we have
\begin{eqnarray}
   C_{\lambda\,r\alpha,s\beta}(t)
      = \sum_{i=\pm}\sum_{\sigma=\pm}
              w^{(i\sigma)}_{r\alpha}\bar{w}_{s\beta}^{(i\sigma)} e^{-E_\lambda^{(i)}t} \,,
\end{eqnarray}
with
\begin{eqnarray}
   w^{(i\sigma)}_{r\alpha} 
      &=& \sum_{\sigma_r} 
            {}_\lambda\langle 0|\hat{B}_{r\alpha}(\vec{0})|
                              B_r(\vec{p}_r,\sigma_r)\rangle_\lambda\,
                                     e^{(i\sigma)}_{\sigma_r r} \,,
                                                           \nonumber  \\
   \bar{w}^{(i\sigma)}_{s\beta} 
      &=& \sum_{\sigma_s} 
            {}_\lambda\langle B_s(\vec{p}_s,\sigma_s)|
                             \hat{\bar{B}}_{s\beta}(\vec{0}) |0\rangle_\lambda
                                     e^{(i\sigma)\,*}_{\sigma_s s} \,.
\end{eqnarray}
This is the general result. In the simplification of 
section~\ref{spin_index} using eq.~(\ref{wf_def}) gives
\begin{eqnarray}
   C^\Gamma_{\lambda\,rs}(t)
      = Z_r\bar{Z}_s \sum_{i=\pm}\sum_{\sigma=\pm} \sum_{\sigma_r\sigma_s}
               \bar{u}^{(s)}(\vec{p}_s,\sigma_s)\Gamma 
                              u^{(r)}(\vec{p}_r,\sigma_r)
                e^{(i\sigma)}_{\sigma_r r}e^{(i\sigma)\,*}_{\sigma_s s} 
                      e^{-E_\lambda^{(i)}t} \,.
\label{C_with_spin}
\end{eqnarray}
However using eq.~(\ref{eigenv_dirac}) we have
\begin{eqnarray}
   \sum_{\sigma=\pm} e^{(\pm\sigma)}_{\sigma_r r}e^{(\pm\sigma)\,*}_{\sigma_s s}
     = {1 \over \Delta E_\lambda}
       \left( \begin{array}{cc}
                 \kappa_\pm\delta_{\sigma_r\sigma_s} &
                       \pm\lambda a^\dagger_{\sigma_r\sigma_s} \\
                 \pm\lambda a_{\sigma_r\sigma_s}    &
                       \kappa_\mp \delta_{\sigma_r\sigma_s}
               \end{array}
        \right)_{rs} \,.
\label{eigenv_out_dirac}
\end{eqnarray}
With no spin index we use the result of eq.~(\ref{eigenvec}) to give
\begin{eqnarray}
   e^{(\pm)}_r\,e^{(\pm)\,*}_s 
      = {1 \over \Delta E_\lambda}
          \left( \begin{array}{cc}
                    \kappa_\pm       & \pm\lambda a_\pm^* \\
                    \pm\lambda a_\pm &  \kappa_\mp
                 \end{array}
           \right)_{rs} \,.
\label{eigenv_out}
\end{eqnarray}
which with the substitutions of eqs.~(\ref{a_with_spin}), (\ref{deta}) 
gives the spin case result in eq.~(\ref{eigenv_out_dirac}).

More concretely if we set $\Gamma = \Gamma^{\rm unpol}$ and use
eq.~(\ref{ubar_u_unpol_explicit}) and eq.~(\ref{eigenv_out_dirac})
in eq.~(\ref{C_with_spin}) and then re-write it using eq.~(\ref{eigenv_out})
this soon leads to eq.~(\ref{C_rs_nodirac}) with $a \to (a_{++}+a_{--})/2$ 
as found there.
A similar result holds for $\Gamma = \Gamma_3^{\rm pol}$. However
this equivalence between the results with spin and without spin is
because as mentioned previously both eqs.~(\ref{ubar_u_unpol}) and 
(\ref{ubar_u_pol3}) are diagonal in $\sigma_r$, $\sigma_s$. 
If we consider a case where this is not true, 
for example $\Gamma^{\rm pol}_\pm$, eq.~(\ref{ubar_u_polpm})
then we soon find that%
\footnote{Again we have redefined $Z_r$ and $\bar{Z}_s$ as in 
footnote~\ref{redef_Z}.}
\begin{eqnarray}
   C^{\Gamma^{\rm pol}_+}_{rs}(t)
      = Z_r \bar{Z}_s {\lambda \over \Delta E_\lambda}
           \left( \begin{array}{cc}
                     0     & a^*_{+-} \\
                     a_{+-} & 0     
                  \end{array}
           \right)_{rs} \left( e^{-E_\lambda^{(+)}t} - e^{-E_\lambda^{(-)}t} \right) \,,
\end{eqnarray}
(and similarly for $\Gamma^{\rm pol}_-$ with $a_{+-}$ replaced by $a_{-+}$).
The diagonal terms have now vanished, so it cannot be re-written as
for the spinless case. Not only that but we now have a difference of 
two exponentials (rather than a sum). Expanding gives
\begin{eqnarray}
   C^{\Gamma^{\rm pol}_+}_{rs}(t)
      = Z_r \bar{Z}_s \lambda
           \left( \begin{array}{cc}
                     0     & a^*_{+-} \\
                     a_{+-} & 0     
                  \end{array}
           \right)_{rs} t e^{-\bar{E}t} \,,
\end{eqnarray}
close to the form of the original Dyson expansion as discussed in
section~\ref{spin_index}.


\section{Correlation functions}
\label{corr_fun}


The correlation functions in eq.~(\ref{correl_fun_decay_mat}) are defined by
\begin{eqnarray}
   C_{\lambda\,\Sigma\Sigma}(t)
     &=&  {\rm tr}_D \Gamma
          \langle \tilde{B}_{\Sigma}(t;\vec{p}) 
                      \bar{B}_{\Sigma}(0,\vec{0}) \rangle_\lambda \,,
                                                        \nonumber \\
   C_{\lambda\,\Sigma N}(t)
     &=&  {\rm tr}_D \Gamma
          \langle \tilde{B}_{\Sigma}(t;\vec{p}) 
                      \bar{B}_{N}(0,\vec{0}) \rangle_\lambda \,,
                                                        \nonumber \\
   C_{\lambda\,N\Sigma}(t)
     &=&  {\rm tr}_D \Gamma
          \langle \tilde{B}_{N}(t;\vec{p}+\vec{q}) 
                          \bar{B}_{\Sigma}(0,\vec{0}) \rangle_\lambda \,,
                                                        \nonumber \\
   C_{\lambda\,NN}(t)
     &=&  {\rm tr}_D \Gamma
          \langle \tilde{B}_{N}(t;\vec{p}+\vec{q}) 
                          \bar{B}_{N}(0,\vec{0}) \rangle_\lambda \,,
\end{eqnarray}
with baryon wavefunctions given by
\begin{eqnarray}
   \tilde{B}_{\Sigma_\alpha}(t;\vec{p})
      &=& \int_{\vec{x}} e^{-i\vec{p}\cdot\vec{x}} B_{\Sigma_\alpha}(t,\vec{x})
       = \sum_{\vec{x}} e^{-i\vec{p}\cdot\vec{x}}
             \epsilon^{abc} d_\alpha^a(x)
                            \left[ d^b(x)^{T_D}C\gamma_5 s^c(x) \right] \,,
                                                             \nonumber \\
   \tilde{B}_{N_\alpha}(t;\vec{p})
      &=& \int_{\vec{x}} e^{-i\vec{p}\cdot\vec{x}} B_{N_\alpha}(t,\vec{x})
       = \sum_{\vec{x}} e^{-i\vec{p}\cdot\vec{x}}
             \epsilon^{abc} d_\alpha^a(x)
                            \left[ d^b(x)^{T_D}C\gamma_5 u^c(x) \right] ,
\end{eqnarray}
($\alpha$ is a Dirac index, $a$ is a colour index and $C = \gamma_4\gamma_2$).
As in eq.~(\ref{correl_fun_decay_mat}) we have taken a trace over the 
Dirac indices with $\Gamma = \Gamma^{\rm unpol}$.
For the diagonal correlation functions this gives
\begin{eqnarray}
   \lefteqn{C_{\lambda\,\Sigma\Sigma}(t) =
            \sum_{\vec{x}} e^{-i\vec{p}\cdot\vec{x}} \epsilon_{abc} 
                                      \epsilon_{a^\prime b^\prime c^\prime} }
         &                                    \nonumber \\
         &                \left\langle
    \mbox{tr}_D \left[ \Gamma G^{(dd)aa^{\prime}}(\vec{x},t;\vec{0},0) \right]
    \mbox{tr}_D \left[ \tilde{G}^{(ss)bb^{\prime}}(\vec{x},t;\vec{0},0)
                              G^{(dd)cc^{\prime}}(\vec{x},t;\vec{0},0)
                                                           \right] \right.
                                              \nonumber \\
         & \qquad \quad + \left.
    \mbox{tr}_D \left[ \Gamma G^{(dd)aa^{\prime}}(\vec{x},t;\vec{0},0)
                       \tilde{G}^{(ss)bb^{\prime}}(\vec{x},t;\vec{0},0)
                              G^{(dd)cc^{\prime}}(\vec{x},t;\vec{0},0) \right]
                         \right\rangle \,,
\label{Sigma_Sigma_gory_detail}
\end{eqnarray}
and
\begin{eqnarray}
   \lefteqn{C_{\lambda\,NN}(t) =
            \sum_{\vec{x}} e^{-i(\vec{p}+\vec{q})\cdot\vec{x}} \epsilon_{abc} 
                                      \epsilon_{a^\prime b^\prime c^\prime} }
         &                                   \nonumber \\
         &                \left\langle
    \mbox{tr}_D \left[ \Gamma G^{(dd)aa^{\prime}}(\vec{x},t;\vec{0},0) \right]
    \mbox{tr}_D \left[ \tilde{G}^{(uu)bb^{\prime}}(\vec{x},t;\vec{0},0)
                              G^{(dd)cc^{\prime}}(\vec{x},t;\vec{0},0)
                                                           \right] \right.
                                              \nonumber \\
         & \qquad \quad + \left.
    \mbox{tr}_D \left[ \Gamma G^{(dd)aa^{\prime}}(\vec{x},t;\vec{0},0)
                       \tilde{G}^{(uu)bb^{\prime}}(\vec{x},t;\vec{0},0)
                              G^{(dd)cc^{\prime}}(\vec{x},t;\vec{0},0) \right]
                         \right\rangle \,,
\label{NN_gory_detail}
\end{eqnarray}
where we have defined a tilde by $\tilde{X} = (C\gamma_5 X \gamma_5)^{T_D}$.
For the off-diagonal correlation functions we have
\begin{eqnarray}
   \lefteqn{C_{\lambda\,\Sigma N}(t) =
              \sum_{\vec{x}} e^{-i\vec{p}\cdot\vec{x}} 
                           \epsilon^{abc} \epsilon^{a^\prime b^\prime c^\prime}}
         & &                                              \nonumber \\
         & &   \left\langle
    {\rm tr}_D \left[ \Gamma G^{(dd)aa^{\prime}}(\vec{x},t;\vec{0},0) \right]
    {\rm tr}_D \left[ \tilde{G}^{(su)bb^{\prime}}(\vec{x},t;\vec{0},0)
                              G^{(dd)cc^{\prime}}(\vec{x},t;\vec{0},0)
                                                           \right] \right.
                                         \label{Sigma_N_gory_detail} \\
         & & \qquad \quad + \left.
    {\rm tr}_D \left[ \Gamma G^{(dd)aa^{\prime}}(\vec{x},t;\vec{0},0)
                       \tilde{G}^{(su)bb^{\prime}}(\vec{x},t;\vec{0},0)
                              G^{(dd)cc^{\prime}}(\vec{x},t;\vec{0},0) \right]
                         \right\rangle \,,
                                                          \nonumber
\end{eqnarray}
and similarly
\begin{eqnarray}
   \lefteqn{C_{\lambda\,N\Sigma}(t) =
              \sum_{\vec{x}} e^{-i(\vec{p}+\vec{q})\cdot\vec{x}} 
                           \epsilon^{abc} \epsilon^{a^\prime b^\prime c^\prime}}
         & &                                             \nonumber \\
         & &   \left\langle
    {\rm tr}_D \left[ \Gamma G^{(dd)aa^{\prime}}(\vec{x},t;\vec{0},0) \right]
    {\rm tr}_D \left[ \tilde{G}^{(us)bb^{\prime}}(\vec{x},t;\vec{0},0)
                              G^{(dd)cc^{\prime}}(\vec{x},t;\vec{0},0)
                                                           \right] \right.
                                       \label{N_Sigma_gory_detail} \\
         & & \qquad \quad + \left.
    {\rm tr}_D \left[ \Gamma G^{(dd)aa^{\prime}}(\vec{x},t;\vec{0},0)
                       \tilde{G}^{(us)bb^{\prime}}(\vec{x},t;\vec{0},0)
                              G^{(dd)cc^{\prime}}(\vec{x},t;\vec{0},0) \right]
                         \right\rangle \,.
                                                         \nonumber
\end{eqnarray}
For simplicity we have taken the source for the Green's functions at 
$(\vec{0},0)$. For the more general smeared sources considered here
we have
\begin{eqnarray}
   \sum_{\vec{x}} e^{-i\vec{p}\cdot\vec{x}} \ldots G(\vec{x},t;\vec{0},0) \ldots
   \to 
   \sum_{\vec{x}_0} f(\vec{x_0}) \sum_{\vec{x}} e^{-i\vec{p}\cdot(\vec{x}-\vec{x}_0)} 
                                    \ldots G(\vec{x},t;\vec{x}_0,0) \ldots\,.
\end{eqnarray}


\section{The fermion matrix inversion}
\label{fermion_inversion}


We give here some more details of the procedure described in 
section~\ref{inversion}.


\subsection{General}


To invert $\cal M$ in general we have
\begin{eqnarray}
   \left( \begin{array}{cc}
             A & C \\
             D & B 
          \end{array}
   \right)
   =  
   \left( \begin{array}{cc}
             A & 0 \\
             D & I
          \end{array}
   \right)
   \left( \begin{array}{cc}
             I & A^{-1}C \\
             D & B-DA^{-1}C
          \end{array}
   \right) \,,
\label{M_splitting}
\end{eqnarray}
which gives
\begin{eqnarray}
   \left( \begin{array}{cc}
             A & C \\
             D & B 
          \end{array}
   \right)^{-1}
      &=& \left( \begin{array}{rr}
                    (A - CB^{-1}D)^{-1}         &  - A^{-1}C(B-DA^{-1}C)^{-1}  \\
                    -B^{-1}D(A - CB^{-1}D)^{-1}  &  (B-DA^{-1}C)^{-1}
                 \end{array}
          \right) \,.
\end{eqnarray}
Equivalent forms, as can be seen by expanding the off-diagonal
elements as a power series, is to re-write them as
\begin{eqnarray}
   B^{-1}D(A-CB^{-1}D)^{-1} &=& (B-DA^{-1}C)^{-1}DA^{-1} \,,
                                                             \nonumber  \\
   A^{-1}C(B-DA^{-1}C)^{-1} &=& (A-CB^{-1}D)^{-1}CB^{-1} \,.
\end{eqnarray}
(Other variations are possible.) Note that we never need that 
$C^{-1}$, $D^{-1}$ exist.


\subsection{Specific}


Thus here we have
\begin{eqnarray}
   A \to D_u\,, \quad B \to D_s\,, \quad 
   C \to -\lambda {\cal T} \,, \quad 
               D \to -\lambda \gamma_5{\cal T}^\dagger\gamma_5 \,,
\end{eqnarray}
giving
\begin{eqnarray}
   \lefteqn{ {\cal M}^{-1} }
      & &                                             \nonumber       \\
      &=& \left( \begin{array}{cc}
                   ({\cal M}^{-1})_{uu}  & ({\cal M}^{-1})_{us}   \\
                   ({\cal M}^{-1})_{su}  & ({\cal M}^{-1})_{ss}   \\
                 \end{array}
          \right)
                                                                      \\
      &=& \left( \begin{array}{rr}
                    (D_u - \lambda^2 {\cal T} D_s^{-1} 
                              \gamma_5 {\cal T}^\dagger \gamma_5)^{-1}   &  
                    \lambda D_u^{-1}{\cal T}
                    (D_s-\lambda^2\gamma_5{\cal T}^\dagger\gamma_5 
                              D_u^{-1}{\cal T})^{-1}  \\
                    \lambda D_s^{-1}\gamma_5{\cal T}^\dagger\gamma_5
                    (D_u - \lambda^2 {\cal T} D_s^{-1} 
                              \gamma_5 {\cal T}^\dagger \gamma_5)^{-1}   &  
                    (D_s-\lambda^2\gamma_5{\cal T}^\dagger\gamma_5 
                              D_u^{-1}{\cal T})^{-1}  \\
                 \end{array}
          \right) \,.
                                                             \nonumber 
\end{eqnarray}
Hence we have, upon re-writing
\begin{eqnarray}
   G^{(uu)} &=& (1 - \lambda^2  D_u^{-1}{\cal T}
                    D_s^{-1}\gamma_5{\cal T}^\dagger\gamma_5)^{-1} D_u^{-1} \,,
                                                          \nonumber  \\
   G^{(ss)} &=& (1- \lambda^2 D_s^{-1}\gamma_5{\cal T}^\dagger\gamma_5 D_u^{-1} 
                             {\cal T})^{-1} D_s^{-1} \,,
\end{eqnarray}
and
\begin{eqnarray}
   G^{(us)} &=& \lambda D_u^{-1}{\cal T} G^{(ss)} \,,
                                                          \nonumber  \\
   G^{(su)} &=& \lambda D_s^{-1}\gamma_5{\cal T}^\dagger\gamma_5 G^{(uu)} \,,
\end{eqnarray}
as given in the main text.
Note that, as built in, we have
\begin{eqnarray}
   \gamma_5 G^{(su)\,\dagger} \gamma_5 = G^{(us)} \,.
\end{eqnarray}




\begin{thebibliography}{99}

\bibitem{Aoki:2021kgd}
   Y.~Aoki et al.,
   {\it FLAG Review 2021},
   \href{https://doi.org/10.1140/epjc/s10052-022-10536-1}
        {Eur. Phys. J. C \textbf{82} (2022) 869},
   [\href{http://arXiv.org/abs/arXiv:2111.09849}{arXiv:2111.09849 [hep-lat]}].

\bibitem{Meyer:2022mix}
   A.~S.~Meyer, A.~Walker-Loud and C.~Wilkinson,
   {\it Status of Lattice QCD Determination of Nucleon Form Factors and 
   their Relevance for the Few-GeV Neutrino Program},
   \href{https://doi.org/10.1146/annurev-nucl-010622-120608}
        {Ann. Rev. Nucl. Part. Sci. 72 (2022) 205},
   [\href{http://arXiv.org/abs/arXiv:2201.01839}{arXiv:2201.01839 [hep-lat]}].

\bibitem{Djukanovic:2021qxp}
   D.~Djukanovic,
   {\it Recent progress on nucleon form factors},
   \href{https://doi.org/10.22323/1.396.0009}
        {PoS \textbf{LATTICE2021} (2022) 009},
   \href{http://arXiv.org/abs/arXiv:2112.00128}{arXiv:2112.00128 [hep-lat]}.

\bibitem{Can:2021ehb}
   K.~U.~Can,
   {\it Lattice QCD study of the elastic and transition form factors of 
   charmed baryons},
   \href{https://doi.org/10.1142/S0217751X21300131}
        {Int. J. Mod. Phys. A \textbf{36} (2021) 2130013},
   [\href{http://arXiv.org/abs/arXiv:2107.13159}{arXiv:2107.13159 [hep-lat]}].

\bibitem{Constantinou:2020hdm}
   M.~Constantinou et al., 
   {\it Parton distributions and lattice-QCD calculations: 
   Toward 3D structure},
   \href{https://doi.org/10.1016/j.ppnp.2021.103908}
        {Prog. Part. Nucl. Phys. \textbf{121} (2021) 103908},
   [\href{http://arXiv.org/abs/arXiv:2006.08636}{arXiv:2006.08636 [hep-ph]}].

\bibitem{Cichy:2018mum}
   K.~Cichy and M.~Constantinou,
   {\it A guide to light-cone PDFs from Lattice QCD: an overview of approaches,
   techniques and results},
   \href{https://doi.org/10.1155/2019/3036904}
        {Adv. High Energy Phys. \textbf{2019} (2019) 3036904},
   [\href{http://arXiv.org/abs/1811.07248}{arXiv:1811.07248 [hep-lat]}].

\bibitem{Gambino:2022dvu}
   P.~Gambino, S.~Hashimoto, S.~M\"achler, M.~Panero, F.~Sanfilippo, 
   S.~Simula, A.~Smecca and N.~Tantalo,
   {\it Lattice QCD study of inclusive semileptonic decays of heavy mesons},
   \href{https://doi.org/10.1007/JHEP07(2022)083}
        {J. High Energy Phys. \textbf{07} (2022) 083)},
   [\href{http://arXiv.org/abs/2203.11762}{arXiv:2203.11762 [hep-lat]}].

\bibitem{Chambers:2017dov}
   A.~J.~Chambers, R.~Horsley, Y.~Nakamura, H.~Perlt, P.~E.~L.~Rakow, 
   G.~Schierholz, A.~Schiller, K.~Somfleth, R.~D.~Young and J.~M.~Zanotti,
   {\it Nucleon Structure Functions from Operator Product Expansion on 
   the Lattice}, [QCDSF Collaboration],
   \href{https://doi.org/10.1103/PhysRevLett.118.242001}
        {Phys. Rev. Lett. \textbf{118} (2017) 242001},
   [\href{http://arXiv.org/abs/1703.01153}{arXiv:1703.01153 [hep-lat]}].

\bibitem{Liu:2016djw}
   K.~F.~Liu,
   {\it Parton Distribution Function from the Hadronic Tensor on the Lattice},
   \href{https://doi.org/10.22323/1.251.0115}
        {PoS \textbf{LATTICE2015} (2016) 115},
   \href{http://arXiv.org/abs/1603.07352}{arXiv:1603.07352 [hep-ph]}.

\bibitem{Seng:2021nar}
   C.~Y.~Seng, D.~Galviz, W.~J.~Marciano and U.~G.~Mei{\ss}ner,
   {\it Update on $|V_{us}|$ and $|V_{us}/V_{ud}|$ from semileptonic kaon 
   and pion decays},
   \href{https://doi.org/10.1103/PhysRevD.105.013005}
        {Phys. Rev. D \textbf{105} (2022) 013005},
   [\href{http://arXiv.org/abs/2107.14708}{arXiv:2107.14708 [hep-ph]}].

\bibitem{Gottlieb:2020zsa}
   S.~Gottlieb,
   {\it Lattice QCD Impact on Determination of CKM Matrix: Status and 
   Prospects},
   \href{https://doi.org/10.22323/1.363.0275}
        {PoS \textbf{LATTICE2019} (2020) 275},
   \href{http://arXiv.org/abs/2002.09013}{arXiv:2002.09013 [hep-lat]}.

\bibitem{Davoudi:2020ngi}
   Z.~Davoudi, W.~Detmold, K.~Orginos, A.~Parre\~no, M.~J.~Savage, 
   P.~Shanahan and M.~L.~Wagman,
   {\it Nuclear matrix elements from lattice QCD for electroweak and 
   beyond-Standard-Model processes},
   \href{https://doi.org/10.1016/j.physrep.2020.10.004}
        {Phys. Rept. \textbf{900} (2021) 1},
   [\href{http://arXiv.org/abs/2008.11160}{arXiv:2008.11160 [hep-lat]}].

\bibitem{Severijns:2006dr}
   N.~Severijns, M.~Beck and O.~Naviliat-Cuncic,
   {\it Tests of the standard electroweak model in beta decay},
   \href{https://doi.org/10.1103/RevModPhys.78.991}
        {Rev. Mod. Phys. \textbf{78} (2006) 991},
   [\href{http://arXiv.org/abs/nucl-ex/0605029}
         {arXiv:nucl-ex/0605029 [nucl-ex]}].

\bibitem{Gonzalez-Alonso:2018omy}
   M.~Gonz\'alez-Alonso, O.~Naviliat-Cuncic and N.~Severijns,
   {\it New physics searches in nuclear and neutron $\beta$ decay},
   \href{https://doi.org/10.1016/j.ppnp.2018.08.002}
        {Prog. Part. Nucl. Phys. \textbf{104} (2019) 165},
   [\href{http://arXiv.org/abs/1803.08732}{arXiv:1803.08732 [hep-ph]}].

\bibitem{Cirigliano:2020yhp}
   V.~Cirigliano, W.~Detmold, A.~Nicholson and P.~Shanahan,
   {\it Lattice QCD Inputs for Nuclear Double Beta Decay},
   \href{https://doi.org/10.1016/j.ppnp.2020.103771} 
        {Prog. Part. Nucl. Phys. \textbf{112} (2020) 103771}
   [\href{http://arXiv.org/abs/2003.08493}{arXiv:2003.08493 [nucl-th]}].

\bibitem{Smail:2023eyk}
   R.~E.~Smail, M.~Batelaan, R.~Horsley, Y.~Nakamura, H.~Perlt, 
   D.~Pleiter, P.~E.~L.~Rakow, G.~Schierholz, H.~St\"uben, R.~D.~Young
   and J.~M.~Zanotti, [QCDSF/UKQCD/CSSM Collaboration],
   {\it Constraining beyond the Standard Model nucleon isovector charges},
   [\href{http://arXiv.org/abs/2304.02866}
         {arXiv:2304.02866 [hep-lat]}].

\bibitem{QCDSF:2012mkm}
    R.~Horsley, R.~Millo, Y.~Nakamura, H.~Perlt, D.~Pleiter,
    P.~E.~L.~Rakow, G.~Schierholz, A.~Schiller, F.~Winter 
    and J.~M.~ Zanotti,  [QCDSF-UKQCD Collaboration],
    {\it A Lattice Study of the Glue in the Nucleon},
    \href{https://doi.org/10.1016/j.physletb.2012.07.004}
         {Phys. Lett. B \textbf{714} (2012) 312},
    [\href{http://arXiv.org/abs/1205.6410}
          {arXiv:1205.6410 [hep-lat]}].

\bibitem{CSSM:2014uyt}
   A.~J.~Chambers, R.~Horsley, Y.~Nakamura, H.~Perlt, D.~Pleiter,
   P.~E.~L.~Rakow, G.~Schierholz, A.~Schiller, H.~St\"uben,
   R.~D.~Young and J.~M.~Zanotti, [CSSM and QCDSF/UKQCD Collaborations],
   {\it Feynman--Hellmann approach to the spin structure of hadrons},
   \href{https://doi.org/10.1103/10.1103/PhysRevD.90.014510}
   {Phys. Rev. D \textbf{90} (2014) 014510},
   [\href{http://arXiv.org/abs/1405.3019}{arXiv:1405.3019 [hep-lat]}].

\bibitem{Chambers:2015bka}
   A.~J.~Chambers, R.~Horsley, Y.~Nakamura, H.~Perlt, D.~Pleiter, 
   P.~E.~L.~Rakow, G.~Schierholz, A.~Schiller, H.~St\"uben,
   R.~D.~Young and J.~M.~Zanotti, [CSSM and QCDSF/UKQCD Collaborations],
   {\it Disconnected contributions to the spin of the nucleon},
   \href{https://doi.org/10.1103/PhysRevD.92.114517}
        {Phys. Rev. D \textbf{92} (2015) 114517},
   [\href{http://arXiv.org/abs/1508.06856}{arXiv:1508.06856 [hep-lat]}].

\bibitem{Bouchard:2016heu}
   C.~Bouchard, C.~C.~Chang, T.~Kurth, K.~Orginos and A.~Walker-Loud,
   {\it On the Feynman--Hellmann Theorem in Quantum Field Theory and 
   the Calculation of Matrix Elements},
   \href{https://doi.org/10.1103/PhysRevD.96.014504}
        {Phys. Rev. D \textbf{96} (2017) 014504},
   [\href{http://arXiv.org/abs/1612.06963}{arXiv:1612.06963 [hep-lat]}].

\bibitem{Chambers:2017tuf}
   A.~J.~Chambers, J.~Dragos, R.~Horsley, Y.~Nakamura, H.~Perlt, D.~Pleiter, 
   P.~E.~L.~Rakow, G.~Schierholz, A.~Schiller, K.~Somfleth, H.~St\"uben, 
   R.~D.~Young and J.~Zanotti [QCDSF/UKQCD/CSSM Collaborations],
   {\it Electromagnetic form factors at large momenta from lattice QCD},
   \href{https://doi.org/10.1103/PhysRevD.96.114509}
        {Phys. Rev. D \textbf{96} (2017) 114509},
   [\href{http://arXiv.org/abs/1702.01513}{arXiv:1702.01513 [hep-lat]}].

\bibitem{Luscher:1990ck}
   M.~L\"uscher and U.~Wolff,
   {\it How to Calculate the Elastic Scattering Matrix in Two-dimensional 
   Quantum Field Theories by Numerical Simulation},
   \href{https://doi.org/10.1016/0550-3213(90)90540-T}
        {Nucl. Phys. B \textbf{339} (1990) 222}.

\bibitem{Blossier:2009kd}
   B.~Blossier, M.~Della~Morte, G.~von~Hippel, T.~Mendes and R.~Sommer,
   [ALPHA Collaboration],
   {\it On the generalized eigenvalue method for energies and matrix 
   elements in lattice field theory},
   \href{https://doi.org/10.1088/1126-6708/2009/04/094}
        {JHEP \textbf{04} (2009) 094},
   [\href{http://arXiv.org/abs/0902.1265}{arXiv:0902.1265 [hep-lat]}].

\bibitem{Owen:2012ts}
  B.~J.~Owen, J.~Dragos, W.~Kamleh, D.~B.~Leinweber, M.~S.~Mahbub, 
  B.~J.~Menadue and J.~M.~Zanotti,
  {\it Variational Approach to the Calculation of $g_A$},
  \href{https://doi.org/10.1016/j.physletb.2013.04.063}
       {Phys. Lett. B \textbf{723} (2013) 217},
  [\href{http://arXiv.org/abs/1212.4668}{arXiv:1212.4668 [hep-lat]}].

\bibitem{Guadagnoli:2006gj}
   D.~Guadagnoli, V.~Lubicz, M.~Papinutto and S.~Simula,
   {\it First Lattice QCD Study of the $\Sigma^- \to n$ Axial and Vector 
   Form Factors with $SU(3)$ Breaking Corrections},
   \href{https://doi.org/10.1016/j.nuclphysb.2006.10.022}
        {Nucl. Phys. B \textbf{761} (2007) 63},
   [\href{http://arXiv.org/abs/hep-ph/0606181}{arXiv:hep-ph/0606181}].

\bibitem{Shanahan:2015dka}
   P.~E.~Shanahan, A.~N.~Cooke, R.~Horsley, Y.~Nakamura, P.~E.~L.~Rakow, 
   G.~Schierholz, A.~W.~Thomas, R.~D.~Young and J.~M.~Zanotti,
   {\it SU(3) breaking in hyperon transition vector form factors},
   \href{https://doi.org/10.1103/PhysRevD.92.074029}
        {Phys. Rev. D \textbf{92} (2015) 074029},
   [\href{http://arXiv.org/abs/1508.06923}{arXiv:1508.06923 [nucl-th]}].

\bibitem{Sasaki:2017jue}
   S.~Sasaki,
   {\it Continuum limit of hyperon vector coupling $f_1(0)$ from $2+1$
   flavor domain wall QCD},
   \href{https://doi.org/10.1103/PhysRevD.96.074509}
        {Phys. Rev. D \textbf{96} (2017) 074509},
   [\href{http://arXiv.org/abs/1708.04008}{arXiv:1708.04008 [hep-lat]}].

\bibitem{Bickerton:2021yzn}
   J.~M.~Bickerton, A.~N.~Cooke, R.~Horsley, Y.~Nakamura, H.~Perlt, 
   D.~Pleiter, P.~E.~L.~Rakow, G.~Schierholz, H.~St\"uben, R.~D.~Young, 
   and J.~M.~Zanotti,
   [QCDSF-UKQCD-CSSM Collaboration],
   {\it Patterns of flavour symmetry breaking in hadron matrix elements 
   involving $u$, $d$ and $s$ quarks},
   \href{https://doi.org/10.22323/1.396.0490}
        {PoS \textbf{LATTICE2021} (2022) 490},
   \href{http://arXiv.org/abs/2112.04445}{arXiv:2112.04445 [hep-lat]}.

\bibitem{Horsley:2022ouc}
   R.~Horsley, M.~Batelaan, K.~U.~Can, Y.~Nakamura, H.~Perlt, 
   P.~E.~L.~Rakow, G.~Schierholz, H.~St\"uben, R.~D.~Young and J.~M.~Zanotti
   [QCDSF-UKQCD-CSSM Collaborations],
   {\it Quasi-degenerate baryon energy states, the Feynman-Hellmann theorem 
   and transition matrix elements},
   \href{https://doi.org/10.22323/1.430.0412}
         {PoS \textbf{LATTICE2022} (2022) 412},
   \href{http://arXiv.org/abs/2302.04911}{arXiv:2302.04911 [hep-lat]}.

\bibitem{Luscher:1984is}
   M.~L\"uscher and P.~Weisz,
   {\it Definition and General Properties of the Transfer Matrix in 
   Continuum Limit Improved Lattice Gauge Theories},
   \href{https://doi.org/10.1016/0550-3213(84)90270-0}
        {Nucl. Phys. B \textbf{240} (1984) 349}.

\bibitem{Can:2020sxc}
   K.~U.~Can, A.~Hannaford-Gunn, R.~Horsley, Y.~Nakamura, H.~Perlt, 
   P.~E.~L.~Rakow, G.~Schierholz, K.~Y.~Somfleth, H.~St\"uben, 
   R.~D.~Young and J.~M.~Zanotti [QCDSF/UKQCD/CSSM Collaborations],
   {\it Lattice evaluation of the Compton amplitude employing the 
   Feynman--Hellmann theorem},
   \href{https://doi.org/10.1103/PhysRevD.102.114505}
        {Phys. Rev. D \textbf{102} (2020) 114505},
   [\href{http://arXiv.org/abs/2007.01523}{arXiv:2007.01523 [hep-lat]}].

\bibitem{Best:1997qp}
   C.~Best, M.~G\"ockeler, R.~Horsley, E.~M.~Ilgenfritz, H.~Perlt, 
   P.~E.~L.~Rakow, A.~Sch\"afer, G.~Schierholz, A.~Schiller and S.~Schramm,
   {\it Pion and rho structure functions from lattice QCD},
   \href{https://doi.org/10.1103/PhysRevD.56.2743}
        {Phys. Rev. D \textbf{56} (1997) 2743},
   [\href{http://arXiv.org/abs/9703014}{arXiv:hep-lat/9703014 [hep-lat]}].

\bibitem{Can:2022chd}
   K.~U.~Can [QCDSF/UKQCD Collaboration],
   {\it The Compton amplitude and nucleon structure functions},
   \href{https://doi.org/10.22323/1.430.0237}
        {PoS \textbf{LATTICE2022} (2023) 237},
   \href{http://arXiv.org/abs/2212.09197}
        {arXiv:2212.09197 [hep-lat]}.

\bibitem{Can:2022rgi}
   K.~U.~Can, A.~Hannaford-Gunn, R.~Horsley, Y.~Nakamura, H.~Perlt, 
   P.~E.~L.~Rakow, E.~Sankey, G.~Schierholz, H.~St\"uben,
   R.~D.~Young and J.~M.~Zanotti, [QCDSF-UKQCD Collaboration],
   {\it The Compton Amplitude, lattice QCD and the Feynman--Hellmann approach},
   \href{https://doi.org/10.21468/SciPostPhysProc.6.003}
        {SciPost Phys. Proc. \textbf{6} (2022) 003},  
   \href{http://arXiv.org/2201.08367}{arXiv:2201.08367 [hep-lat]}.

\bibitem{Bickerton:2019nyz}
   J.~M.~Bickerton, R.~Horsley, Y.~Nakamura, H.~Perlt, D.~Pleiter, 
   P.~E.~L.~Rakow, G.~Schierholz, H.~St\"uben, R.~D.~Young and J.~M.~Zanotti
   [QCDSF-UKQCD-CSSM Collaboration],
   {\it Patterns of flavor symmetry breaking in hadron matrix elements 
   involving $u$ , $d$ , and $s$ quarks},
   \href{https://doi.org/10.1103/PhysRevD.100.114516}
        {Phys. Rev. D \textbf{100} (2019) 114516},
   [\href{http://arXiv.org/abs/1909.02521}{arXiv:1909.02521 [hep-lat]}].

\bibitem{Cundy:2009yy}
   N.~Cundy, M.~G\"ockeler, R.~Horsley, T.~Kaltenbrunner, A.~D.~Kennedy, 
   Y.~Nakamura, H.~Perlt, D.~Pleiter, P.~E.~L.~Rakow, A.~Sch\"afer, 
   G.~Schierholz, A.~Schiller, H.~St\"uben and J.~M.~Zanotti,
   [QCDSF-UKQCD Collaboration],
   {\it Non-perturbative improvement of stout-smeared three flavour 
   clover fermions},
   \href{https://doi.org/10.1103/PhysRevD.79.094507}
        {Phys. Rev. D \textbf{79} (2009) 094507},
   [\href{http://arXiv.org/abs/0901.3302}{arXiv:0901.3302 [hep-lat]}].

\bibitem{Bietenholz:2011qq}
   W.~Bietenholz, V.~Bornyakov, M.~G\"ockeler, R.~Horsley, W.~G.~Lockhart,
   Y.~Nakamura, H.~Perlt, D.~Pleiter, P.~E.~L.~Rakow, G.~Schierholz,
   A.~Schiller, T.~Streuer, H.~St\"uben, F.~Winter and J.~M.~Zanotti,
   [QCDSF-UKQCD Collaboration],
   {\it Flavour blindness and patterns of flavour symmetry breaking in 
   lattice simulations of up, down and strange quarks},
   \href{https://doi.org/10.1103/PhysRevD.84.054509}
        {Phys. Rev. D {\textbf 84} (2011) 054509},
   [\href{http://arXiv.org/abs/1102.5300}{arXiv:1102.5300 [hep-lat]}].

\bibitem{Bedaque:2004kc}
   P.~F.~Bedaque,
   {\it Aharonov-Bohm effect and nucleon nucleon phase shifts on the lattice},
   \href{https://doi.org/10.1016/j.physletb.2004.04.045}
        {Phys. Lett. B \textbf{593} (2004) 82},
   [\href{http://arXiv.org/abs/nucl-th/0402051}
         {arXiv:nucl-th/0402051 [nucl-th]}].

\bibitem{Sachrajda:2004mi}
   C.~T.~Sachrajda and G.~Villadoro,
   {\it Twisted boundary conditions in lattice simulations},
   \href{https://doi.org/10.1016/j.physletb.2005.01.033}
   {Phys. Lett. B \textbf{609} (2005) 73},
   [\href{http://arXiv.org/abs/hep-lat/0411033}
         {arXiv:hep-lat/0411033 [hep-lat]}].

\bibitem{Bedaque:2004ax}
  P.~F.~Bedaque and J.~W.~Chen,
  {\it Twisted valence quarks and hadron interactions on the lattice},
  \href{https://doi.org/10.1016/j.physletb.2005.04.045}
       {Phys. Lett. B \textbf{616} (2005) 208},
  [\href{http://arXiv.org/abs/hep-lat/0412023}
        {arXiv:hep-lat/0412023 [hep-lat]}].

\bibitem{Flynn:2007ess}
   J.~M.~Flynn, A.~J\"uttner, C.~T.~Sachrajda, P.~A.~Boyle and J.~M.~Zanotti,
   {\it Hadronic form factors in Lattice QCD at small and vanishing momentum 
        transfer},
   \href{https://doi.org/10.1088/1126-6708/2007/05/016}
        {JHEP \textbf{05} (2007) 016},
   [\href{http://arXiv.org/abs/hep-lat/0703005}
         {arXiv:hep-lat/0703005 [hep-lat]}].

\bibitem{Boyle:2008yd}
   P.~A.~Boyle, J.~M.~Flynn, A.~J\"uttner, C.~Kelly, H.~P.~de~Lima, 
   C.~M.~Maynard, C.~T.~Sachrajda and J.~M.~Zanotti,
   [UKQCD Collaboration],
   {\it The Pion's electromagnetic form-factor at small momentum transfer
        in full lattice QCD},
   \href{https://doi.org/10.1088/1126-6708/2008/07/112}
        {JHEP \textbf{07} (2008) 112},
   [\href{http://arXiv.org/abs/0804.3971}{arXiv:0804.3971 [hep-lat]}].

\bibitem{Yoon:2016dij}
   B.~Yoon, R.~Gupta, T.~Bhattacharya, M.~Engelhardt, J.~Green, 
   B.~Jo\'o, H.~W.~Lin, J.~Negele, K.~Orginos and A.~Pochinsky, 
   D.~Richards, S.~Syritsyn and F.~Winter,
   [NME Collaboration],
   {\it Controlling Excited-State Contamination in Nucleon Matrix Elements},
   \href{https://doi:10.1103/PhysRevD.93.114506}
        {Phys. Rev. D {\textbf 93} (2016) 114506},
   [\href{http://arXiv:1602.07737}{arXiv:1602.07737 [hep-lat]}].

\bibitem{Gockeler:2003ay}
    M.~G\"ockeler, T.~R.~Hemmert, R. Horsley, D.~Pleiter, P.~E.~L.~Rakow,
    A.~Sch\"afer and G. Schierholz,
    [QCDSF Collaboration],
    {\it Nucleon electromagnetic form-factors on the lattice and in 
         chiral effective field theory},
    \href{https://doi:10.1103/PhysRevD.71.034508}
         {Phys. Rev. D \textbf{71} (2005) 034508},
    [\href{http://arXiv:hep-lat/0303019}{arXiv:hep-lat/0303019 [hep-lat]}].

\bibitem{Wightman:2001zj}
   A.~S.~Wightman,
   {\it Three formulas for the off-diagonal density matrix of a Dirac spinor
        with an application},
   \href{https://doi.org/10.1063/1.1332784}
        {J. Math. Phys. 42, \textbf{674} (2001) 674}.

\bibitem{Pal:2007dc}
   P.~B.~Pal,
   {\it Representation-independent manipulations with Dirac spinors},
   [\href{http://arXiv.org/abs/physics/0703214}
         {arXiv:physics/0703214 [physics.ed-ph]}].

\bibitem{Lorce:2017isp}
   C.~Lorc\'e, 
   {\it New explicit expressions for Dirac bilinears}, 
   \href{https://doi.org/10.1103/PhysRevD.97.016005}
        {Phys. Rev. D \textbf{97} (2018) 016005},
   [\href{http://arXiv.org/abs/1705.08370}{arXiv:1705.08370 [hep-ph]}].

\bibitem{Olpak:2019cth}
   M.~A.~Olpak and A.~Ozpineci,
   {\it On the calculation of covariant expressions for Dirac bilinears},
   \href{https://doi.org/10.1140/epjc/s10052-021-09592-w}
        {Eur. Phys. J. C \textbf{81} (2021) 798},
   [\href{http://arXiv.org/abs/1905.10470}{arXiv:1905.10470 [hep-th]}].

\bibitem{Dragos:2016rtx}
   J.~Dragos, R.~Horsley, W.~Kamleh, D.~B.~Leinweber, Y.~Nakamura, 
   P.~E.~L.~Rakow, G.~Schierholz, R.~D.~Young and J.~M.~Zanotti,
   {\it Nucleon matrix elements using the variational method in lattice QCD},
   \href{https://doi.org/10.1103/PhysRevD.94.074505}
        {Phys. Rev. D \textbf{94} (2016) 074505},
   [\href{http://arXiv.org/abs/1606.03195}{[arXiv:1606.03195 [hep-lat]}].

\bibitem{Haar:2017ubh}
   T.~R.~Haar, Y.~Nakamura and H.~St\"uben,
   {\it An update on the BQCD Hybrid Monte Carlo program},
   \href{https://doi.org/10.1051/epjconf/201817514011}
        {EPJ Web Conf. \textbf{175} (2018) 14011},
   [\href{http://arXiv.org/abs/1711.03836}{arXiv:1711.03836 [hep-lat]}].

\bibitem{Edwards:2004sx}
   R.~G.~Edwards and B.~Jo{\'o},
   {\it The Chroma software system for lattice QCD},
   \href{https://doi.org/10.1016/j.nuclphysbps.2004.11.254}
        {Nucl. Phys. B Proc. Suppl. \textbf{140} (2005) 832},
   \href{http://arXiv.org/abs/hep-lat/0409003}{arXiv:hep-lat/0409003}.

\bibitem{Creutz:1976ch}
   M.~Creutz,
   {\it Gauge Fixing, the Transfer Matrix, and Confinement on a Lattice},
   \href{https://doi.org/10.1103/PhysRevD.15.1128}
        {Phys. Rev. D \textbf{15} (1977) 1128}.

\bibitem{montvay+munster}
   I.~Montvay and G. M\"unster
   {\it Quantum Fields on a Lattice},
   \href{https://doi.org/10.1017/CBO9780511470783}
         {Cambridge University Press 1994}.

\end{thebibliography}
\end{document}